\documentclass[fleqn,usenatbib]{mnras}

\usepackage{newtxtext,newtxmath}
\usepackage[T1]{fontenc}
\usepackage{ae,aecompl}
\usepackage{blindtext}
\usepackage{lipsum}
\usepackage{amsmath}			% advanced maths commands
\usepackage{amssymb}			% extra maths symbols
\usepackage{array}				% array handling
\usepackage{graphicx}			% including figure files
\usepackage[squaren]{SIunits}	% handling of values with units -> command unit
\usepackage{xcolor}				% different text colors

\makeatletter					% make @ useable for table options
\graphicspath{{./plots/}}		% define folder for graphics

\newcolumntype{L}[1]{>{\vspace{-\f@size pt}{\vspace{1pt}}\raggedright\let\newline\\\arraybackslash\hspace{0pt}}p{#1}}	%left alinged table cell with linebreaks
\newcolumntype{R}[1]{>{\vspace{-\f@size pt}{\vspace{1pt}}\raggedleft\let\newline\\\arraybackslash\hspace{0pt}}p{#1}}	%right alinged table cell with linebreaks
\newcommand{\Msun}{{\rm M}_{\sun}}				% solar mass
\newcommand{\Rsun}{{\rm R}_{\sun}}				% solar radius
\newcommand{\yr}{{\rm yr}}						% year
\newcommand{\days}{{\rm d}}						% days
\newcommand{\pc}{{\rm pc}}						% parsec
\newcommand{\mCO}{m_{\rm cobj}}					% mass of a compact object
\newcommand{\mini}{m_{\rm ini}}					% initial mass
\newcommand{\tmerge}{t_{\rm merge}}				% gravitational wave merger delay time
\newcommand{\trel}{t_{\rm rel}}					% relative time
\newcommand{\ratio}[1]{r_{\rm #1}}				% ratio type
\newcommand{\maximum}[1]{#1_{\rm max}{}}		% subscript max
\newcommand{\up}[1]{#1_{\rm up}{}}				% subscript up
\newcommand{\low}[1]{#1_{\rm low}{}}			% subscript low
\newcommand{\OurCode}{{\sc ComBinE}\xspace}		% code name ComBinE
\newcommand{\AppendixRef}[1]{Appendix~\ref{#1}}		% reference to an appendix section (mnras-guide: first small, but usually capital)
\newcommand{\EquationRef}[1]{Equation~\eqref{#1}}	% reference to an equation (mnras-guide: not specified)
\newcommand{\EquationsRef}[1]{Equations~\eqref{#1}}	% reference to several equations
\newcommand{\FigureRef}[1]{Fig.~\ref{#1}}			% reference to a figure (mnras-guide: first capital, but small when referring to a figure of an other paper)
\newcommand{\FiguresRef}[1]{Figs~\ref{#1}}			% reference to several figures
\newcommand{\SectionRef}[1]{Section~\ref{#1}}		% reference to a section (mnras-guide: first small, but usually capital)
\newcommand{\SectionsRef}[1]{Sections~\ref{#1}}		% reference to several sections (mnras-guide: first small, but usually capital)
\newcommand{\TableRef}[1]{Table~\ref{#1}}			% reference to a table (mnras-guide: first capital, but small when referring to a table of an other paper)
\newcommand{\TablesRef}[1]{Tables~\ref{#1}}			% reference to several tables
\newcommand{\product}{\,}							% product sign (small space)
\newcommand{\diff}{\mathop{}\!\mathrm{d}}			% differential d

\title[Binary evolution and LIGO-Virgo rates]{Progenitors of gravitational wave mergers:\\Binary evolution with the stellar grid-based code \OurCode}
\author[M.U. Kruckow et al.]
{Matthias U. Kruckow,$^{1}$\thanks{E-mail: mkruckow@astro.uni-bonn.de}
Thomas M. Tauris,$^{2,1}$
Norbert Langer,$^{1,2}$
Michael Kramer,$^{2}$
\newauthor
Robert G. Izzard$^{3}$
\\
$^{1}$Argelander-Institut f\"ur Astronomie, Universit\"at Bonn, Auf dem H\"ugel 71, 53121 Bonn, Germany\\
$^{2}$Max-Planck-Institut f\"ur Radioastronomie, Auf dem H\"ugel 69, 53121 Bonn, Germany\\
$^{3}$Astrophysics Research Group, Faculty of Engineering and Physical Sciences, University of Surrey, Guildford, Surrey, GU2 7XH,\\ United Kingdom
}

\date{Accepted 2017. Received 2017; in original form ZZZ}

\pubyear{2017}

\begin{document}
\label{firstpage}
\pagerange{\pageref{firstpage}--\pageref{lastpage}}
\maketitle

% Abstract of the paper
\begin{abstract}
  The first gravitational wave detections of mergers between black holes and neutron stars represent a remarkable new regime of high-energy transient astrophysics. The signals observed with LIGO-Virgo detectors come from mergers of extreme physical objects which are the end products of stellar evolution in close binary systems. To better understand their origin and merger rates, we have performed binary population syntheses at different metallicities using the new grid-based binary population synthesis code \OurCode. Starting from newborn pairs of stars, we follow their evolution including mass loss, mass transfer and accretion, common envelopes and supernova explosions. We apply the binding energies of common envelopes based on dense grids of detailed stellar structure models, make use of improved investigations of the subsequent Case~BB Roche-lobe overflow and scale supernova kicks according to the stripping of the exploding stars. We demonstrate that all the double black hole mergers, GW150914, LVT151012, GW151226, GW170104, GW170608 and GW170814, as well as the double neutron star merger GW170817, are accounted for in our models in the appropriate metallicity regime. Our binary interaction parameters are calibrated to match the accurately determined properties of Galactic double neutron star systems, and we discuss their masses and types of supernova origin. Using our default values for the input physics parameters, we find a double neutron star merger rate of $\unit{3.0}{\mega\yr^{-1}}$ for Milky-Way equivalent galaxies. Our upper limit to the merger-rate density of double neutron stars is \mbox{$R\simeq\unit{400}{\yr^{-1}\usk\giga{\rm pc}^{-3}}$} in the local Universe (\mbox{$z=0$}).
\end{abstract}


%%%%%%%%%%%%%%%%% BODY OF PAPER %%%%%%%%%%%%%%%%%%
\section{Introduction}\label{sec:introduction}
The evolution of massive binary stars and subsequent production of pairs of compact objects in tight orbits plays a central role in many areas of modern astrophysics, including: the origin of different types of supernova (SN) explosions \citep{ywl10}, accretion processes in X-ray binaries \citep{lv06} and the formation of radio millisecond pulsars \citep[MSPs,][]{bv91}. Furthermore, the final outcome of massive binary evolution may in some cases be fatal collisions between neutron stars (NSs) and/or black holes (BHs). These events give rise to powerful emission of gravitational waves (GWs), as recently detected by advanced LIGO and Virgo \citep{aaa+16,aaa+16b,aaa+17,aaa+17b,aaa+17c}. Such NS mergers also lead to chemical enrichment of the interstellar medium by heavy $r$-process elements which decay and power an electromagnetic transient termed `macronova' or `kilonova' \citep[e.g.][]{ls74,kul05,mmd+10,ros15,jbp+15,aaa+17c}. In addition, they can also produce short gamma-ray bursts \citep[GRBs,][]{elps89,ber14,aaa+17c}.

Double compact objects (DCOs) -- in the following defined as binary systems with a pair of NSs, BHs or one of each type -- represent an end point of massive binary stellar evolution. According to their various formation channels, the progenitor systems have survived two SN explosions and multiple stages of mass transfer, often with one or more common-envelope (CE) episodes \citep[e.g.][]{vt03,tv06,bkr+08,dbf+12,mlp+16,md16b,bhbo16,tkf+17}. Their observed properties are fossil records of their past evolutionary history and DCOs can therefore be used as key probes of binary stellar astrophysics. For a recent review of the formation of double NS systems, see \citet{tkf+17}. For general investigations and reviews of massive star evolution in pre-SN binaries, see e.g. \citet{pjh92,wl99,lan12,di17}.

DCOs hosting a radio pulsar are also of special interest since their ultra-stable spin-down nature allows for precise timing of their motion in relativistic orbits and thereby tests of gravitational theories in the strong-field regime \citep{dt92,ksm+06,wex14}. Finally, observations of NS binaries help to constrain the long-sought-after equation-of-state (EoS) of nuclear matter at high densities \citep{afw+13,of16}.

%%%%%%%%%%%%%%%%%%%%%%%%%%%%%%%%%%%%%%%%%%%%%%%%%%
\subsection{Double compact object merger rates}\label{sec:merger_rates}
The Galactic formation and merger rate of DCO systems has been estimated for almost four decades \citep[e.g.][]{cvs79}. As will be described below, the standard formation scenario of DCO binaries involves a number of highly uncertain aspects of binary interactions. The main uncertainties include, in particular, the treatment of CE evolution \citep{ijc+13,ktl+16} and SN kicks \citep{jan12,jan17}. Together, these processes lead to an uncertainty in the expected merger rates of several orders of magnitude. As an example, the simulated values of the double NS merger rate based on binary population synthesis covers a broad range of about $\unit{1-1\,000}{\mega\yr^{-1}}$ per Milky~Way equivalent galaxy \citep{aaa+10}. After the recent success in also detecting GW signals from merging NSs \citep{aaa+17c}, it is expected that GWs from a large number of colliding systems will soon determine the double NS merger-rate density in the local Universe. It is also anticipated that collisions of mixed BH/NS systems will be detected in the near future. Thus, it will soon be possible to obtain broad DCO merger rate constraints from GW detectors like advanced LIGO, Virgo, KAGRA and LIGO India, and finally determine which species of DCO binaries dominate the detection rate. A (perhaps slightly na\"{i}ve) hope, but a difficult task due to the degeneracy involved, is that the empirical detection rate from advanced LIGO and sister observatories can be inverted to infer constraints on CE physics and SN momentum kicks \citep[e.g.][]{dbf+12,duvs17,bmn+17}.

%%%%%%%%%%%%%%%%%%%%%%%%%%%%%%%%%%%%%%%%%%%%%%%%%%
\subsection{R\'{e}sum\'{e} of double compact object formation}\label{sec:standard_scenario}
\begin{figure}
  \includegraphics[width=\columnwidth]{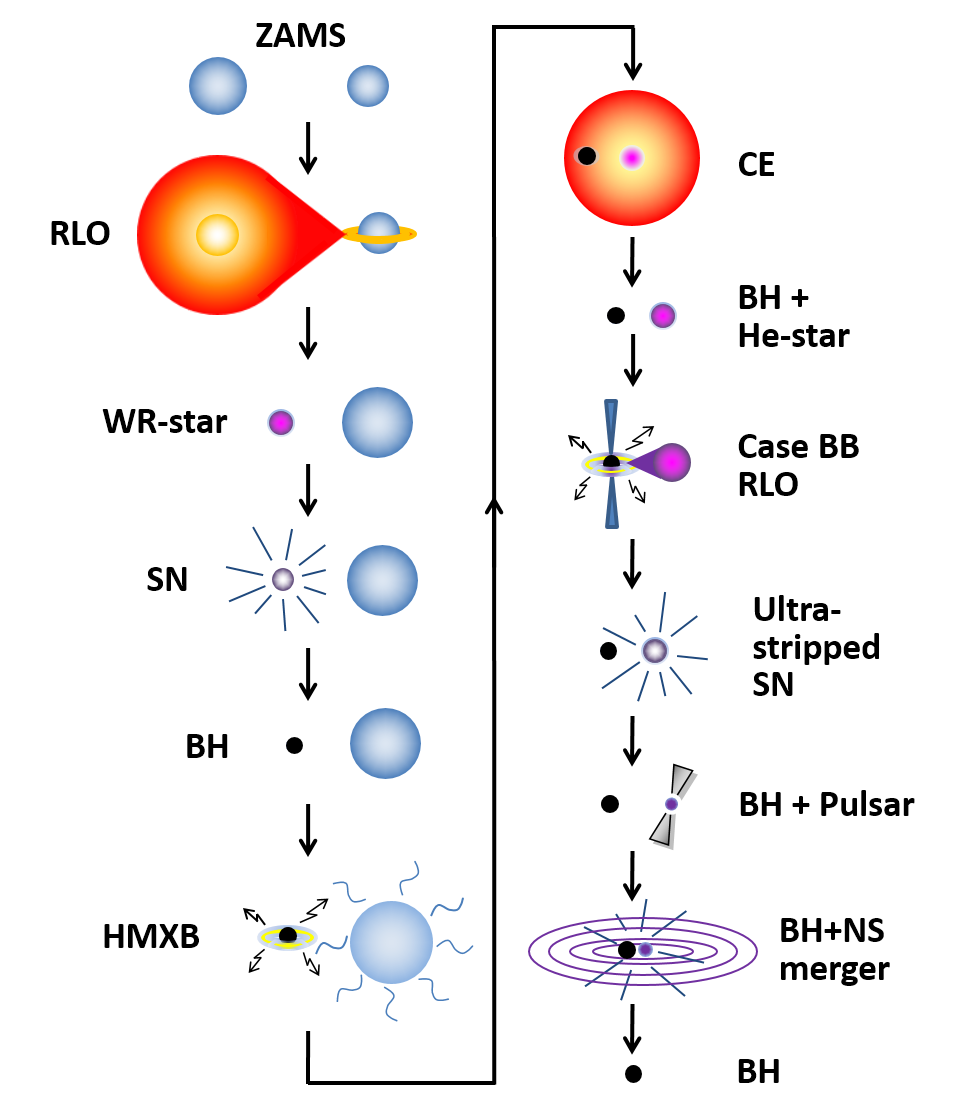}
  \caption{\label{fig:vdHcartoon} Illustration of the formation of a BH-NS system which merges within a Hubble time and produces a single BH, following a powerful burst of GWs and a short~GRB.
           Acronyms used in this figure: ZAMS: zero-age main sequence; RLO: Roche-lobe overflow (mass transfer); He-star: helium star; SN: supernova; BH: black hole; HMXB: high-mass X-ray binary; CE: common envelope; NS: neutron star.}
\end{figure}

Previous theoretical work on the physics of DCO formation includes: \citet{bk74,wml74,fv75,sv82,vdh94a,ibk+03,dp03,plp+04,vdh04,dpp05,bkr+08,tlp15,tkf+17}. From these papers, a {\it standard scenario}\footnote{See \SectionRef{sec:alternative_scenario} for discussions on alternative formation scenarios and \TableRef{tab:formationchannel_MW_beta75} for several further sub-channels.} has emerged \citep[e.g.][]{bv91,tv06,bkr+08} which we now summarize in more detail.

In \FigureRef{fig:vdHcartoon}, we illustrate the formation of a DCO system. The initial system contains a pair of OB-stars which are massive enough to terminate their lives in a core-collapse SN. The secondary (initially least massive) star may, in principle, be a $\unit{5-7}{\Msun}$ star which accretes mass from the primary (initially most massive) star to reach the threshold limit for core collapse at $\sim\unit{8-12}{\Msun}$ \citep{jhn+13,wh15}. The donor (primary) star loses its hydrogen-rich envelope, via Roche-lobe overflow (RLO) to the secondary star, and becomes a helium star. If such a star is more massive than about $\unit{8}{\Msun}$ it is often observable as a Wolf-Rayet star \citep{cro07}. Whether or not the system survives the following SN explosion depends on the amount of ejected mass, the orbital separation and the kick imparted onto the newborn NS or BH \citep{fv75,hil83,tt98}. If the binary system remains bound after the first SN explosion \citep[which is of type~Ib/c,][]{ywl10}, the system eventually becomes observable as a high-mass X-ray binary (HMXB). Before this stage, and if the first-born compact object is a NS, the system may also be detectable as a radio pulsar orbiting an OB-star, e.g. as in PSRs~B1259$-$63 and J0045$-$7319 \citep{jml+92,kjb+94}. 

When the secondary star expands and initiates RLO during the HMXB stage, the system may eventually become dynamically unstable on a timescale that could be as short as a few $\unit{100}{\yr}$ \citep{sav78}. This leads to the formation of a CE \citep{pac76,ijc+13} where dynamical friction of the motion of the compact object (NS or BH) inside the giant star's envelope often causes extreme loss of orbital angular momentum and energy. In case the hydrogen-rich envelope is successfully ejected and the binary system survives the CE~phase, it consists of a NS or BH orbiting a helium star (the naked core of the former giant star). Depending on the orbital separation and the mass of the helium star, an additional phase of mass transfer may be initiated \citep[Case~BB RLO,][]{hab86a,tlp15}, which mostly leads to further shrinkage of the orbit. This stage of mass transfer is important since it enables a relatively long phase of accretion onto the compact object, whereby the first-formed compact object is recycled to a high spin rate. In addition, it allows for extreme stripping of the helium star prior to its explosion \citep[in a so-called {\it ultra-stripped} SN,][]{tlm+13,tlp15,sys+15,mmt+17,mgh+18}.

If the post-SN orbital period after the second explosion is short enough (and especially if the eccentricity is large) the DCO system will eventually merge due to GW radiation. The final remnant is in most cases a BH, although for double NS mergers a massive NS (or, at least, a meta-stable NS) may be left behind instead depending on the EoS \citep{vs98,fr14,mtq18}.

%%%%%%%%%%%%%%%%%%%%%%%%%%%%%%%%%%%%%%%%%%%%%%%%%%
\subsection{Binary population synthesis studies}\label{sec:population_synthesis_studies}
To estimate the formation and merger rates of DCO binaries, the nature of the merging compact objects and their delay timescales, and thus the offset of the associated short~GRBs and kilonovae from their birth places, it is necessary to evolve a larger number of binary systems. Many binary star interactions, however, are uncertain and various input distributions are often used to quantify key physical parameters by the use of e.g. Monte Carlo techniques. This is the essence of binary population synthesis.

Based on observational evidence and theoretical development of the necessary input physics from stellar evolution and binary interactions, a large number of binary population synthesis studies have emerged over the last two decades to investigate the formation and evolution of DCO binaries. Examples include \citet{bsp99,bkb02,vt03,bkr+08,dbf+12,mv14,bhbo16,es16,svm+17,cbkb18,vns+18}.

Here, we present new results based on a significantly improved version of the binary population synthesis code applied by \citet{vt03}. The innovative aspect comes in when applying updated stellar evolution models (at different metallicities) with high resolution of the stellar structure, and a proper treatment of Case BB RLO. For example, in the last couple of years the calculations of the final stage of close binary evolution with an accreting NS has been advanced to a new stage, towards the end of oxygen burning, providing evidence for the existence of ultra-stripped SNe \citep{tlm+13,tlp15}. It was demonstrated that for such SNe the total envelope mass surrounding the metal core can even be $<\unit{0.1}{\Msun}$, which results in very little mass ejection during the SN. This is important for the subsequent calculations of the resulting NS kicks \citep{tkf+17} which affect the estimated number of mergers that GW observatories will detect and the offsets from their host galaxies \citep[relevant for short~GRBs and electromagnetic follow-up observations,][]{bpb+06,fb13}.

\bigskip
The applied binary population synthesis code \OurCode and our newly developed upgrades to this code will be described in \SectionRef{sec:code}. In \SectionRef{sec:stellar_grids}, our default grid of stellar models extracted from detailed stellar evolution calculations is presented. The results and the first comparison to observations are given in \SectionRef{sec:results}. This is followed in \SectionRef{sec:discussion} with further discussions with respect to observations, the influence of the different binary population synthesis input parameters and a comparison to other publications. We conclude our findings in \SectionRef{sec:conclusions}. Additional material can be found in \AppendixRef{app:stellar_grid}--\ref{app:further_parameter_variations}.

%%%%%%%%%%%%%%%%%%%%%%%%%%%%%%%%%%%%%%%%%%%%%%%%%%
\section{The \OurCode code}\label{sec:code}
\OurCode is a rapid binary population synthesis code. It is a significantly upgraded version of the code developed by \citet{vt03}, which again is based on the original code of \citet{tb96}. Another version of this code has been applied in e.g. \citet{tt98,tfv+99,ts00}. Compared to the older versions, the new one is faster and allows for e.g. simultaneous evolution of the two stars, as well as RLO from the secondary star to the primary star before the primary star has terminated its nuclear evolution (i.e. mass-transfer reversals). Several updates on the input physics are included as well. It is faster than other binary population synthesis codes like {\it StarTrack} \citep{bkr+08} or {\it binary\_c} \citep{itk+04,idk+06,igs+09,ipj+18}. \OurCode does not rely on fitting formulae for the stellar evolution \citep[e.g.][]{htp02} which often leads to usage outside their range of validity when studying formation of massive DCO binaries. Our code interpolates in tabulated data from a dense grid of detailed stellar models (for more details see \SectionRef{sec:stellar_grids}). This allows for a more accurate treatment of, for example, the CE evolution which is still the most uncertain part in binary star population synthesis. Additionally, we have implemented the latest results of detailed numerical Case~BB RLO calculations \citep{tlp15}.

%%%%%%%%%%%%%%%%%%%%%%%%%%%%%%%%%%%%%%%%%%%%%%%%%%
\subsection{Initial conditions}\label{sec:initial_conditions}
Usually stellar or binary population synthesis start with stars on their zero-age main sequence (ZAMS). There are several distribution functions in the literature to describe the statistical distribution of the most important parameters of a binary system. We now discuss the most important ones applied in \OurCode.

%%%%%%%%%%%%%%%%%%%%%%%%%%%%%%%%%%%%%%%%%%%%%%%%%%
\subsubsection{Stellar masses}\label{sec:masses}
The first fundamental parameters of a binary system are the two masses of the stars which build the binary.

The primary mass, $m_{\rm p}$, is defined to be the mass of the initially more massive star. It is selected from an initial-mass function (IMF) for single stars. By default the Salpeter-like IMF \citep{sal55,sca86} is used,
\begin{equation}
  \xi(m_{\rm p}) \propto m_{\rm p}^{-\alpha_{\rm IMF}} \quad \text{with} \quad \alpha_{\rm IMF}=2.7 .
  \label{eq:IMF}
\end{equation}
Different IMF slopes are considered in \SectionRef{sec:alpha_IMF}. Other IMFs, \citep[e.g.][where $\alpha_{\rm IMF}$ depends on the mass range]{kro08} are implemented as well for comparison. According to the chosen IMF, the primary mass is selected randomly.

The minimum and maximum masses of the primary and secondary stars can be varied to fix the desired mass ranges for the ZAMS stars. When considering NS and BH progenitors at solar metallicity, our primary star mass range starts at $8$ and $\unit{22}{\Msun}$, respectively. As a result of mass transfer, however, the initial ZAMS mass of the secondary star can be smaller (see below). All our stars with initial masses above $\unit{30}{\Msun}$ typically produce BHs (\SectionRef{sec:progenitor_ZAMS_masses}).

The secondary mass, $m_{\rm s}$, is chosen based on the primary mass and a distribution function for the mass ratio, \mbox{$q\equiv m_{\rm s}\product {m_{\rm p}}^{-1}$}. Observational selection effects are still not known sufficiently well to choose which $q$-distribution is the closest to reality. Our default distribution is therefore that of \citet{kui35},
\begin{equation}
  f(q) = \frac{2}{(1+q)^2},
  \label{eq:massratio}
\end{equation}
which simply reflects the statistical distribution of mass ratios if nature divides a gas cloud in two pieces with proportions randomly chosen on a linear scale between 0 and 1. Alternatively, we can apply e.g. a flat distribution for $q$ or investigate other distributions based on recent empirical data, such as \citet{sdd+12} and \citet{md17}. It is also possible to set a range for the secondary mass and \OurCode will automatically calculate the relevant range for the mass ratio. However, \citet{db15} demonstrated that their DCO merger rates are almost independent of the initial distributions of mass ratios and orbital periods, compared to the strong dependence on input physics parameters governing binary interactions and SNe. Our test simulations yield a similar conclusion. We compared our DCO merger-rate results using \EquationRef{eq:massratio} for the mass ratios and a flat distribution of initial orbital periods (see below) to the results obtained using the input distributions of \citet{sdd+12}, and we find that our merger rates only change by a factor of $2-3$, which is a relatively small change compared to the effects of changing various input physics parameters (see \SectionRef{sec:parameter_studies}).

If the secondary star gains mass as a result of mass-transfer processes in a binary, even lower ZAMS masses than needed in an isolated evolution must be considered for the secondary star when producing selected compact objects. As an example, it has been demonstrated that secondary stars with ZAMS masses of e.g. $\unit{6-7}{\Msun}$ may accrete sufficient material to end up producing a NS \citep{ts00,zdi+17}. 

%%%%%%%%%%%%%%%%%%%%%%%%%%%%%%%%%%%%%%%%%%%%%%%%%%
\subsubsection{Orbital parameters and binary fraction}\label{sec:initial_orbital_parameters}
Orbital period distributions often used in the literature are flat in $\log(P)$ \citep{opi24,abt83}. Alternative distributions have been proposed \citep[e.g.][]{kro08,sdd+12} and are also included for usage in \OurCode. For all our simulations, we assume a binary fraction of $100$~per~cent. We do not account for a suggested correlation between binary fraction and orbital period \citep{md17}. Again, we emphasize that the properties of the final DCO binaries and their merger rates are only weak functions of the initial input distributions \citep[][and \SectionRef{sec:masses}]{db15}.

The minimum and maximum orbital separations can also be specified. Our default interval is between $2$ and $\unit{10\,000}{\Rsun}$ (\TableRef{tab:standard} in \SectionRef{sec:results}). The lower value is further limited by the condition that none of the stars must fill their Roche~lobe (see \SectionRef{sec:roche-lobe_overflow}) on the ZAMS.

Another orbital parameter is the eccentricity, $e$. Close systems circularise with time due to tidal effects \citep{zah77}, especially when they evolve to fill their Roche lobes \citep{vp95}. Therefore, our default simulations are always initiated with circular orbits, \mbox{$e=0$}. Other possibilities for an eccentricity distribution included in \OurCode are a thermal distribution \citep{heg75}, a flat distribution or the distribution of \citet{sdd+12} which favours low eccentricities. Another option is a flat distribution in orbital angular momentum.

%%%%%%%%%%%%%%%%%%%%%%%%%%%%%%%%%%%%%%%%%%%%%%%%%%
\subsubsection{Further parameters}\label{sec:further_parameters}
In addition to the above, there are further parameters which can influence the evolution of a binary system. Stars usually possess rotation. In a close binary tidal forces tend to synchronise the stellar spin with the orbit \citep{zah77}. The influence of the rotation on the evolution of a star is very limited as long as the star spins much more slowly than its break-up velocity {\citep[e.g.][]{bdc+11}, where the centrifugal force fully compensates the gravity. Hence, for most purposes a differentiation between slow and very fast rotating stars is sufficient. Here we focus on non-rotating stars, with the caveat that efficient accretion in some cases may produce very rapidly rotating stars \citep{pac81} whose evolution can be rather different.

The metallicity may also have an important effect on the evolution of a star \citep{lan12}. It varies between galaxies and for different generations of star formation within each galaxy. Finally, depending on the stellar density, there could be dynamical interactions with other stars or binary systems \citep{pm00,rcr16,ban17,pkl+17} which can change the orbital and stellar evolution compared to that of an isolated binary. Such dynamical interactions are not considered in this investigation.

%%%%%%%%%%%%%%%%%%%%%%%%%%%%%%%%%%%%%%%%%%%%%%%%%%
\subsection{Evolutionary phases}\label{sec:evolutionary_phases}
In the following, we highlight our treatment of various evolutionary phases in \OurCode, including: stellar winds, tides, mass transfer/loss (RLO, CE), SNe and GWs. Further details on the orbital evolution can be found in \citet{vdh94a,spv97}. For a general review, see e.g. \citet{tv06}. 

If a binary system is initially very wide (and remains wide throughout its evolution), the two stars evolve as if they where isolated. They just follow their evolutionary tracks of single stars taken from the stellar grids (see \SectionRef{sec:computed_quantities}, and also \AppendixRef{app:interpolation} for a list of all stellar quantities calculated). These tracks are assumed to be extended by a short-lasting phase of burning elements heavier than helium prior to the core collapse (typically of order $\unit{10^5}{\yr}$, see \SectionRef{sec:stellar_grids} for more details on the grids). In this case of isolated star evolution, the only changes to the orbital separation are caused by stellar winds and SN explosions. 

For a close binary system, or a system initially in a wide orbit that later becomes tight after the first SN \citep{kal98}, however, we consider in each individual case whether some or all of the phases described below apply.

%%%%%%%%%%%%%%%%%%%%%%%%%%%%%%%%%%%%%%%%%%%%%%%%%%
\subsubsection{Stellar winds}\label{sec:stellar_winds}
Following the stellar wind models prescribed for our applied stellar grids (\SectionRef{sec:stellar_grids}), we calculate the orbital widening by assuming that the average angular momentum (per unit mass) carried away by a spherically symmetric wind at high velocity is the same as the average orbital angular momentum of the mass-losing star. This leads to a simple expression for the orbital widening given by
\begin{equation}
  \frac{a}{a_0} = \frac{M_0}{M},
  \label{eq:windmassloss}
\end{equation}
where $M$ is the total mass of the binary, $a$ is the semi-major axis, and indices ``$0$'' refer to the values prior to wind mass loss. The above expression also holds in the case of simultaneous wind mass loss from both stars.

%%%%%%%%%%%%%%%%%%%%%%%%%%%%%%%%%%%%%%%%%%%%%%%%%%
\subsubsection{Circularisation}\label{sec:circularisation}
In close systems, the tidal friction in the stars will circularise the orbit and synchronise their spins with the orbital phase \citep{sut74}. Prior to mass transfer, when the donor star is close to filling its Roche lobe, the tidal effects are particularly strong and the orbit is likely to circularise on a short timescale. For this process, angular momentum conservation yields the orbital changes for a given eccentricity, $e_0$,
\begin{equation}
  \frac{a}{a_0} = 1-e_0^2.
  \label{eq:circularise}
\end{equation}
The orbit is assumed to be fully circularised at the onset of mass transfer.

%%%%%%%%%%%%%%%%%%%%%%%%%%%%%%%%%%%%%%%%%%%%%%%%%%
\subsubsection{Roche-lobe overflow}\label{sec:roche-lobe_overflow}
The change in orbital separation upon non-conservative\footnote{Meaning that total mass and orbital angular momentum are not conserved within the binary system.} mass transfer depends crucially on the specific angular momentum of the matter lost (which is rather poorly known). 

When the donor star expands and fills its Roche lobe \citep{egg83}, large scale mass transfer to its companion star initiates. This transfer of matter will continue in a stable or unstable way depending on the reaction of the two stars upon the mass transfer. We refer to stable mass transfer as RLO, while dynamically unstable mass transfer is assumed to result in a CE (\SectionRef{sec:common_envelope}).

For stellar components on the main sequence (i.e. during core hydrogen burning) the stability of mass transfer is evaluated in \OurCode by comparing their mass ratio, $q$ at the onset of the mass transfer to a threshold value, $q_{\rm limit}$ \citep[e.g.][and see discussion in \SectionRef{sec:q_limit}]{ne01,dcp08}. Here, and in the following, we define \mbox{$q\equiv m_{2}\product {m_{1}}^{-1}$}, where $m_2$ is the donor star mass and $m_1$ is the accretor star mass. Hence, for \mbox{$q<q_{\rm limit}$} we assume that the RLO remains dynamically stable, while a CE is assumed if \mbox{$q\geq q_{\rm limit}$}. Our default \mbox{$q_{\rm limit}=2.5$} is motivated by a $q_{\rm limit}$ of $3.8$ \citep{ghw+10} or between $1.5$ and $2.2$ \citep{pi15} for stars with radiative or convective envelopes, respectively (see also discussions in \SectionRef{sec:q_limit}). Additionally, for massive OB-stars, to avoid a Darwin instability \citep{dar79} and ensure dynamically stable mass transfer, we require that the system has a minimum orbital period $\ga\unit{3}{\days}$ (Pablo Marchant, priv.~comm.).

Giant donor stars have a deep convective envelope which often leads to unstable mass transfer \citep{hw87,ts99,tvs00,prp02,pibv17}, in many cases resulting in a CE. For giant star donors, \OurCode checks for the depth of the convective envelope and assumes a CE forms if RLO is initiated for donor stars which have convective envelops exceeding $10$~per~cent in mass coordinate (see further discussions in \SectionRef{sec:q_limit}).

For helium star donors (Case~BB RLO), we distinguish between a non-degenerate accretor and a compact object accretor. For non-degenerate accretors, we apply the same stability criterion based on the mass ratio as for the hydrogen-rich donors mentioned above. If a helium star transfers mass onto a compact object, however, we apply the numerical results and the recipe from \citet{tlp15}.

In order to calculate the orbital period changes due to RLO, we adopt the isotropic re-emission model \citep[][and references therein]{tv06}. Here, the change in orbital angular momentum caused by mass loss, $\dot{J}_{\rm ml}$, in terms of the orbital angular momentum, $J_{\rm orb}$}, from the binary system (usually the dominant term in the orbital angular momentum balance equation) is given by
\begin{equation}
  \frac{\dot{J}_{\rm ml}}{J_{\rm orb}} = \frac{\alpha_{\rm RLO} + \beta_{\rm RLO}\product q^2 + \delta_{\rm RLO}\product\gamma\product(1+q)^2}{1+q}\product\frac{\dot{m}_2}{m_2},
  \label{eq:angular_momentum}
\end{equation}
where $\alpha_{\rm RLO}$, $\beta_{\rm RLO}$ and $\delta_{\rm RLO}$ are the fractions of mass lost from the donor in the form of a direct fast wind, the mass ejected from the vicinity of the accretor and from a circumbinary coplanar toroid (with radius, \mbox{$a_{\rm r} = \gamma^2\product a$}), respectively \citep{vdh94a,spv97}. The accretion efficiency of the accreting star (here index 1) is thus given by: \mbox{$\epsilon = 1 -\alpha_{\rm RLO} -\beta_{\rm RLO} -\delta_{\rm RLO}$}, or equivalently
\begin{equation}
  \partial m_1 = -\epsilon\product\partial m_2,
  \label{eq:epsilon}
\end{equation}
where $m_2$ refers to the donor star mass and \mbox{$\partial m_2 < 0$} is its mass loss. These factors are functions of time as the binary system evolves during the mass-transfer phase.

In \OurCode, we assume $\alpha_{\rm RLO}$, $\beta_{\rm RLO}$ and $\delta_{\rm RLO}$ to be a constant, reflecting a time average during the entire RLO. Hence, their values determine the orbital angular momentum budget. The value of \mbox{$\alpha_{\rm RLO}=0.20$} is based on the possibility of rather intense stellar winds among donor stars. It is a simplification to assume a constant value independent of the evolutionary status of the donor star. A study of the dependency of our GW merger results on $\alpha_{\rm RLO}$ is presented in \SectionRef{sec:alpha_RLO} and \AppendixRef{app:mass-transfer_efficiency}. The parameter $\beta_{\rm RLO}$ is motivated by expected mass loss from the system if either the mass-transfer rate is high and/or the accreting star evolves close to critical rotation. The latter effect is particular important in high-mass, non-degenerate binaries where $\beta$ up to 0.90 is achieved \citep{plv05}. For the evolution of HMXBs, the Eddington limit of the accretion rate onto the compact object may cause mass loss in the form of a disk wind or a jet \citep[][e.g. as seen in SS\,433]{bh11}. Furthermore, propeller effects \citep{is75} might be at work too. For simplicity, we assume here that $\beta_{\rm RLO}$ is the same in all RLO systems irrespective of the nature of the accretor. Our default value is \mbox{$\beta_{\rm RLO}\ge 0.75$} (i.e. \mbox{$\beta_{\rm min}=0.75$}, see below). This choice is based on our ability to reproduce NS masses in double NS systems (\SectionRef{sec:compact_object_masses}). However, we also explore scenarios with \mbox{$\beta_{\rm min}=0$} in detail (\AppendixRef{app:efficient_mass_transfer}) and \mbox{$0.00\leq\beta_{\rm min}\leq 0.80$} (\SectionRef{sec:beta_min}).

We note that detailed numerical stellar evolution modelling of each binary is required to obtain the mass-transfer evolution $\dot{m}_2(t)$. However, for our purposes, it suffices to estimate an average $\dot{m}_2$ based on the ratio between the total amount of material transferred (i.e. the envelope mass of the donor star) and the characteristic timescale of mass transfer (see below).

To limit the number of free parameters, we neglect the possibility of mass outflow to a circumbinary disk and thus we apply \mbox{$\delta_{\rm RLO}=0$}.  In this case, we obtain for the change in the orbital separation,
\begin{equation}
  \frac{a}{a_0} = \left(\frac{q}{q_0}\right)^{2\product\alpha_{\rm RLO}-2}\product\left(\frac{1+q}{1+q_0}\right)^{-1}\product\left(\frac{1+\epsilon\product q}{1+\epsilon\product q_0}\right)^{2\product{\textstyle\frac{\alpha_{\rm RLO}\product\epsilon^2+\beta_{\rm RLO}}{\epsilon\product(1-\epsilon)}}+3}.
  \label{eq:RLO}
\end{equation}

\subsubsection{Response of the accretor to Roche-lobe overflow}\label{sec:RLO_response}
To evaluate the response of the accreting star, in \OurCode we first compare the resultant average mass-transfer rate, estimated from our initially adopted value of $\beta_{\rm RLO}$, to the Eddington limit of the accreting star\footnote{In addition, the thermal timescale of the accretor should be compared to the timescale of the RLO as well. This is not included here, because for relatively massive stars the thermal timescale of the accretor is comparable to that of the RLO -- thus avoiding a contact phase.},
\begin{equation}
  \dot{m}_{\rm Edd} = 4\product\pi\product c\product G\product m_1 \frac{m_{\rm H}\product\mu_{\rm e}}{\varepsilon\product\sigma_{\rm T}},
  \label{eq:mEdd}
\end{equation}
where $c$ is the speed of light, $G$ is the gravitational constant, $m_1$ is the accretor mass, $m_{\rm H}$ is the proton mass, $\mu_{\rm e}$ is the mean molecular weight per electron of the accreted material, $\sigma_{\rm T}$ is the Thomson cross section and $\varepsilon$ is the sum of released energy per unit mass from released gravitational binding energy and nuclear burning of accreted material. 

In systems with compact object accretors, and for which \mbox{$\dot{m}_1>\dot{m}_{\rm Edd}$}, we then reduce the mass-accretion efficiency, $\epsilon$, by increasing the re-emission fraction of the accretor, $\beta_{\rm RLO}$, to ensure that the accretion rate does not exceed \mbox{$\dot{m}_1=\dot{m}_{\rm Edd}$}. This means that in practice, rather than applying a fixed $\beta_{\rm RLO}$, we operate with a minimum value which we refer to as $\beta_{\rm min}$. As discussed in \SectionRef{sec:compact_object_masses_double_neutron_star_binaries}, based on the observed masses of double NS components, we adopt \mbox{$\beta_{\rm min}=0.75$} as our default. For simplicity, we assume that $\beta_{\rm min}$ is the same for all accreting stars (non-degenerate stars and compact objects).

Systems with non-degenerate accretors will usually not have their mass-accretion rate limited by $\dot{m}_{\rm Edd}$ because in this case $\dot{m}_{\rm Edd}$ is typically about $\unit{10^{-3}}{\Msun\usk\yr^{-1}}$. In the rare case of super-Eddington accretion onto a non-degenerate accretor \OurCode assumes the formation of a CE.

After mass transfer, the accreting star is assumed to relax and become rejuvenated. Hence, we attach it to a new evolutionary track (beyond the ZAMS) depending on its core mass and total mass (\SectionRef{sec:stellar_grids} and \AppendixRef{app:place_into_grid}). As a result of the mass transfer, the accretor's envelope mass increases. Hence, the new track has a larger ZAMS mass compared to the old track and the star appears younger. Only if processed elements (He, C, O) are transferred, some material is added to the core mass. If the rejuvenated accretor, or the relaxed core of the donor, fills its Roche~lobe, then the system remains attached and we assume that it coalesces.

The donor star is always assumed to lose its mass down to its core-envelope boundary, see \SectionRef{sec:core-envelope_boundary}. Therefore, every hydrogen-rich donor becomes a helium star (\SectionRef{sec:helium_stars}) and a helium rich donor leaves behind a naked metal core which is composed of carbon and heavier elements. In the latter case, we assume it terminates its life and becomes a compact object, see \SectionRef{sec:formation_of_compact_objects}, before any other binary interaction is dealt with in \OurCode.

We assume that any stable RLO typically proceeds on the thermal timescale of the donor star which depends on its internal structure at the onset of RLO, multiplied by a factor of 3 or 1 for hydrogen- or helium-rich donor stars, respectively. This correction factor for the timescale of the RLO is found from test cases in which we studied detailed models of binary evolution using the binary stellar evolution code BEC. We are aware that some systems with initial mass ratios close to unity or mass transfer from a low-mass star to a more massive compact object (i.e. a low-mass X-ray binary system), may proceed on a nuclear timescale. However, for this investigation such systems are either rare or not relevant. We plan to improve on this aspect in a future version of \OurCode.

%%%%%%%%%%%%%%%%%%%%%%%%%%%%%%%%%%%%%%%%%%%%%%%%%%
\subsubsection{Common-envelope evolution}\label{sec:common_envelope}
If the mass transfer in a binary is unstable, we assume a CE will be formed, engulfing the two stars. In the standard formation channel for DCO binaries which become GW mergers (\FigureRef{fig:vdHcartoon}), the systems often enter a CE phase during RLO in the HMXB stage as a result of enhanced orbital shrinking due to a large mass ratio between the donor star and the accretor. Here, at the latest when it becomes a red supergiant, the massive donor star captures its NS/BH companion and causes it to spiral in. For a successful CE ejection, it is believed that the envelope will be lost from the system on a short timescale of $\la\unit{1\,000}{\yr}$ \citep{pod01}.

There are many uncertainties in calculations of the in-spiral process and the subsequent ejection of the CE. This causes large uncertainties in the predicted rates for GW merger events obtained from binary population synthesis \citep{aaa+10}. A full understanding of the CE phase requires detailed multi-dimensional hydrodynamical calculations. Nevertheless, studies in this direction still have difficulties ejecting the envelope and securing deep in-spiral \citep{tas00,pdf+12,rt12,nil14,orps16}.

As a result of the current limited knowledge of CE physics \citep{ijc+13} a simple but robust prescription is implemented in \OurCode. The outcome of the CE ejection is calculated according to the $(\alpha,\lambda)$-formalism \citep{web84,dek90}. In this framework, it is assumed that a certain fraction, $\alpha_{\rm CE}$, of the released orbital energy, caused by frictional torques acting on the in-spiralling star, is converted into kinetic energy in the envelope. We assume $\alpha_{\rm CE}$ to be a fixed parameter for all stars and our default value is \mbox{$\alpha_{\rm CE}=0.5$}. The influence of this parameter is further discussed in \SectionRef{sec:alpha_CE}.

The released orbital energy from in-spiral, $|\Delta E_{\rm orb}|$, is given by
\begin{equation}
  \Delta E_{\rm orb} = -\frac{G\product m_{\rm 2,core}\product m_{1}}{2\product a} + \frac{G\product m_{2}\product m_{1}}{2\product a_{0}},
  \label{eq:Eorb}
\end{equation}
where $m_{1}$ and $m_{2}$ denote the pre-CE companion star mass and donor star mass, respectively, while $m_{\rm 2,core}$ is the core mass of the donor star (\SectionRef{sec:core-envelope_boundary}). The pre- and post-CE orbital separations are denoted by $a_0$ and $a$.

A successful CE ejection can only occur in systems where \mbox{$|E_{\rm bind}|\le\alpha_{\rm CE}\product|\Delta E_{\rm orb}|$}. Here, the binding energy of the envelope of the donor star is given by
\begin{equation}
  E_{\rm bind}\equiv -\frac{G\product m_{2}\product m_{\rm 2,env}}{\lambda\product R},
  \label{eq:lambda}
\end{equation}
where $R$ denotes the pre-CE donor star radius and \mbox{$m_{\rm 2,env}=m_{2}-m_{\rm 2,core}$} is its envelope mass. The value of $\lambda$ is not a constant but depends strongly on the mass and the evolutionary status of the donor star \citep{dt00,dt01,ktl+16}. From our detailed stellar structure models (\SectionRef{sec:stellar_grids}) we calculate the relevant $\lambda$ values and determine the outcome of the CE event from the energy budget.

The total binding energy of the envelope is the sum of the gravitational binding energy and the internal thermodynamic energy, 
\begin{equation}
  E_{\rm bind}  =  -\int_{m_{\rm 2,core}}^{m_{2}}\frac{G\product m(r)}{r}\diff m + \alpha_{\rm th}\product\int_{m_{\rm 2,core}}^{m_{2}} U\diff m,
  \label{eq:Ebind}
\end{equation}
where $m(r)$ is the mass within the radius coordinate $r$ and $U$ is calculated following \citet{hpe95}. The latter involves the basic thermal energy for a simple perfect gas, the energy of radiation, as well as terms due to ionization of atoms and dissociation of molecules. The value of $\alpha_{\rm th}$ depends on the details of the ejection process. In our default model we assume \mbox{$\alpha_{\rm th}=0.5$}; other values are discussed in \AppendixRef{app:alpha_th}.

If the binary system is not able to eject the CE, we assume that its stellar components merge and we disregard further evolution of the product. If the system survives and ejects the CE, the change in orbital separation according to the in-spiral is calculated from
\begin{equation}
  \frac{a}{a_0} = \frac{m_{\rm 2,core}}{m_{2}}\product\left(1+\frac{2}{\alpha_{\rm CE}\product\lambda}\product\frac{m_{\rm 2,env}}{m_1}\product\frac{a_0}{R}\right)^{-1},
  \label{eq:CE}
\end{equation}
where $R$ is taken as the Roche-lobe radius of the donor star at the onset of the CE. Anyway, stellar or binary evolution between the onset of RLO and the formation of a CE is disregarded.

\OurCode also takes the released energy of accretion onto a compact object, $E_{\rm acc}$, and the associated additional nuclear-burning energy, $E_{\rm nuc}$, into account. This rescales (increases) the final semi-major axis by a factor \mbox{$1+\left(E_{\rm acc}+E_{\rm nuc}\right)\product|E_{\rm bind}|^{-1}$}. For detailed discussions on CE ejection from massive stars (NS and BH progenitors, including the conditions of detachment and core size), we refer to \citet{ktl+16}.

If the binary system survives the CE phase, the donor star becomes a naked core similar to in the case of stable RLO. If the pre-CE donor star is a hydrogen-rich star, \OurCode places the exposed stellar core on the ZAMS helium-star track because the grid of hydrogen-rich stars does not in all cases contain information about the CO-core (some of the stars are not evolved far enough). If the pre-CE donor star is a helium-rich donor, then the exposed metal core is assumed to end its life in a SN or becomes a white dwarf (WD) before any further binary interactions will occur.

%%%%%%%%%%%%%%%%%%%%%%%%%%%%%%%%%%%%%%%%%%%%%%%%%%
\subsubsection{Formation of compact objects}\label{sec:formation_of_compact_objects}
\begin{figure*}
  \includegraphics[width=\textwidth]{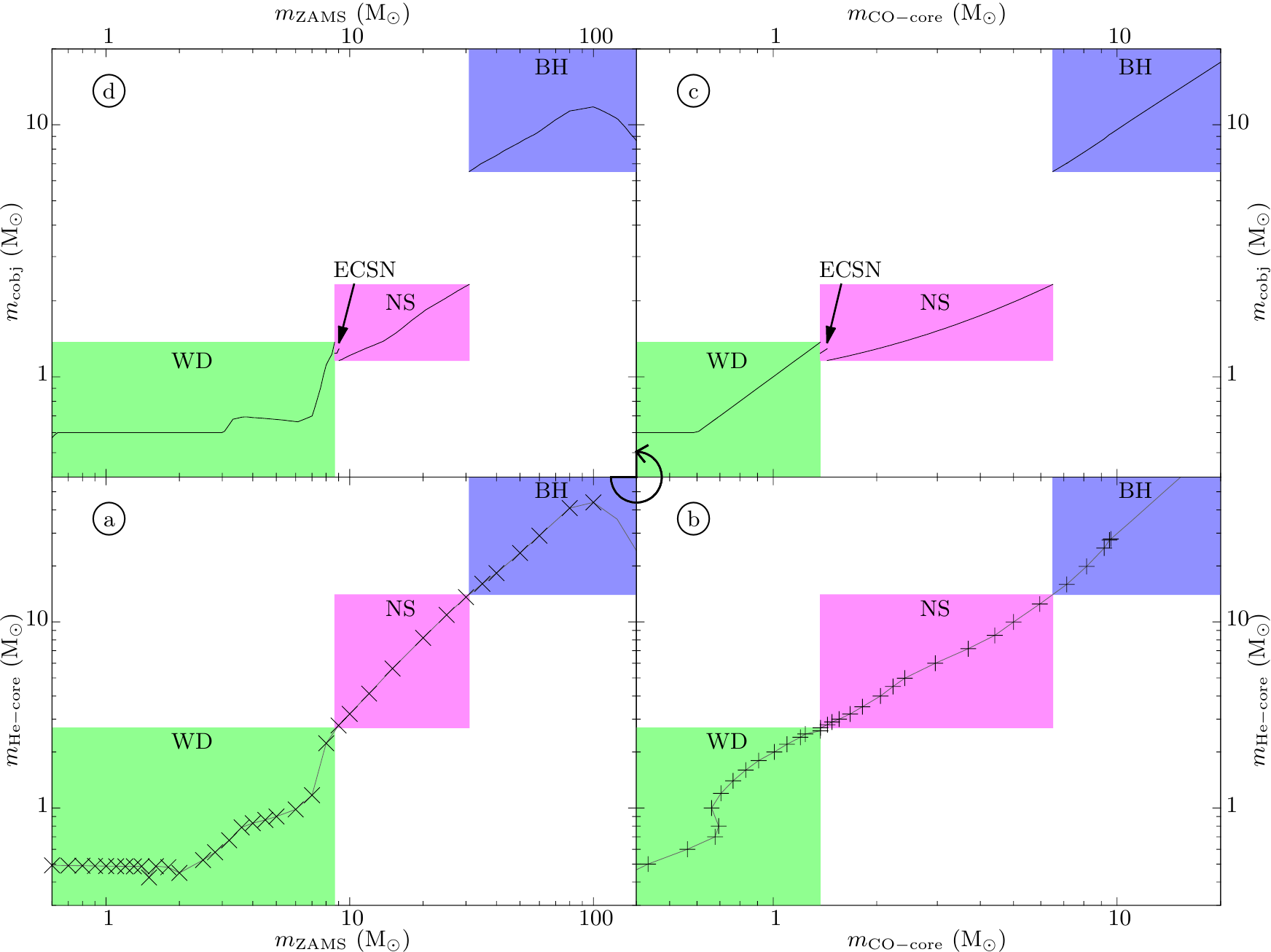}
  \caption{\label{fig:ZAMS-CompactObject_Mass}Mapping of ZAMS mass, \mbox{$m_{\rm ZAMS}\in\unit{[0.6:150]}{\Msun}$}, to final stellar remnant gravitational mass, $\mCO$, for single star stellar tracks (data points of \FiguresRef{fig:HRD_MW} and \ref{fig:HRD_He} are marked by $\times$ and $+$, respectively) at Milky~Way metallicity (panel d is inferred from panels a to c). As intermediate steps, the He-core mass, $m_{\rm He-core}$, (panel a) and CO-core mass, $m_{\rm CO-core}$, (panel b) are calculated to obtain the compact object gravitational masses (panel c). The shallow increase in WD mass as a function of ZAMS mass arises from incomplete stellar tracks of intermediate-mass stars (these are not relevant for massive DCOs with NS and BH remnants). In binary stars, the remnant masses differ because of additional mass loss or mass gain from the companion star. Acronyms: ZAMS: zero-age main sequence; WD: white dwarf; NS: neutron star; BH: black hole; ECSN: electron capture supernova.}
\end{figure*}

When a star finishes nuclear burning its core contracts and it either forms a WD remnant or it produces a NS or a BH. The nature and mass of the stellar remnant depends on the mass of its progenitor star. \FigureRef{fig:ZAMS-CompactObject_Mass} shows the relations between ZAMS mass, He-core mass, CO-core mass (based on the stellar grids applied in {\OurCode}) and estimated compact object gravitational mass.

The compact remnants resulting from our single star evolution grids (\SectionRef{sec:stellar_grids}) are assumed to be WDs with masses \mbox{$m_{\rm WD}<\unit{1.37}{\Msun}$}, NSs with masses \mbox{$\unit{1.16}{\Msun}<m_{\rm NS}<\unit{2.33}{\Msun}$} and BHs with masses \mbox{$m_{\rm BH}>\unit{6.52}{\Msun}$}, depending on the collapsing core mass. In binary star systems, however, \OurCode can produce BHs with masses down to $\unit{5.2}{\Msun}$ if their helium envelope is stripped off prior to core collapse.

WDs produced in our binaries can be either He~WDs, CO~WDs or ONeMg~WDs, depending on the mass of their progenitor cores, \mbox{$m_{\rm CO-core}<\unit{1.37}{\Msun}$} (hereafter, simply referred to as WDs independent of chemical composition). Our NSs result from either an electron capture SN \citep[EC~SN,][]{nom87} if \mbox{$\unit{1.37}{\Msun}\leq m_{\rm CO-core}<\unit{1.435}{\Msun}$} or an iron-core collapse SN (FeCC~SN) if \mbox{$\unit{1.435}{\Msun}\leq m_{\rm CO-core}<\unit{6.5}{\Msun}$}, following binary star calculations in \citet{tlp15}. In the EC~SN case, a reduction of 10~per~cent in gravitational mass is assumed during formation \citep{hmj+10}, leading to NS masses in the range $\unit{1.24\text{ to }1.29}{\Msun}$. For NSs produced in FeCC~SNe from non-stripped and partially stripped stars, we use the relation
\begin{equation}
  \mCO = 0.23\product m_{\rm CO-core}+\unit{0.83}{\Msun},
  \label{eq:NSmass}
\end{equation}
while
\begin{equation}
  \frac{\mCO}{\Msun} = -\frac{1}{0.168}+\sqrt{\frac{1}{0.168^2}+\product\frac{1.06}{0.084}\product\left(\frac{m_{\rm CO-core}}{\Msun}\right)^{0.454}}
  \label{eq:ultrastrippedNSmass}
\end{equation}
is used for ultra-stripped NS progenitors \citep[by combining the results of][]{ly89,tww96,tlp15}. We note that the NSs produced with the very smallest masses (see \FigureRef{fig:ZAMS-CompactObject_Mass}) come from FeCC~SNe \citep{tww96}. We are aware that many detailed studies of SN explosions and their progenitors find evidence for a non-monotonic mapping of (ZAMS) progenitor star mass and final compact object mass \citep{ujma12,pt15}, which is not included here.

Our BHs are assumed to be formed from the collapsing CO-cores with \mbox{$m_{\rm CO-core}>\unit{6.5}{\Msun}$} and receive in addition a partial fallback of 80~per~cent of the mass of their surrounding helium envelopes. Studies of BH formation via hydrodynamical calculations \citep[e.g.][]{fwh99,fry06} show that the fallback fraction ranges from 0 (no fallback) to 1 (complete fallback) and may depend on the mass of the collapsing star in a complex manner. Here, for simplicity, we assume a constant partial fallback for the formation of all BHs. We account for the release of gravitational binding energy during BH formation and calculate the resulting gravitational mass by lowering it by 20~per~cent. By analogy, the release of gravitational binding energy of an accreting BH is between 6 to 42~per~cent, depending on the spin of the BH \citep{fkr02}.

In the regime of pair-instability SNe (PISNe), no remnant will be left behind \citep{hw02}. While this effect is included in \OurCode, the occurrence of pulsational PISNe \citep{woo17} is not considered in the current version. Pulsational PISNe may eject the outer layers and thus reduce the mass of the star prior to its final core collapse. This reduces the mass of the BH remnant, because the amount of potential fallback material is decreased.

%%%%%%%%%%%%%%%%%%%%%%%%%%%%%%%%%%%%%%%%%%%%%%%%%%
\subsubsection{Supernova kicks in \OurCode}\label{sec:supernovae}
\begin{table}
 \caption{\label{tab:kick_values}Projected 1-dimensional root-mean-square kick velocities $\left(w_{\rm rms}^{\rm 1D}\right)$, or 3-dimensional kick velocity ranges$\,^{\ast}$, applied to various exploding stars in the first and second SN in a binary.}
 \begin{tabular}{lr<{$\unit{}{\kilo\meter\usk\reciprocal\second}$}r<{$\unit{}{\kilo\meter\usk\reciprocal\second}$}}
  \hline
  {\normalsize\rule{0pt}{\f@size pt}}SN type & \multicolumn{1}{c}{first SN} & \multicolumn{1}{c}{second SN}\\
  \hline
  {\normalsize\rule{0pt}{\f@size pt}}Electron capture SN$\,^{\ast}$ & $0-50$ & $0-50$\\
  \hline
  \multicolumn{3}{l}{{\normalsize\rule{0pt}{\f@size pt}}Iron-core collapse SN depending on the NS progenitor:}\\
  -- Isolated star or wide binary & $265$ & $265$\\
  -- Close binary, no H env.$\,^{\ast\ast}$ & $120$ & $120$\\
  -- Close binary, no He env.$\,^{\ast\ast}$ & $60$ & $30$\\
  \hline
  {\normalsize\rule{0pt}{\f@size pt}}Formation of BH$\,^{\ast}$ & $0-200$ & $0-200$\\  
  \hline
 \end{tabular}\\
 {{\normalsize\rule{0pt}{\f@size pt}}
  $^{\ast}$ The stated velocity interval corresponds to 3-dimensional velocities and we applied a flat probability distribution rather than applying a Maxwellian distribution.\\
  $^{\ast\ast}$ For FeCC~SNe in close systems, we applied a bimodal kick distribution such that the above $w_{\rm rms}^{\rm 1D}$ values for a Maxwellian distribution account for 80~per~cent of the cases and in the remaining 20~per~cent of the cases we applied a larger kick using $w_{\rm rms}^{\rm 1D}=\unit{200}{\kilo\meter\usk\reciprocal\second}$ (see \SectionRef{sec:supernovae}).\\
 }
\end{table}

Massive stars usually end their life in a SN. A core-collapse SN ejects the envelope while the core collapses to become a NS or a BH. As the SN is not spherically symmetric, the explosion usually leads to a kick imparted on the newborn compact remnant \citep{jan12}. In the following, we discuss the treatment of SN kicks in \OurCode. A summary is given in \TableRef{tab:kick_values}.

For an EC~SN we apply a flat 3-dimensional kick magnitude distribution up to $\unit{50}{\kilo\meter\usk\reciprocal\second}$. This choice of a small kick is motivated by arguments based on the pre-SN stellar structure as well as SN simulations showing that such SNe usually result in small kicks \citep{plp+04,kjh06,dbo+06}. EC~SNe account for the small population of NSs shown in \FigureRef{fig:ZAMS-CompactObject_Mass} which are produced from the lowest mass ZAMS stars.

For an FeCC~SN a Maxwell-Boltzmann distribution,
\begin{equation}
  f(w) = \sqrt{\frac{54}{\pi}}\product\frac{w^2}{w_{\rm rms}^3}\product\exp\left(-\frac{3}{2}\product\frac{w^2}{w_{\rm rms}^2}\right),
  \label{eq:maxwell_boltzmann_distribution}
\end{equation}
is used for the kick magnitude, $w$. The default value for the projected 1-dimensional root-mean-square velocity is \mbox{$w_{\rm rms}^{\rm 1D}=\unit{265}{\kilo\meter\usk\reciprocal\second}$}, taken from observations of radio pulsars \citep{hllk05}. The 3-dimensional $w_{\rm rms}$ of the Maxwell-Boltzmann distribution is then found by \mbox{$w_{\rm rms}=\sqrt{3}\product w_{\rm rms}^{\rm 1D}$}. We select the kick magnitude and orientation for each SN event using Monte Carlo techniques. The kick orientation is assumed to be isotropic.

If the progenitor of the exploding star loses its hydrogen envelope as a consequence of mass loss to its companion star, $w_{\rm rms}^{\rm 1D}$ is reduced to $\unit{120}{\kilo\meter\usk\reciprocal\second}$. This reduction in kick velocity is motived by a combination of the hypothesis of reduced kicks for stripped stars, see e.g. discussions in \citet{tb96,tkf+17}, and a detailed investigation of the locations and ages of Galactic HMXBs \citep{cc13}. An even more stripped progenitor star, which also loses its helium envelope prior to the core collapse \citep[via Case~BB RLO,][]{hab86a}, is treated with $w_{\rm rms}^{\rm 1D}=\unit{60}{\kilo\meter\usk\reciprocal\second}$, unless this star has a compact object companion. If the exploding star forms the second-born compact star of the binary, it undergoes an ultra-stripped SN \citep{tlm+13,tlp15,sys+15,mmt+17} because of severe mass stripping by the nearby compact object, leaving an almost naked metal core at the time of the explosion. For such ultra-stripped SNe we apply $w_{\rm rms}^{\rm 1D}=\unit{30}{\kilo\meter\usk\reciprocal\second}$, in accordance with the many double NS systems observed with small eccentricities of $e\la 0.2$ \citep{tkf+17}. 

Although the kick magnitudes are generally believed to be smaller for the core collapse of stripped stars, there is evidence from observations that a minor fraction of the FeCC~SNe still produce rather large kicks. For ultra-stripped SNe this is motived by the kinematics of known double NS systems; see detailed discussions in \citet{tkf+17}. According to this work, such a difference in kick magnitudes is possibly related to the mass of the final iron core and thus to the mass of the resulting NS. As a result, for these stripped and ultra-stripped stars we apply in \OurCode a bimodal kick distribution with 80~per~cent of the explosions receiving reduced kick magnitudes as stated above, and the remaining 20~per~cent receiving a large kick of \mbox{$w_{\rm rms}^{\rm 1D}=\unit{200}{\kilo\meter\usk\reciprocal\second}$}. We remind the reader that in a Maxwellian distribution the average 3-dimensional kick magnitude is \mbox{$\sqrt{8/\pi}\product w_{\rm rms}^{\rm 1D}\simeq 1.60\product w_{\rm rms}^{\rm 1D}$}.

Regarding kicks on newly formed BHs less is known \citep{ntv99,jan13,rn15,man16}. Therefore, a simple flat 3-dimensional distribution up to $\unit{200}{\kilo\meter\usk\reciprocal\second}$ is used as our default distribution.

To solve for the post-SN orbital dynamics (also including the SN shell impact on the companion star) we apply the formulae of \citet{tt98}, where a circular pre-SN orbit is assumed. For very wide orbits prior to the second SN (which arise if there is no RLO from the last evolved star) the orbit remains eccentric. Here, the pre-SN separation and orbital velocity are taken at a random orbital phase from a flat distribution of the mean anomaly of the orbit. Any resulting changes of the mass of the companion star from the SN shell impact leads to a new stellar track following the prescription in \AppendixRef{app:place_into_grid}. Depending on the SN mass loss and the kick velocity (magnitude and direction) the system may survive, disrupt or coalesce. A post-SN binary is assumed to coalesce in case the companion star fills its Roche lobe directly after the explosion.

The exact criterion for coalescence at post-SN periastron (at distance $a\product(1-e)$) is somewhat unclear, however, as any periodic mass transfer to the newborn compact object in each orbit may cause the periastron separation to widen while the orbit may circularizes and the semi-major axis decreases (at least for an approximate conservation of orbital angular momentum). Therefore, we take as a criterion for post-SN coalescence a limiting separation of $a\product(1-e^2)$, such that systems will coalesce if the companion star fills its Roche lobe at that separation. We tested the consequence of applying a critical separation of $a\product(1-e)$ vs. $a\product(1-e^2)$ and the difference in final DCO merger rates is only at a few percent level.

%%%%%%%%%%%%%%%%%%%%%%%%%%%%%%%%%%%%%%%%%%%%%%%%%%
\subsubsection{Gravitational wave radiation}\label{sec:gravitational_wave_radiation}
GW radiation leads to a shrinking of the binary orbit. Thus for a tight DCO binary, this may eventually lead to a merger event. \OurCode uses the prescription by \citet{pet64} to calculate the delay time of a merger after the formation of the two compact objects: 
\begin{equation}
  \tmerge = \frac{15}{304}\product\frac{a_0^4\product c^5}{G^3\product m_1\product m_2\product M}\product\Xi(e_0),
  \label{eq:tmerge}
\end{equation}
where $a_0$ is the semi-major axis after the formation of the DCO binary, $m_1$ and $m_2$ are the two component masses, \mbox{$M=m_1+m_2$} is the total mass and
\begin{equation}
 \begin{split}
  \Xi(e_0)\equiv &\left[\left(1-e_0^2\right)\product e_0^{-\frac{12}{19}}\product\left(1+\frac{121}{304}\product e_0^2\right)^{-\frac{870}{2299}}\right]^4\\
           \cdot &\int_0^{e_0}\frac{e^{\frac{29}{19}}\product\left(1+\frac{121}{304}\product e^2\right)^{\frac{1181}{2299}}}{\left(1-e^2\right)^{\frac{3}{2}}}\diff e.
  \label{eq:Xi}
 \end{split}
\end{equation}
This time delay strongly depends on the separation of the two compact objects and the orbital eccentricity, $e_0$. To follow the evolution in separation and eccentricity as a function of time for a given in-spiralling binary would slow down \OurCode considerably. Therefore, such calculations, which are needed for some of the plots presented later (e.g. \FigureRef{fig:observed_NSNS_orbit_MW_beta75}), are done postprocessing.

%%%%%%%%%%%%%%%%%%%%%%%%%%%%%%%%%%%%%%%%%%%%%%%%%%
\subsection{Galactic motion}\label{sec:galactic_motion}
Because a DCO binary often needs a long time to merge it can move a significant distance from its birth site within its host galaxy before the merger event. To follow the motion of a binary within a galaxy, a simple Runge-Kutta 4 integrator is used. As a default gravitational potential we apply a Milky Way-like potential by \citet{as91} which contains a central mass, a disk and a halo. The initial distribution of the binaries simulated with \OurCode follows the mass-density distribution of the Galactic disk component. The initial velocities are set to be in co-rotation with the disk by default. 

%%%%%%%%%%%%%%%%%%%%%%%%%%%%%%%%%%%%%%%%%%%%%%%%%%
\subsection{Computed quantities}\label{sec:computed_quantities}
To keep track of the evolution, \OurCode calculates the age, mass, core mass, radius, luminosity, effective temperature and the envelope structure parameter, $\lambda$, for both stars in the binary. The semi-major axis (orbital period), eccentricity, galactic position and velocity of each binary system is also tracked. Finally, after the DCO is formed, the time until the merger of the two compact objects is determined.

%%%%%%%%%%%%%%%%%%%%%%%%%%%%%%%%%%%%%%%%%%%%%%%%%%
\section{Stellar grids}\label{sec:stellar_grids}
\OurCode interpolates in dense girds of detailed stellar models. The underlying stellar models are calculated with the stellar evolution code BEC \citep[e.g.][and references therein]{ywl10}. In the following, we describe how we calculate the stellar grids from hydrogen-rich stars and helium stars, and how we determine the core-envelope boundary.

%%%%%%%%%%%%%%%%%%%%%%%%%%%%%%%%%%%%%%%%%%%%%%%%%%
\subsection{Hydrogen-rich stars}\label{sec:hydrogen_stars}
On the ZAMS, stars consists mainly of hydrogen. \citet{bdc+11} calculated grids of massive stars at various metallicities. Our computed grid takes its basis in similar calculations, performed with the same stellar code (BEC) but having more frequent full structure output to calculate the structure parameter of the envelope, $\lambda$.

%%%%%%%%%%%%%%%%%%%%%%%%%%%%%%%%%%%%%%%%%%%%%%%%%%
\subsubsection{Milky Way metallicity}\label{sec:MW_metallicity}
\begin{figure}
  \includegraphics[width=\columnwidth]{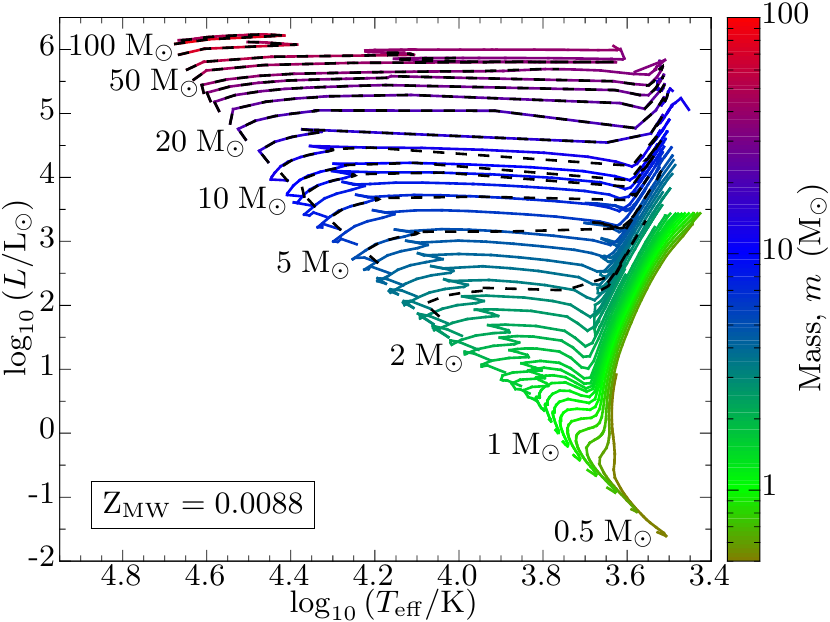}
  \caption{\label{fig:HRD_MW}Hertzsprung-Russell diagram of our non-rotating stars at Milky Way metallicity, see \SectionRef{sec:MW_metallicity}. The mass along the tracks is colour-coded and the black dashed lines are the original models of \citet{bdc+11} which contain less full structure data. For low-metallicity tracks, see \citet{bdc+11} and \citet{sly+15}.}
\end{figure}

\begin{figure}
  \includegraphics[width=\columnwidth]{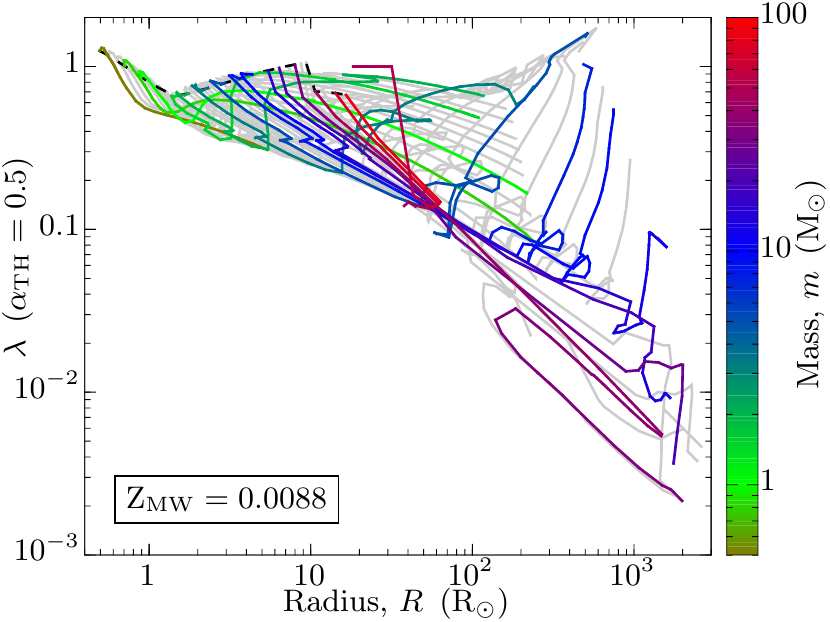}
  \caption{\label{fig:Lambda_MW}Dependence of the envelope binding energy parameter, $\lambda$, on stellar mass, $m$, and  stellar radius, $R$. The grey models are the same as in \FigureRef{fig:HRD_MW}. The models with an initial mass of $0.5$, $0.8$, $1.0$, $1.5$, $2.0$, $3.2$, $5.0$, $8.0$, $10.0$, $15.0$, $20.0$, $30.0$, $50.0$, $80.0$ and $\unit{100}{\Msun}$ are shown in colour according to their mass at a given point along the tracks. The dashed line marks the ZAMS values (where $\lambda$, strictly speaking, is not well defined).}
\end{figure}

For a Milky Way (MW)-like metallicity (\mbox{$Z={\rm Z}_{\rm MW}\equiv0.0088$}), the grid calculated by \citep{bdc+11} contains evolved models starting from ZAMS stars with masses from $\unit{3}{\Msun}$ to $\unit{100}{\Msun}$. To make use of this grid for the \OurCode code, we extended it to lower ZAMS masses (down to $\unit{0.5}{\Msun}$) and added some more intermediate-mass tracks as well. The grid is also refined by evolving some existing models of \citet{bdc+11} to a more advanced nuclear burning stage\footnote{The end of the calculation of the detailed stellar models is caused by numerical instabilities during carbon burning or when the density of the outermost envelope becomes too low.} and by adding some intermediate data points with full information about the stellar structure. In all these cases, we used the same parameters and version of BEC as \citet{bdc+11}. An extension to higher masses with similar parameters were numerically unstable with this version of BEC. Therefore, the grid is simply extrapolated to higher masses if needed; see e.g. the curve beyond the last data point in the lower left panel of \FigureRef{fig:ZAMS-CompactObject_Mass} where the wind mass loss becomes very strong. We caution that this technique implies some uncertainty, although this is only relevant in a few of our simulated binaries where a very massive star accretes a substantial amount of material. In the simulations presented here, we limit the initial masses to those covered by our grid.

\FigureRef{fig:HRD_MW} shows a Hertzsprung-Russell diagram (HRD) of the models at MW metallicity for non-rotating stars. Compared to the HRDs shown in \citet{bdc+11}, \FigureRef{fig:HRD_MW} only contains line-connected data points where information on the full stellar structure is saved. Before using the grid data, the required quantities are extracted for \OurCode. These are the total mass, $m$, the time since ZAMS, $t$, the photospheric radius, $R$, the core mass, $m_{\rm core}$ -- the chosen core-envelope boundary is discussed in \SectionRef{sec:core-envelope_boundary} -- the luminosity, $L$, the effective temperature, $T_{\rm eff}$ and the structure parameter of the envelope, $\lambda$ (\EquationRef{eq:lambda}). Two $\lambda$-parameters are saved: one only accounting for the gravitational binding energy, and one taking the additional internal energy (including recombination energy) into account. In this way, one can choose the considered amount of internal energy when running \OurCode. To speed-up our code, with the knowledge that it uses linear interpolations (see \AppendixRef{app:interpolation}), the tables of the stellar tracks with full stellar structure models are reduced to a smaller number of supporting data points (typically about 100, or less, along each stellar track) such that all extracted stellar quantities are at all times precise to within 2~per~cent of a model based on the full amount of calculated data.

\FigureRef{fig:Lambda_MW} shows how $\lambda$ depends on stellar mass and radius. This structure parameter is very crucial for the all-important CE prescription. It is clear that treating CE evolution for all stellar masses and at all evolutionary stages using a constant $\lambda$ is a poor approximation given that $\lambda$ varies by more than two orders of magnitude \citep[see e.g.][]{dt00,dt01,ktl+16}. In this respect, it is surprising to see the use of a constant $\lambda$-value in several recent papers on binary population synthesis, e.g. on DCO merger rates for LIGO, and which therefore are quite likely to lead to erroneous results.

The $\lambda$--values in \FigureRef{fig:Lambda_MW} are calculated for \mbox{$\alpha _{\rm th}=0.5$}, i.e. by taking $50$~per~cent of the internal energy of the envelope into account as in our default setup. The influence of the amount of internal energy included on our results is discussed in \AppendixRef{app:alpha_th}.

We caution against the method of applying calculated $\lambda$-values from the literature to stellar grids based on a different stellar evolution code. For example, when combining our stellar tables based on the BEC code with $\lambda$-values based on the Eggleton code \citep[taken from][]{dt00,dt01}, the GW merger rates change significantly compared to the self-consistent treatment applied in \OurCode. In the former case, the rates for double NS and BH-NS binaries increase by roughly 1 dex. The NS-BH systems and double BH binaries increase by roughly 2 dex. These large discrepancies demonstrate the importance of $\lambda$ and stellar tracks being calculated with the same stellar evolution code.

The mass of the stellar envelope usually decreases as nuclear burning shifts material from the inner edge of the hydrogen envelope to the helium core and, at the same time, stellar wind material is lost from the outer edge of the envelope. In the most massive stars, very strong winds strip the whole envelope. The only way to increase the mass of the envelope in single star models is to mix some core material into the envelope. The radius, on the other hand, increases during the expansion phases on the giant branch(es). Usually the binding energy of the envelope, $|E_{\rm bind}|$, decreases during stellar evolution. Whether $\lambda$ decreases or increases depends on the dominating term in the change of mass, radius or binding energy \citep[e.g.][]{ktl+16}.

Besides our default, non-rotating simulations, a data set for rapidly rotating stars at MW-like metallicity from \citet{bdc+11} is also available in \OurCode. In this case, however, the mass range is limited to \mbox{$\unit{3}{\Msun}\leq m_{\rm ZAMS}\leq\unit{100}{\Msun}$}. Rotating stars are not included in the study presented here.

%%%%%%%%%%%%%%%%%%%%%%%%%%%%%%%%%%%%%%%%%%%%%%%%%%
\subsubsection{Lower metallicities}\label{sec:lower_metallicity}
In the present investigation of GW merger sources, it is important also to consider binaries in low-metallicity environments. We thus include stellar tracks for metallicities equal to those of the Large Magellanic Cloud (LMC, \mbox{$Z={\rm Z}_{\rm LMC}\equiv0.0047$}) and the Small Magellanic Cloud (SMC, \mbox{$Z={\rm Z}_{\rm SMC}\equiv0.0021$}), which are taken form \citet{bdc+11}. Our lowest metallicity included is at a similar level as that of the dwarf galaxy IZwicky18 (IZw18, \mbox{$Z={\rm Z}_{\rm IZw18}\equiv0.0002\simeq\unit{0.02}{{\rm Z}_{\sun}}$}) and these stellar tracks are adopted from \citet{sly+15} which cover a mass range from $\unit{4}{\Msun}$ to $\unit{294}{\Msun}$. Note, the metal distribution among the chemical elements is slightly different between the different metallicity tracks mentioned above.

%%%%%%%%%%%%%%%%%%%%%%%%%%%%%%%%%%%%%%%%%%%%%%%%%%
\subsection{Helium stars}\label{sec:helium_stars}
\begin{figure}
  \includegraphics[width=\columnwidth]{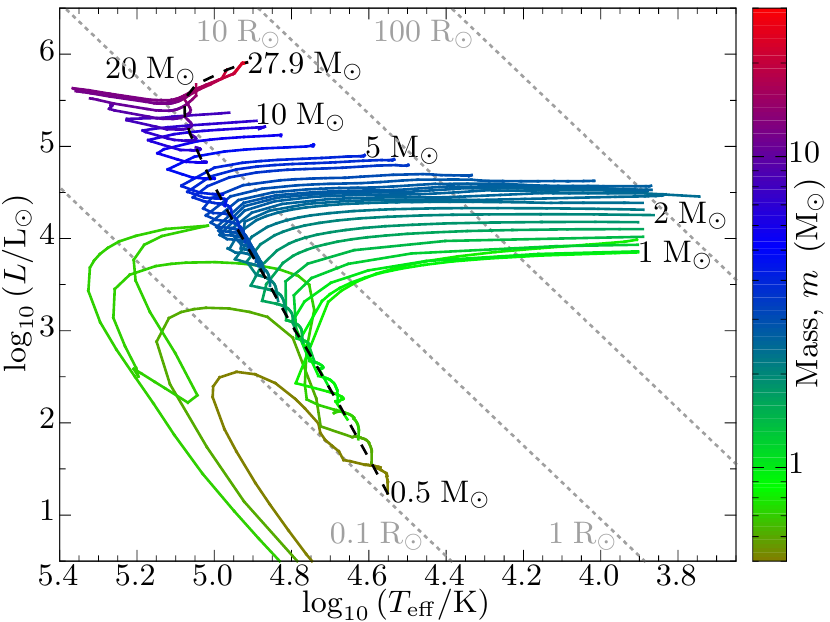}
  \caption{\label{fig:HRD_He}Evolutionary tracks in the HRD of non-rotating naked helium stars at MW-like metallicity, see \SectionRef{sec:helium_stars}. The mass along the tracks is colour-coded. The grey dotted lines indicate He-star radii of $0.1$, $1$, $10$ and $\unit{100}{\Rsun}$. The black dashed line marks the He-ZAMS.}
\end{figure}

\OurCode assumes that the mass-transfer stage peels off the whole hydrogen-rich envelope and leaves a naked helium-rich core as remnant. \FigureRef{fig:HRD_He} shows the tracks of naked helium stars evolved with BEC. The plotted region with an over-density of tracks contains the isolated helium star models of \citet{tlp15}. We calculated additional models following the same recipe. Here again, we only plot the data points with full stellar structure data. The helium stars with the lowest masses do not evolve into a giant stage and evolve directly onto the WD cooling track during the helium shell burning. The $\unit{0.7}{\Msun}$ model shown in \FigureRef{fig:HRD_He} evolves through a helium shell flash and a thermal pulse. This phenomenon is similar to the thermal-pulse driven hydrogen shell flashes \citep[e.g.][]{gau13,imt+16}.

For the most massive of the helium stars in \FigureRef{fig:HRD_He}, the ZAMS turns to lower effective temperatures. At the same time they develop large inflated envelopes \citep{ppl06,gfl+16}. Such an inflated envelope is efficiently ejected by the winds of these stars. Therefore, their photospheric radius decreases during the first part of their evolution (evolving to higher effective temperature in \FigureRef{fig:HRD_He}). When a star with such an inflated envelope overfills its Roche lobe, the mass loss is not significantly increased compared to the wind mass loss \citep{ktl+16}. Therefore, mass-transfer algorithms like those described in \SectionsRef{sec:roche-lobe_overflow} and \ref{sec:common_envelope} should only be applied when the non-inflated part of the star fills its Roche lobe. To determine the boundary between the inflated and non-inflated part we follow \citet{sglb15}. 

For helium stars with ZAMS masses up to $\unit{3.5}{\Msun}$ and evolving to the stage of Case~BB RLO with a compact object accretor (i.e. post-HMXB/CE evolution), \OurCode uses the recent results of \citep{tlp15} to determine their evolution.

We use the same helium star tracks for all the different metallicities. However, to account for the major effects of metallicity, we rescale the wind mass-loss rates according to \citet{hpt+15}. Applying different mass-loss rates affects the final properties of the post-helium stars \citep[e.g.][]{dt03}.

%%%%%%%%%%%%%%%%%%%%%%%%%%%%%%%%%%%%%%%%%%%%%%%%%%
\subsection{Core-envelope boundary}\label{sec:core-envelope_boundary}
During a mass-transfer phase (RLO or CE), it is always assumed that the whole envelope is lost and only the core remains. Therefore, a robust definition of the boundary which separates the core from the envelope is needed. Many different criteria exist to define this boundary and they often yield different results \citep{td01}.

On the one hand, the boundary should be located outside the hydrogen-depleted core. On the other hand, there should only be a small amount of hydrogen left within the core otherwise the exposed core would expand further and continue the mass transfer. In \OurCode, we apply the simple criterion that the hydrogen abundance in mass is \mbox{$X=0.1$} at the bifurcation point at the onset of the mass transfer \citep[e.g.][]{dt00}. For further discussions on this topic we refer to \citet{td01,iva11,ktl+16}.

%%%%%%%%%%%%%%%%%%%%%%%%%%%%%%%%%%%%%%%%%%%%%%%%%%
\section{Results}\label{sec:results}
\begin{table*}
 \caption{\label{tab:standard}Initial values and default settings of key input physics parameters.}
 \begin{tabular}{lll}
  \hline
  name & value & note\\
  \hline
  {\normalsize\rule{0pt}{\f@size pt}}number of simulated binaries, $N$ & $10^{9}$ & our results converge for $N\geq 3\times 10^8$\\
  primary mass, $m^{\rm p}_{\rm ZAMS}$ & $\in\unit{[4:100]}{\Msun}$ & Salpeter IMF, see \SectionRef{sec:masses}\\
  secondary mass, $m^{\rm s}_{\rm ZAMS}$ & $\in\unit{[1:100]}{\Msun}$ & from mass ratio, see \EquationRef{eq:massratio} and \SectionRef{sec:masses}\\
  semi-major axis, $a$ & $\in\unit{[2:10\,000]}{\Rsun}$ & flat in $\log(P)$, see \SectionRef{sec:initial_orbital_parameters}\\
  eccentricity, $e$ & $\enspace0$ & initially circular orbit, see \SectionRef{sec:initial_orbital_parameters}\\
  metallicity, $Z$ & ${\rm Z}_{\rm MW}$ & Milky Way-like, see \SectionRef{sec:further_parameters}\\
  rotation, $v_{\rm rot}$ & $\enspace\unit{0}{\kilo\meter\usk\reciprocal\second}$ & non-rotating stars, see \SectionRef{sec:further_parameters}\\
  \hline
  {\normalsize\rule{0pt}{\f@size pt}}wind mass loss, $\alpha_{\rm RLO}$ & $\enspace0.20$ & during RLO \citep{spv97}, see \SectionRef{sec:roche-lobe_overflow}\\
  minimum mass ejection by accretor, $\beta_{\rm min}$ & $\enspace0.75$ & during RLO \citep{spv97}, see \SectionRef{sec:roche-lobe_overflow}\\
  circumbinary torus mass transfer, $\delta_{\rm RLO}$ & $\enspace0$ & during RLO \citep{spv97}, see \SectionRef{sec:roche-lobe_overflow}\\
  circumbinary torus size, $\gamma$ & $\enspace2$ & during RLO \citep{spv97}, see \SectionRef{sec:roche-lobe_overflow}\\
  CE efficiency parameter, $\alpha_{\rm CE}$ & $\enspace0.50$ & during CE, see \SectionRef{sec:common_envelope}\\
  fraction of internal energy, $\alpha_{\rm th}$ & $\enspace0.50$ & during CE, see \SectionRef{sec:common_envelope}\\
  mass ratio limit, $q_{\rm limit}$ & $\enspace2.5$ & criterion for stable / unstable mass transfer, see \SectionRef{sec:roche-lobe_overflow}\\
  kick velocity, $w$ & $>\unit{0}{\kilo\meter\usk\reciprocal\second}$ & from the distribution of SN kicks, see \TableRef{tab:kick_values} in \SectionRef{sec:supernovae}\\
  \hline
 \end{tabular}
\end{table*}

In this section, we present the outcomes of binary population synthesis runs with \OurCode. The parameters used in our default simulation are summarised in \TableRef{tab:standard}. Our choices of parameters resemble those of many other binary population synthesis investigations discussed in \SectionRef{sec:compare_merger_rates}, with the exception of the accretion efficiency during RLO. We adopt highly inefficient mass transfer with an accretion efficiency of only \mbox{$\epsilon\leq 0.05=1-\alpha_{\rm RLO}-\beta_{\rm min}$}. This choice leads to DCO results which best represent the observational data, especially the double NS systems, see below in \SectionRef{sec:compact_object_masses_double_neutron_star_binaries} and \FigureRef{fig:final_masses_formation_MW_NSNS}. Furthermore, evidence to support highly non-conservative mass transfer during Case~A and Case~B RLO in massive binaries was investigated by \citet{plv05}, and more recently presented by \citet{sl16} who found $\epsilon < 0.20$ to reproduce the observed Galactic population of WR/O-star binaries. Additional observational evidence for such inefficient mass transfer comes from \citet{fdh94} and \citet{dph07}. Some investigations of less massive binaries, however, suggest more conservative mass transfer \citep{pcwh91,sgd+18}. The accretion efficiency may depend on e.g. the stellar masses and the mass ratio, besides the composition of the stars. For further discussions and application of efficient mass transfer in \OurCode, see \SectionRef{sec:mass-transfer_efficiency} and \AppendixRef{app:efficient_mass_transfer}, respectively.

The synthesized data presented here is obtained by simulating \mbox{$N=10^{9}$} binary systems. In general, we find that our DCO merger rates converge with \mbox{$N\geq 3\times 10^8$} -- except for the double BH systems where statistical noise remains at the 2~per~cent level. In a MW-like galaxy, we assume a constant star formation rate of one binary per year with a primary star mass $>\unit{0.8}{\Msun}$ \citep{htp02}. This rate is rescaled according to the adopted primary mass range and the IMF. It is also applied at other metallicities to mimic the properties of both a young MW-like galaxy and a present-day observable MW.

As a comparison to our simulations with a MW-like metallicity, throughout this section we also present the results based on a low-metallicity case. There, the metallicity is set to \mbox{$Z={\rm Z}_{\rm IZw18}=0.0002$} although the initial mass ranges are slightly changed to $\unit{[5:150]}{\Msun}$ for both the primary and the secondary mass to allow more massive stars at lower metallicity \citep{kgc12}. In \SectionRef{sec:comparison_to_other_work}, we also present properties of DCO mergers based on LMC and SMC metallicities, thus simulating a total of four different metallicities for the GW sources.

%%%%%%%%%%%%%%%%%%%%%%%%%%%%%%%%%%%%%%%%%%%%%%%%%%
\subsection{Progenitor zero-age main-sequence masses}\label{sec:progenitor_ZAMS_masses}
\begin{figure}
  \includegraphics[width=\columnwidth]{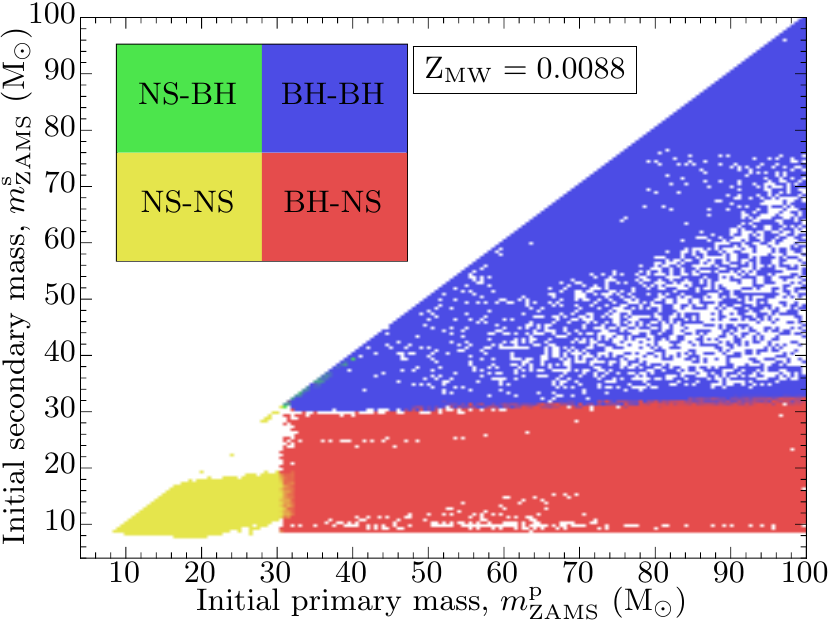}
  \includegraphics[width=\columnwidth]{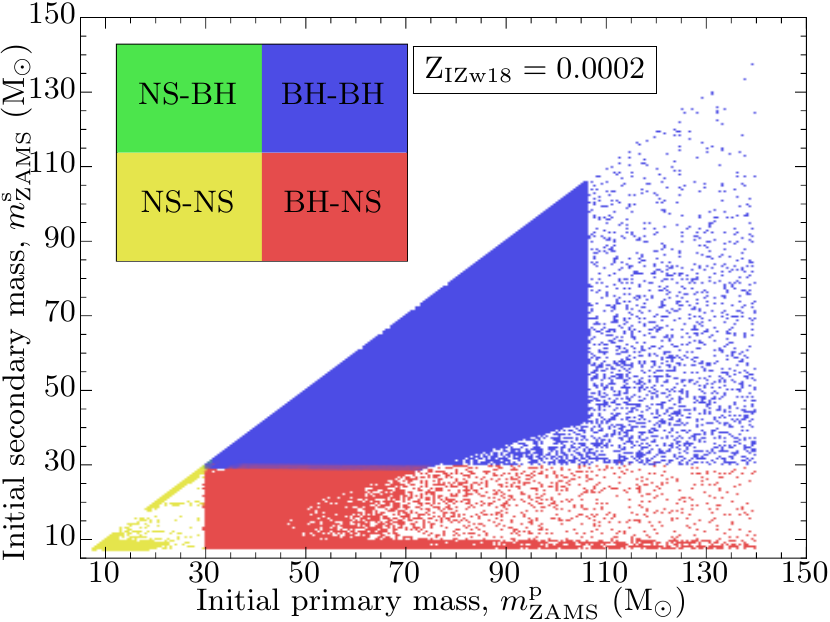}
  \caption{\label{fig:initial_masses_beta75}Progenitors of the systems forming a DCO binary. The upper panel is for a MW-like metallicity \mbox{($Z={\rm Z}_{\rm MW}=0.0088$)} and the lower panel shows systems at low metallicity \mbox{($Z={\rm Z}_{\rm IZw18}=0.0002$)}. Their final compact object masses are shown in \FigureRef{fig:final_masses_formation_beta75}. Colour-coded is the type of the two compact objects formed from the initial primary and secondary star, respectively. This type is not necessarily the formation order. In some cases the secondary star forms a compact object first. The barely visible NS-BH systems originate from ZAMS binaries with a mass ratio close to one and \mbox{$\unit{30}{\Msun}\la m_\mathrm{ZAMS}^\mathrm{p}\la\unit{40}{\Msun}$} (\SectionRef{sec:progenitor_ZAMS_masses}).}
\end{figure}

The progenitor masses of the stellar components of the binaries which form a bound system of two compact objects are shown in \FigureRef{fig:initial_masses_beta75}. In each pixel, the colour is a mixture reflecting the relative formation frequency of the different binary types. White regions indicate where no final DCO binary is formed.

The double NS systems (yellow) originate from binaries where both components are initially less massive than $\unit{33}{\Msun}$ (bottom left). The paucity of these systems produced from secondary stars with masses above $\unit{20}{\Msun}$ is explained by the vast majority of double NS progenitor systems evolving through a CE phase. It is difficult for an in-spiralling NS to successfully eject the envelope of a donor star with \mbox{$m_\mathrm{ZAMS}^\mathrm{s}\ga \unit{20}{\Msun}$} at MW metallicity \citep{ktl+16}.

The more massive the primary star is, the more likely it produces a BH, eventually leading to a BH-NS binary (bottom right, red region). A slight overlap is seen, but less pronounced in the low-metallicity case. In such overlapping areas where the red region becomes yellowish or blueish, the binaries originate from a primary or secondary star too massive to produce a NS in single star evolution. However, as a result of mass transfer and mass loss these systems end up producing NSs anyway. The BH-NS binaries at the very bottom of the population with secondary ZAMS masses less than $\unit{10}{\Msun}$, especially at low metallicty, are mainly wider systems where the NS forms with small kicks, often by EC~SNe.

Few NS-BH systems (green) form in our default simulations. Since mass accretion is needed for the secondary star to produce a BH in cases where the primary star produces a NS, the choice of a small accretion efficiency $\epsilon$ (\SectionRef{sec:results}) hinders the formation of NS-BH systems. Those few NS-BH binaries that do form have an initial mass ratio close to one. In most cases mass transfer slows down the evolution of the primary to the extent that the secondary forms the BH first before the initial primary becomes a NS. Further discussions on the formation of NS-BH binaries are given in \SectionRef{sec:NSBH_rare}.

The largest region in \FigureRef{fig:initial_masses_beta75} is blue indicating the formation of double BH binaries. Note, double BH systems occupy the largest area in the phase space of ZAMS masses but are not the most common DCO systems formed at MW-like metallicity (see \TableRef{tab:formationrates_beta75} in \SectionRef{sec:compact_object_masses} and \FigureRef{fig:Tmerge_LIGO_normalised_beta75} in \SectionRef{sec:gravitational_wave_radiation_and_merger}). The diagonal border separating the dense and the sparse populated regions of both the double BH and the BH-NS binaries is caused by the adopted mass-ratio limit, $q_{\rm limit}$, to differentiate between stable and unstable mass transfer.

In the low-metallicity case, the formation of double BHs dominates for two reasons: i) the stellar winds are less strong and therefore create more massive BH progenitors, resulting in more massive BHs. Massive BHs can more easily eject the CE during in-spiral and thus eventually produce a BH-BH binary. The weaker stellar winds also result in more massive companion stars such that these binaries survive SN kicks more easily. ii) a low-metallicity environment allows for more massive ZAMS stars to form \citep{kgc12} and, assuming the star-formation rate and the IMF slope remain constant, this leads to the formation of more BHs (\TableRef{tab:rate_variations2}).

Finally, no BHs are formed from single stars with an initial mass above $\unit{140}{\Msun}$ in our low-metallicity simulations (empty region in the lower panel of \FigureRef{fig:initial_masses_beta75}) because such massive stars end their life in a PISN with no compact-object remnant \citep{hw02}. In our binary systems, however, PISNe are assumed to occur at smaller initial masses. In \OurCode, stripped stars with initial masses already exceeding $\unit{106}{\Msun}$ produce helium stars with masses above $\unit{64}{\Msun}$ which are expected to obtain lower central densities at high temperatures and result in a PISN \citep{hw02}. This explains the vertical border at $\unit{106}{\Msun}$ separating the dense and the sparse region of BH-BH systems in the lower panel of \FigureRef{fig:initial_masses_beta75}. Pulsational PISNe \citep{woo17} are thought to remove a substantial amount of mass from stars shortly before their final core collapse and therefore lead to a similar effect of further reducing the upper mass limit of the BHs formed below the ordinary pair instability regime.

In \AppendixRef{app:formation_channels} we list in detail the various formation channels leading to the different DCO binaries.

%%%%%%%%%%%%%%%%%%%%%%%%%%%%%%%%%%%%%%%%%%%%%%%%%%
\subsection{Compact object masses}\label{sec:compact_object_masses}
\begin{figure}
  \includegraphics[width=\columnwidth]{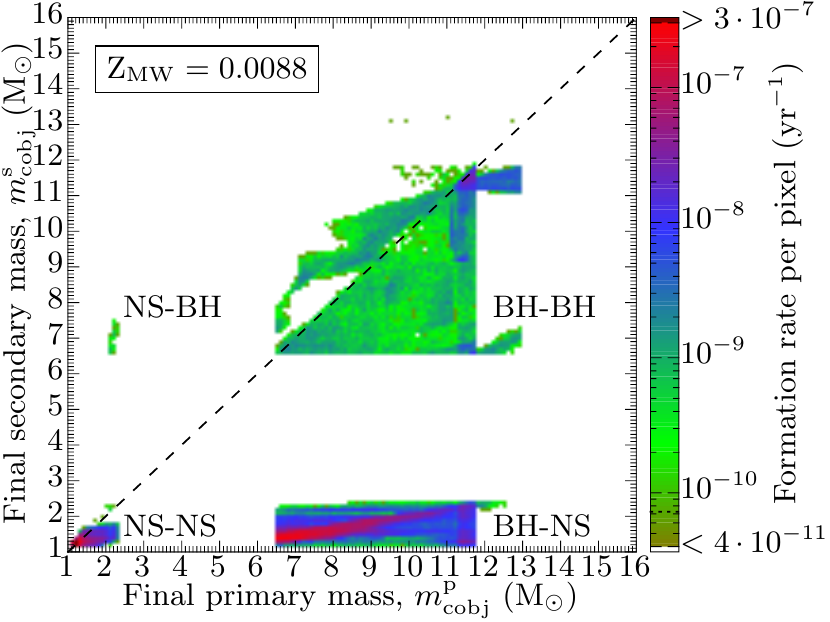}
  \includegraphics[width=\columnwidth]{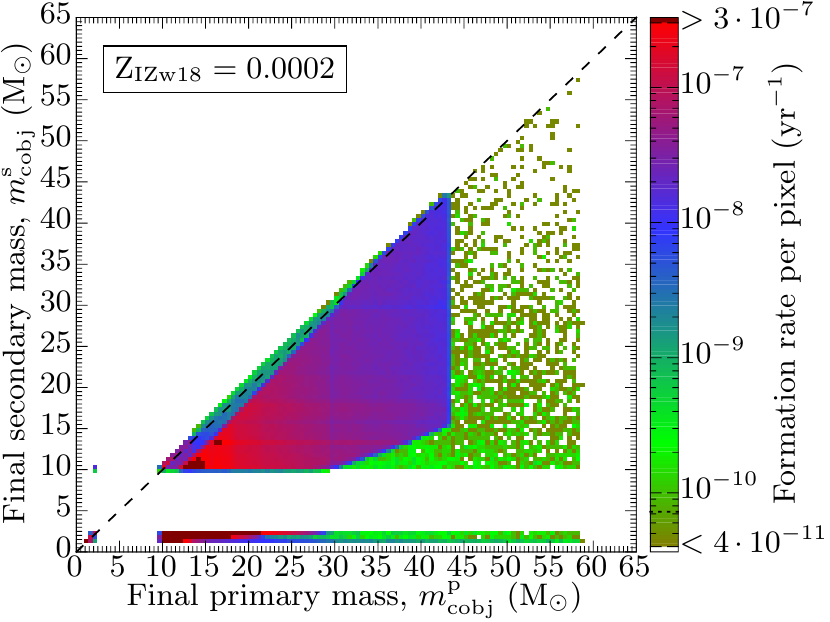}
  \caption{\label{fig:final_masses_formation_beta75}Masses of NSs and BHs in DCO binaries at high metallicity (\mbox{$Z={\rm Z}_{\rm MW}=0.0088$}, upper panel) and low metallicity (\mbox{$Z={\rm Z}_{\rm IZw18}=0.0002$}, lower panel). The initially more massive star produces the primary compact object with the mass, $\mCO^{\rm p}$ while the secondary star produces a compact object with a mass $\mCO^{\rm s}$. The formation rate per grid cell of such systems in a MW-like galaxy is colour-coded. The minor tics show the pixel size. The black dashed line indicates equally massive compact objects. For more details see \SectionRef{sec:compact_object_masses}. A more efficient mass transfer would change this picture, cf. \FigureRef{fig:final_masses_formation}.}
\end{figure}

\FigureRef{fig:final_masses_formation_beta75} shows the plane of the compact object masses ($\mCO$) for NSs and BHs produced by our simulations. There are four distinct regions labelled in the upper panel. Those are the double NS systems in the lower left corner, the mixed BH-NS and NS-BH systems populating the regions just above the x-axis at higher masses and the small population close to the y-axis around $\unit{7}{\Msun}$, and finally the double BH systems occupying the top right part of the plot \mbox{($\mCO>\unit{6}{\Msun}$)}. The double BH systems show two separated populations at high metallicity (upper panel). The majority of the systems have BH masses \mbox{$\mCO\la\unit{12}{\Msun}$} and the more massive of the two BHs is the remnant of the primary star. In addition, a few double BH binaries form with secondary BH masses \mbox{$\mCO^{\rm s}>\unit{13}{\Msun}$}. These systems have a less massive primary BH. For a more efficient mass accretion process this sub-population would be more populated (\AppendixRef{app:compact_object_masses}).

At high metallicity, the masses of the BHs are moderate because of relatively strong wind mass loss during the stellar evolution. Given that the resulting masses of BHs and NSs in our models mostly increase monotonically with the mass of the progenitor stars for isolated evolution, at least all systems above the black dashed line in \FigureRef{fig:final_masses_formation_beta75} had some binary interaction during their evolution.

At low metallicity (\FigureRef{fig:final_masses_formation_beta75}, lower panel) the DCO systems are dominated by systems containing double BHs. These BHs can reach significantly higher masses because the progenitors have weaker winds and lose less mass. This also results in a higher survival rate of CE evolution where such massive BHs more easily strip-off the envelope of their donor star companion by spiral-in \citep{ktl+16}. The lower panel of \FigureRef{fig:final_masses_formation_beta75} thus covers a much larger BH mass range compared to the upper panel at high metallicity.

\begin{table}
 \caption{\label{tab:formationrates_beta75}Formation rates of DCO binaries in a MW-like galaxy at two different metallicities with our default setting. The binary types quote the first and second formed compact object.}
 \begin{tabular}{cr@{$\times$}l<{$\unit{}{\yr^{-1}}$}r@{$\times$}l<{$\unit{}{\yr^{-1}}$}}
  \hline
   Formation rates & \multicolumn{2}{c}{${\rm Z}_{\rm MW}=0.0088$} & \multicolumn{2}{c}{${\rm Z}_{\rm IZw18}=0.0002$}\\
 \hline
  {\normalsize\rule{0pt}{\f@size pt}}NS-NS & $6.81$ & $10^{-6}$ & $1.53$ & $10^{-5}$\\
  {\normalsize\rule{0pt}{\f@size pt}}NS-BH & $5.49$ & $10^{-9}$ & $1.65$ & $10^{-8}$\\
  {\normalsize\rule{0pt}{\f@size pt}}BH-NS & $1.49$ & $10^{-5}$ & $4.27$ & $10^{-5}$\\
  {\normalsize\rule{0pt}{\f@size pt}}BH-BH & $2.27$ & $10^{-6}$ & $9.65$ & $10^{-5}$\\
  \hline
 \end{tabular}
\end{table}

The minimum BH mass originating from a non-stripped star is larger the lower the metallicity, as there is more mass left in the envelope which can contribute to the BH. \TableRef{tab:formationrates_beta75} shows the total formation rates of DCO binaries at two different metallicities of galaxies otherwise similar to the MW. \FigureRef{fig:final_masses_formation_beta75} and succeeding figures differentiate formation and merger rates according to the grid cells of different mass ranges.

In the following subsections we discuss the different binary types formed at high metallicity, as shown in the upper panel of \FigureRef{fig:final_masses_formation_beta75}.

%%%%%%%%%%
\subsubsection{Double black hole binaries}\label{sec:compact_object_masses_double_black_hole_binaries}
First, we look at the remnant masses in the double BH population evolved at a MW-like metallicity as shown in the upper panel of \FigureRef{fig:final_masses_formation_beta75}. The largest visible sub-population creates a triangular shape like the progenitors shown in \FigureRef{fig:initial_masses_beta75}. All binaries without interactions, or only minor interactions, fall into this region. The binaries where either the secondary star becomes the more massive BH or the primary BH is produced with a mass above $\unit{11.8}{\Msun}$ originate from progenitor binaries which experienced intensive interactions. To obtain a secondary BH more massive than the primary BH, at an early stage in the primary's evolution it has to lose or transfer some material to its companion star. In our simulations, mainly Case~A mass transfer is responsible for a final mass ratio reversal. Furthermore, the initial mass ratio should be close to one.

At high metallicity, because of strong winds, the remnant core masses of our very massive single star models ($m_\mathrm{ZAMS}^\mathrm{p}>\unit{80}{\Msun}$) can become less massive than those which are left in stars which have their envelope removed by mass transfer while still on the main sequence. Therefore, our most massive Case~A RLO primaries of MW metallicity can form slightly more massive BHs than compared to single stars.

%%%%%%%%%%
\subsubsection{Mixed binaries}\label{sec:compact_object_masses_mixed_binaries}
Mixed binaries (BH-NS or NS-BH) are the most frequent DCO systems resulting from our default simulations, see \TableRef{tab:formationrates_beta75}. The labels in \FigureRef{fig:final_masses_formation_beta75} refer to the compact objects resulting from the primary and secondary stars, respectively. In most cases this is also the formation order of the two remnants. However, in some rare cases the secondary star (after accretion) evolves faster than the primary star and thereby the formation order reverses. In all the cases where the primary star produces a NS while the secondary star produces a BH, the primary star has to lose or transfer mass to the secondary star. Furthermore, because of our assumption of inefficient accretion, both stars must have ZAMS masses close to the border between forming a NS or a BH.

In the vast majority ($99.96$~per~cent) of mixed systems the primary star produces a BH and the secondary star produces a NS, cf. upper panel of \FigureRef{fig:final_masses_formation_beta75}. In those binaries the parameter range of initial configurations is much larger compared to the NS-BH systems. Here again, the most massive of the primary BHs are produced by Case~A RLO from the primary progenitor star ($m_\mathrm{ZAMS}^\mathrm{p}>\unit{80}{\Msun}$). This is seen as an upper right extension of the red diagonal in the BH-NS binaries on the upper panel of \FigureRef{fig:final_masses_formation_beta75}. Whenever there is a limiting or preferred initial mass ratio, a diagonal structure from the lower left to the upper right is created, as more massive progenitors usually produce more massive NSs or BHs.

%%%%%%%%%%
\subsubsection{Double neutron star binaries}\label{sec:compact_object_masses_double_neutron_star_binaries}
\begin{figure}
  \includegraphics[width=\columnwidth]{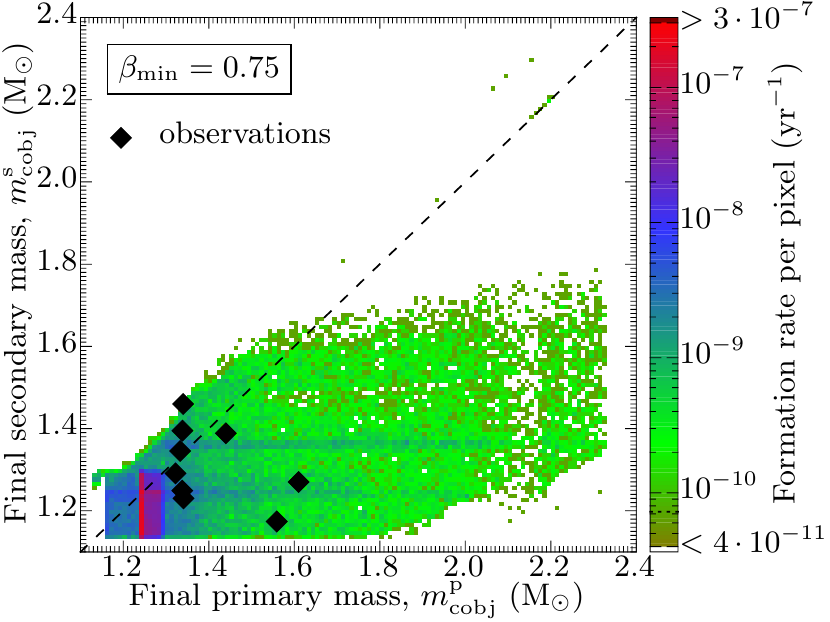}
  \includegraphics[width=\columnwidth]{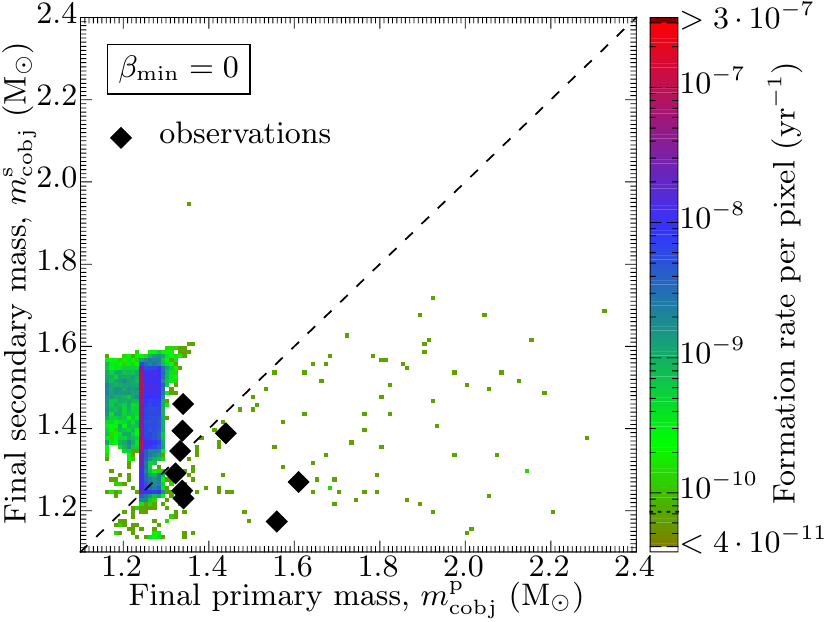}
  \caption{\label{fig:final_masses_formation_MW_NSNS}The upper panel shows a zoom-in on the double NS systems plotted in the upper panel of \FigureRef{fig:final_masses_formation_beta75}. Because the resolution is increased, the formation rate per grid cell is smaller. In the lower panel, efficient mass transfer \mbox{$\beta_{\rm min}=0$} is assumed. For more details see \SectionRef{sec:compact_object_masses_double_neutron_star_binaries}. The observational data shown contain only double NS systems in which the individual masses can be inferred, see \citet[][and references therein]{tkf+17}, \citet{cck+17}, \citet{fer17} and \citet{lsk+18}.}
\end{figure}

\begin{figure}
  \includegraphics[width=\columnwidth]{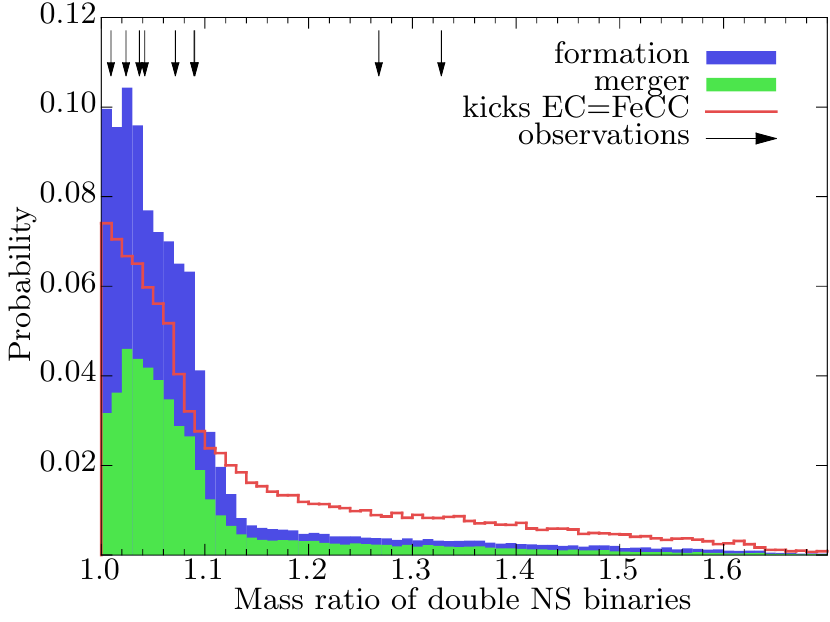}
  \caption{\label{fig:final_massratios_MW_beta75_NSNS}Histogram of the mass ratios of our simulated double NS binaries. In blue colour is shown the probability distribution for all formed a double NS systems, whereas the green colour applies to those which will merge within a Hubble time. The red line shows the mass ratio distribution of double NS systems formed when the kicks applied to EC and FeCC~SNe are similar, see discussion in \SectionRef{sec:ECvsFECCSN}. The observational data of mass ratios from binary radio pulsars \citep{tkf+17,cck+17,fer17,lsk+18} are indicated by arrows. The largest mass ratio of roughly $1.33$ is that of PSR~J0453+1559 \citep{msf+15}. GW170817 (not shown) had a mass ratio somewhere between \mbox{$1.0<q<1.4$} \citep{aaa+17c}.}
\end{figure}

\FigureRef{fig:final_masses_formation_MW_NSNS} shows a zoom-in on the double NS systems plotted in the upper panel of \FiguresRef{fig:final_masses_formation_beta75} and \ref{fig:final_masses_formation}, respectively. The black data points are measured NS masses where we have assumed that the observed recycled pulsars originate from the primary stars. The measurement error bars are much smaller than the symbol sizes. 

The observed data matches the distribution of our theoretical simulations using a highly inefficient mass-transfer process much better than in the case of efficient mass-transfer, cf. upper and lower panel, respectively. This is the main reason for the choice of \mbox{$\beta_{\rm min}=0.75$} as our default. As seen in the lower panel, efficient mass transfer strongly suppresses the formation of primary (recycled) NSs with masses $\ga\unit{1.35}{\Msun}$. The reason for this is that the first mass-transfer phase from the primary to the secondary star is expected to become unstable when the mass-transfer rate exceeds the Eddington limit of the secondary star.

More than 98 per cent of the double NS binaries experience mass transfer from the primary star to its companion prior to the first SN. The majority of these systems undergo Case~B RLO.

Our results in \FigureRef{fig:final_masses_formation_MW_NSNS} show a clear over-density of primary NS masses around $\unit{1.25}{\Msun}$. These NSs are created by an EC~SN instead of an FeCC~SN \citep[e.g.][]{spr10}. This NS formation channel is clearly favoured in our default simulations. EC~SNe are assumed to produce small NS kicks which makes it easier for the post-SN binary to remain bound and yet often tight enough to initiate a subsequent CE~phase. As the binary separation shrinks significantly during a CE and spiral-in phase, the kick of the second formed NS can be much larger without disrupting the binary \citep[e.g.][]{tkf+17}. Nevertheless, a clear discrepancy between our simulated primary NS masses, with a strong over-density of EC~SNe, and observational data is seen and potential explanations are discussed in \SectionRef{sec:ECvsFECCSN}.

\FigureRef{fig:final_massratios_MW_beta75_NSNS} shows a histogram of the mass ratios, $q$, of the simulated double NS systems plotted in the upper panel of \FigureRef{fig:final_masses_formation_MW_NSNS}. The agreement with observed systems looks reasonable, given the small number statistics. According to our simulations, a very few systems are even produced with a mass ratio $>1.7$.

%%%%%%%%%%%%%%%%%%%%%%%%%%%%%%%%%%%%%%%%%%%%%%%%%%
\subsection{Orbital parameters}\label{sec:orbital_parameters}
After the second compact object is formed in an isolated binary no further wind mass loss or mass transfer occurs until the system eventually merges. The orbital parameters will consequently only change slowly because of GW radiation.

%%%%%%%%%%
\subsubsection{Double neutron star binaries}\label{sec:orbital_parameters_double_neutron_star_binaries}
\begin{figure}
  \includegraphics[width=\columnwidth]{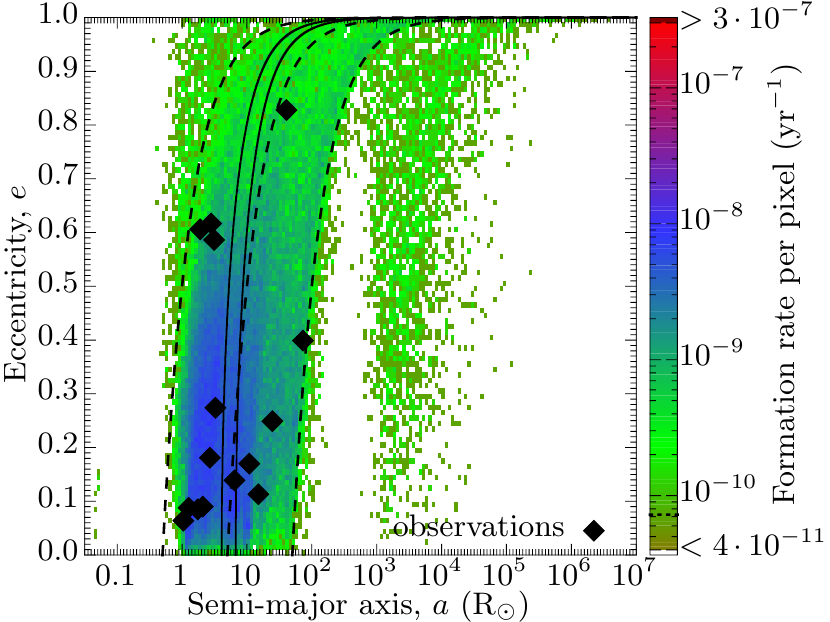}
  \caption{\label{fig:final_orbit_formation_MW_beta75_NSNS}Eccentricity versus semi-major axis of double NS systems after the birth of the second-formed NS. Colour coded is the formation rate per grid cell of such systems in a MW-like galaxy. The minor tick marks show the grid cell size. The dashed black lines show a constant periastron separation of $\unit{0.5}{\Rsun}$, $\unit{5}{\Rsun}$ and $\unit{50}{\Rsun}$. The solid lines are for $\tmerge$ equal to a Hubble time (the left one is for the minimum and the right one for the maximum NS mass configuration). The black diamonds are measurements taken from \citet[][and references therein]{tkf+17}, \citet{cck+17}, \citet{msf+18}, \citet{sfc+18} and \citet{lsk+18}.}
\end{figure}

\begin{figure*}
  \includegraphics[width=\textwidth]{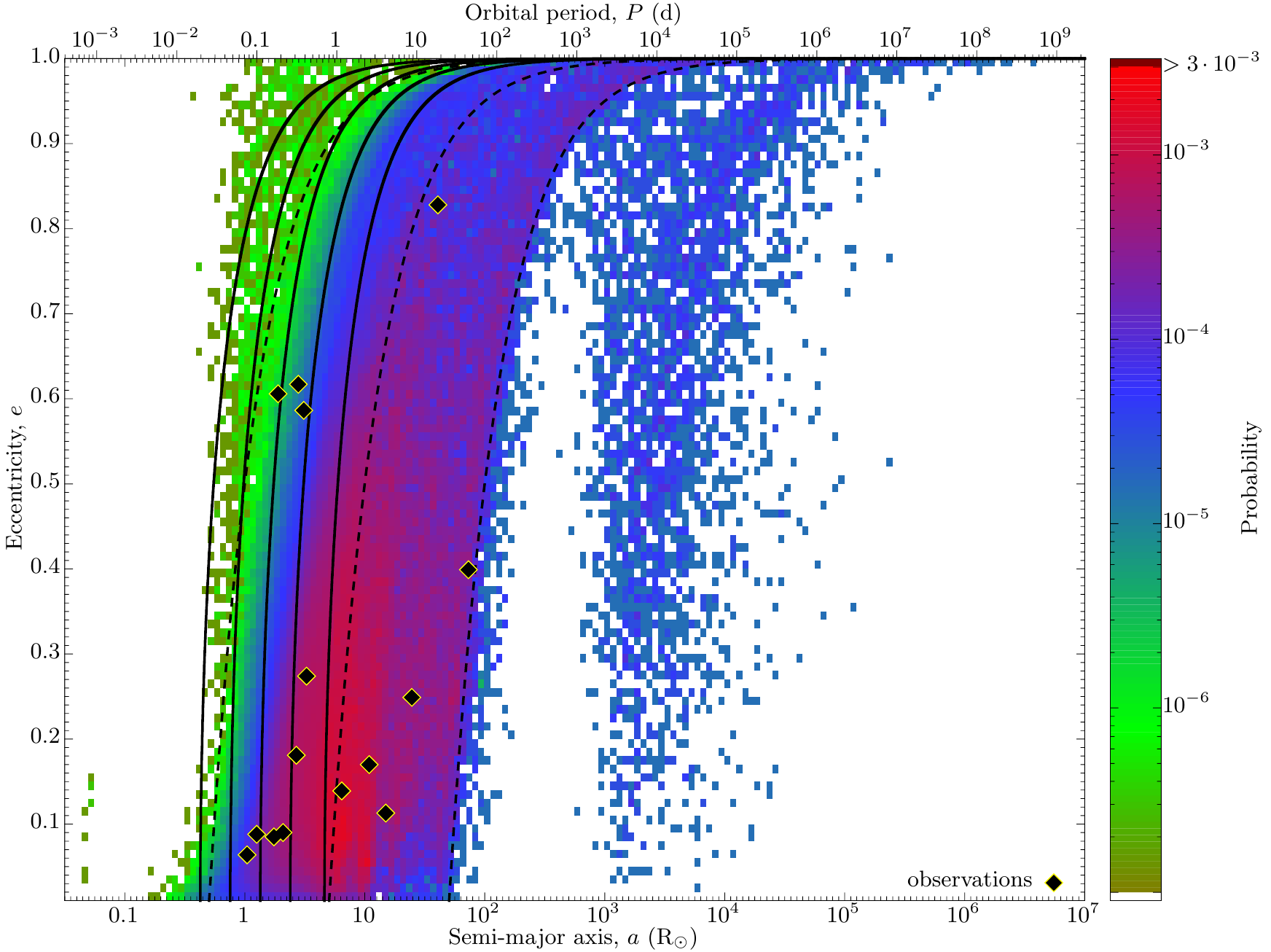}
  \caption{\label{fig:observed_NSNS_orbit_MW_beta75}Eccentricity versus semi-major axis of the simulated double NS binaries where GW evolution over a timespan of $\unit{10}{\giga\yr}$ is included to change the observable parameters compared to their values at formation as displayed in \FigureRef{fig:final_orbit_formation_MW_beta75_NSNS}. The solid black lines show values of constant $\tmerge\in\{\unit{1}{\mega\yr},\unit{10}{\mega\yr},\unit{100}{\mega\yr},\unit{1}{\giga\yr},t_{\rm hubble}\}$ from left to right. The lines of constant $\tmerge$ and the orbital period scale on top assumes a mass of $\unit{1.35}{\Msun}$ for both NSs. The colour code shows the probability of each grid cell. The black diamonds are measurements taken from \citet[][and references therein]{tkf+17}, \citet{cck+17}, \citet{msf+18}, \citet{sfc+18} and \citet{lsk+18}.}
\end{figure*}

\FigureRef{fig:final_orbit_formation_MW_beta75_NSNS} shows the orbital eccentricity and semi-major axis of the double NS binaries plotted in the upper panel of \FigureRef{fig:final_masses_formation_MW_NSNS}. Here again, the colour coding represents the formation rate per grid cell of such systems in a MW-like galaxy. The three thick dashed lines in \FigureRef{fig:final_orbit_formation_MW_beta75_NSNS} show systems with a constant periastron separation of $\unit{0.5}{\Rsun}$, $\unit{5}{\Rsun}$ and $\unit{50}{\Rsun}$, respectively. The two solid lines indicate the boundaries to the left of which systems will merge within a Hubble time by pure GW radiation, using the least and the most massive NS combination from our simulations: \mbox{$(\mCO^{\rm p}, \mCO^{\rm s})=(\unit{1.161}{\Msun}, \unit{1.137}{\Msun})$} and $(\unit{2.158}{\Msun}, \unit{2.299}{\Msun})$, respectively.

Two main sub-populations are visible for eccentricities, \mbox{$e<0.9$} in \FigureRef{fig:final_orbit_formation_MW_beta75_NSNS}. The most dominant one is the sub-population with orbital separations of \mbox{$\unit{1}{\Rsun}\la a\la\unit{100}{\Rsun}$}. All systems in this region survived one CE~phase in their evolution. The widest of these systems usually had a large separation at the onset of the CE~phase and a relatively massive in-spiralling NS. Both of these conditions help to unbind the CE and thus result in a wide orbit after its ejection.

The observed double NS systems coincide nicely with the peak population of our simulated systems in \FigureRef{fig:final_orbit_formation_MW_beta75_NSNS}. In the bluish/purple region, all values of our simulated NS masses \mbox{($\approx 1.14\text{ to }\unit{2.30}{\Msun}$)} are present. One should keep in mind that the observed systems did not necessarily evolve from progenitor binaries with the same metalillicity or mass-transfer efficiency (cf. \FigureRef{fig:final_orbit_formation_MW} first panel).

\FigureRef{fig:observed_NSNS_orbit_MW_beta75} shows a version of \FigureRef{fig:final_orbit_formation_MW_beta75_NSNS} in which all the systems are evolved further for $\unit{10}{\giga\yr}$ after the last SN. For each further $\unit{0.1}{\giga\yr}$ in time, the binaries contribute to this plot if they did not merge in the meantime. With the assumption that the formation rate was constant over time, this plot represents more correctly the observable systems present in our Galaxy today. The match between simulations and observations of double NS binaries looks even better here than in \FigureRef{fig:final_orbit_formation_MW_beta75_NSNS}. We caution that a given observed system could, in principle, have been formed in a dense cluster and ejected afterwards, and therefore in that case it need not be represented by our simulations of isolated binaries. However, from the shown data there are no indications that this is the case. All the observed double NS binaries are located somewhat close to the Galactic disk. The two known double NS binaries found in globular clusters have already been removed from the plotted data.

The wide binaries in \FigureRef{fig:final_orbit_formation_MW_beta75_NSNS} with \mbox{$a\ga\unit{1\,000}{\Rsun}$} are those which, after the first NS formed, avoided a CE and spiral-in phase. They survived the second SN because they  experienced small kicks (often EC~SNe). In \FigureRef{fig:observed_NSNS_orbit_MW_beta75}, this sub-population becomes relatively more dominant as their orbital separations only decay by a marginal amount due to very weak GW damping. Therefore, they remain stationary in this plot, independent of when they were formed. A radio pulsar discovered in a double NS binary with an orbital period of the order $\unit{10}{\yr}$ would thus be a good candidate to originate from an EC~SN. In such wide binaries, both NSs remain non-recycled \citep{tkf+17}, except for some negligible amount of wind accretion from a distant companion progenitor. Their lifetimes as active radio pulsars are expected to be similar to those of normal non-recycled radio pulsars \citep[$\unit{10-50}{\mega\yr}$,][]{lk04,jk17}.

Finally, among our simulated systems there is a very minor, and probably never observed, sub-population of systems formed in very tight orbits with an orbital period of less than $\unit{2}{\minute}$. These extreme systems are descendants of binaries which underwent two CE~phases after the first NS formed. As the progenitor of the second NS becomes ultra-stripped the expected kick is low, leading to only small eccentricities of \mbox{$e<0.2$}. GW radiation will merge these systems within $\unit{1\,000}{\yr}$ after their formation.

%%%%%%%%%%
\subsubsection{Mixed binaries}\label{sec:orbital_parameters_mixed_binaries}
\begin{figure}
  \includegraphics[width=\columnwidth]{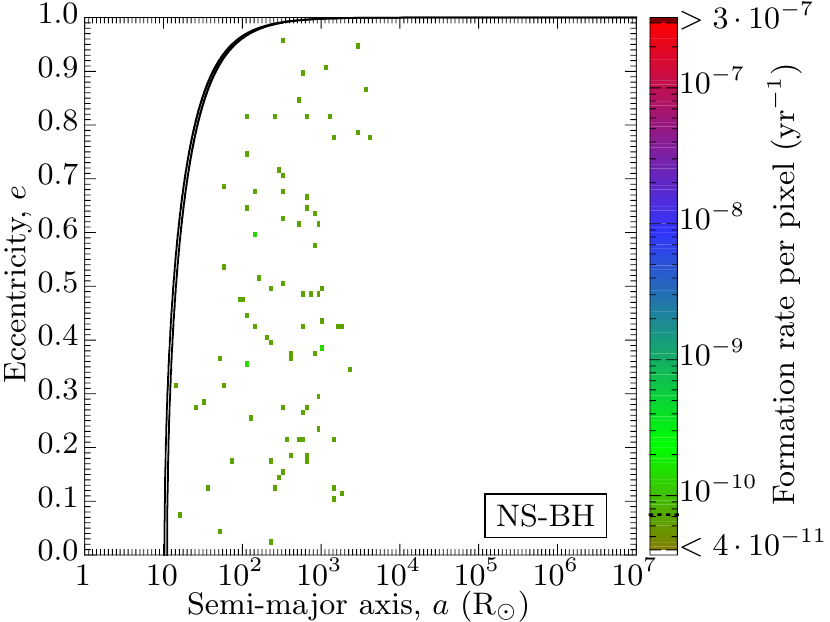}
  \includegraphics[width=\columnwidth]{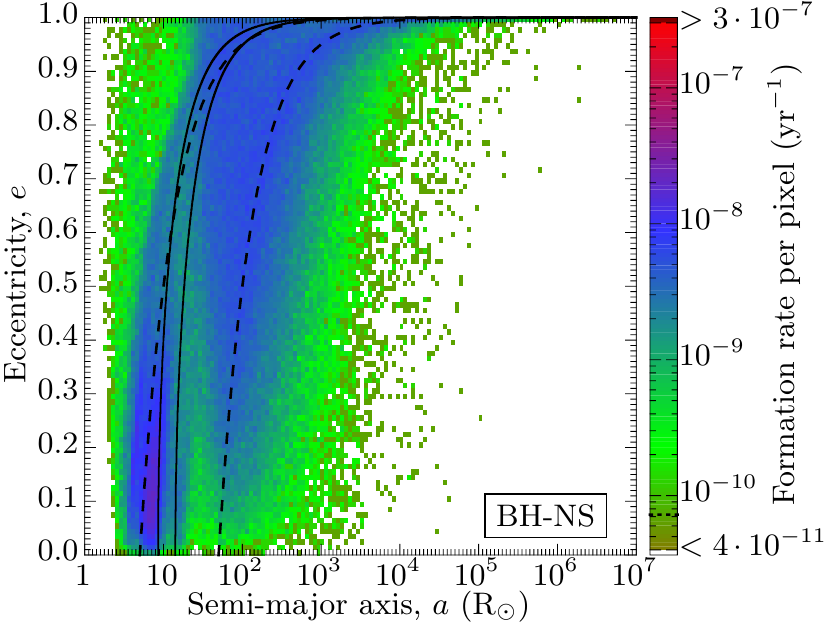}
  \caption{\label{fig:final_orbit_formation_MW_beta75_mixed}As \FigureRef{fig:final_orbit_formation_MW_beta75_NSNS} for the mixed systems consisting of one NS and one BH. In the upper panel the NS formed first, and in the lower panel the BH formed first. The dashed lines are constant periastron separations of $\unit{5}{\Rsun}$ and $\unit{50}{\Rsun}$.} 
\end{figure}

The orbital parameters of the mixed binaries consisting of a BH and a NS are shown in \FigureRef{fig:final_orbit_formation_MW_beta75_mixed}. Compared to the double NS binaries all the different sub-populations overlap into one population.

The upper panel of \FigureRef{fig:final_orbit_formation_MW_beta75_mixed} shows the few NS-BH systems in which the NS forms first. In all cases, there is no mass transfer after the formation of the NS. Therefore, based on our default simulations, we do not produce any recycled radio pulsars orbiting a BH (however, see the discussion in \SectionRef{sec:NSBH_rare} on the formation of such systems). Because the NSs remain non-recycled in our default NS-BH binaries, it is impossible to differentiate them observationally from BH-NS binaries in which the NS forms second.

Among the BH-NS binaries (\FigureRef{fig:final_orbit_formation_MW_beta75_mixed}, lower panel) the widest systems are again the ones which evolved without any mass transfer after the first SN. Avoiding mass transfer from the NS progenitor to the BH means that the orbit is so wide prior to the second SN that this translates into a semi-major axis after the second SN of \mbox{$a>\unit{800}{\Rsun}$}.

The majority of the mixed systems belong to the thick blue band with \mbox{$a\ga\unit{15}{\Rsun}$}. Prior to the second SN, these systems had orbital separations of \mbox{$a\ga\unit{25}{\Rsun}$}. All these binaries only experienced stable mass transfer and no CE~phase.

Only systems which had unstable mass transfer from the NS progenitor onto the BH create tighter binaries, and these contribute to observable GW events of mixed systems resulting from our default simulation. These systems, which evolved through a CE~phase, follow mainly the narrow blue band with a minimum periastron separation of $2.5$ to $\unit{5}{\Rsun}$. The resulting ultra-stripped SNe produce the binaries shown in \FigureRef{fig:final_orbit_formation_MW_beta75_mixed} with large eccentricities and small semi-major axes (depending on the applied NS kick, see \SectionRef{sec:supernovae}). To shrink the orbit to a tight post-SN semi-major axis, those kicks have to be directed backwards, close to a direction opposite to the orbital velocity of the exploding star.

%%%%%%%%%%
\subsubsection{Double black hole binaries}\label{sec:orbital_parameters_double_black_hole_binaries}
\begin{figure}
  \includegraphics[width=\columnwidth]{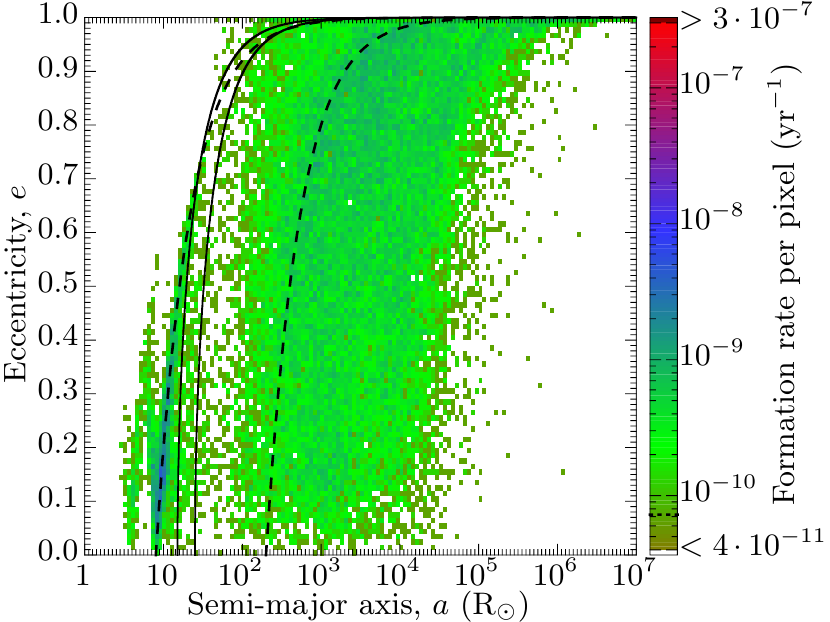}
  \caption{\label{fig:final_orbit_formation_MW_beta75_BHBH}As \FigureRef{fig:final_orbit_formation_MW_beta75_NSNS} for the double BH systems. The dashed lines show constant periastron separations of $\unit{8}{\Rsun}$ and $\unit{200}{\Rsun}$, respectively.}
\end{figure}

%moved here to appear at the right position, originally after subsection{Gravitational wave-driven merger rates}
\begin{figure*}
  \framebox{$\mathrm{Z}_\mathrm{MW}=0.0088$}\\
  \includegraphics[width=\columnwidth]{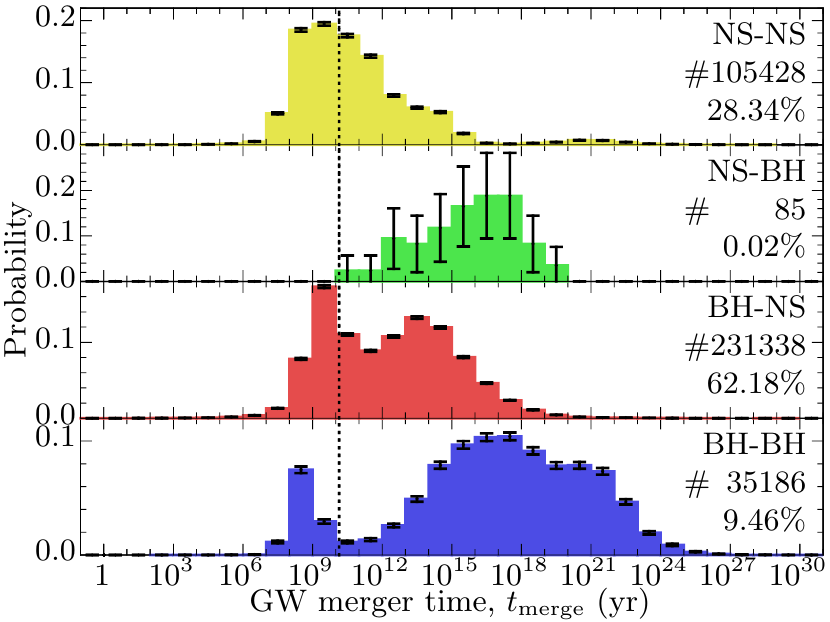}
  \hfill
  \includegraphics[width=\columnwidth]{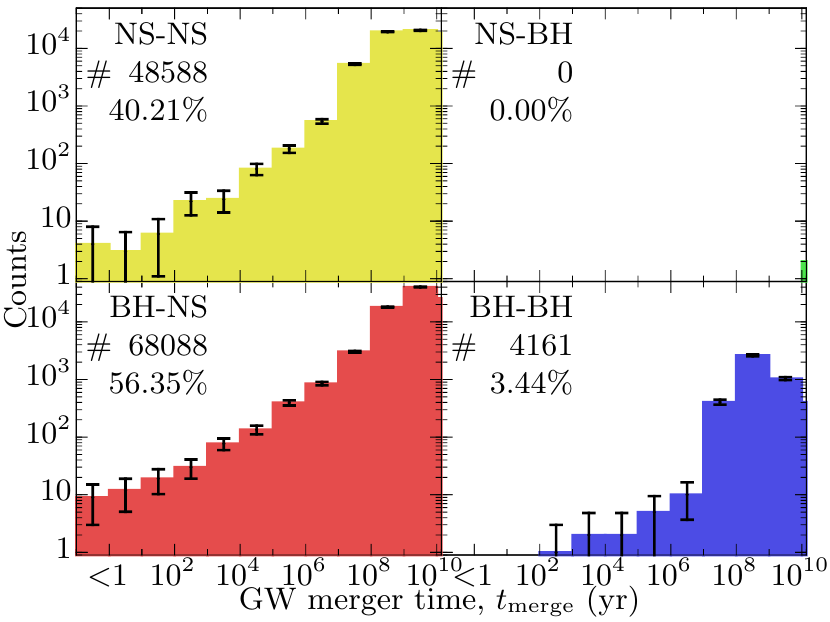}\\
  \framebox{$\mathrm{Z}_\mathrm{IZw18}=0.0002$}\\
  \includegraphics[width=\columnwidth]{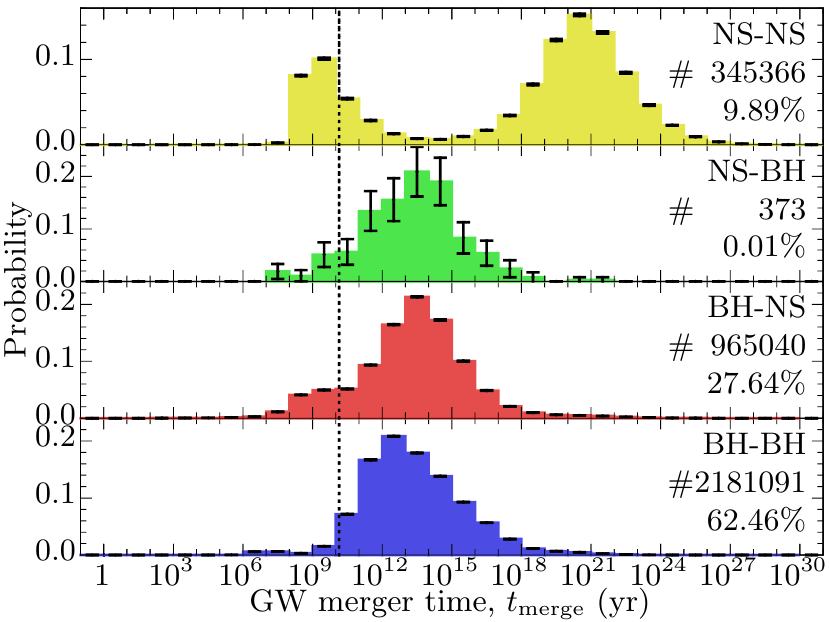}
  \hfill
  \includegraphics[width=\columnwidth]{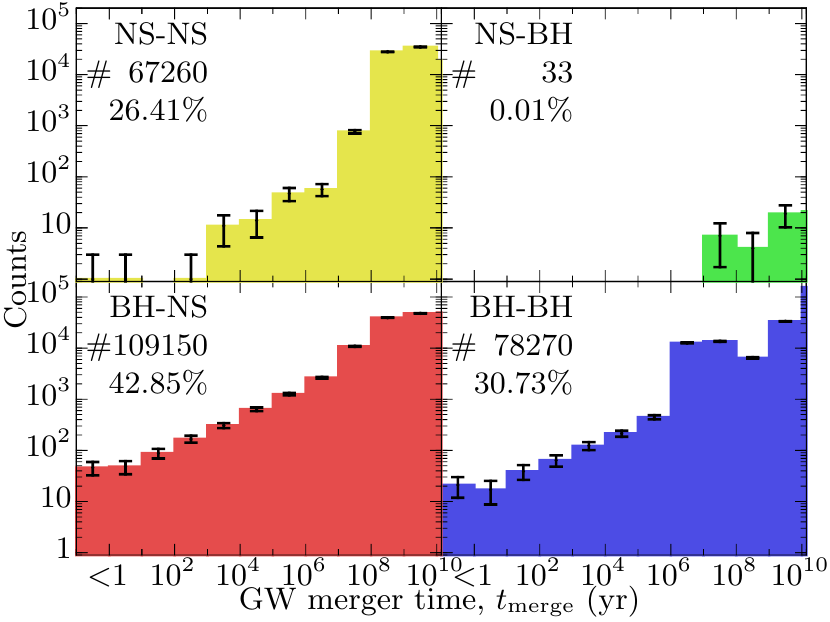}\\
  \caption{\label{fig:Tmerge_LIGO_normalised_beta75}Left column: individually normalized probability distributions of the merger time, $\tmerge$ of our default simulation of DCO binaries. The black dotted line marks the Hubble time.
  Right column: Count of mergers (log scale) with a given value of $\tmerge$ smaller than the Hubble time, for the same simulations shown in the left column.
  The order of the compact object formation is stated in the name of the binary type. Quoted in the legends are also the total and the relative number of systems shown in the sub-figures. The upper and lower rows are for different metallicities of \mbox{${\rm Z}_{\rm MW}=0.0088$} and \mbox{${\rm Z}_{\rm IZw18}=0.0002$}, respectively. The error bars show the pure statistical 95~per~cent confidence level.}
\end{figure*}

Our simulated double BH binaries also split into several sub-populations, as shown in \FigureRef{fig:final_orbit_formation_MW_beta75_BHBH}. The widest systems are those without mass-transfer interactions. The systems at intermediate orbits have a stable mass transfer before the second BH forms. The tightest double BH binaries went through a CE~phase. 

The majority (88~per~cent) of the double BH binaries in our simulations are too wide to merge within a Hubble time. Only the tightest or most eccentric double BH binaries produced will merge via GW radiation. Therefore, most of the BH-BH GW detections from a high-metallicity environment are dominated by systems which evolved through a CE~phase -- unless they are formed by a completely different formation channel not investigated here (chemical homogeneous evolution or dynamical encounters, see \SectionRef{sec:alternative_scenario}).

%%%%%%%%%%%%%%%%%%%%%%%%%%%%%%%%%%%%%%%%%%%%%%%%%%
\subsection{Gravitational wave-driven merger rates}\label{sec:gravitational_wave_radiation_and_merger}
\begin{table}
 \caption{\label{tab:GWmergerrates_beta75}GW merger rates in a MW-like galaxy. The values are based on systems merging within $\unit{10^{10}}{\yr}$. The upper and lower bounds are for systems merging within $\unit{(10\pm3.81)}{\giga\yr}$. The binary types indicate the order of the first and second-formed compact objects.}
 \begin{tabular}{cr@{$\times$}l<{$\unit{}{\yr^{-1}}$}r@{$\times$}l<{$\unit{}{\yr^{-1}}$}}
  \hline
  {\normalsize\rule{0pt}{\f@size pt}}GW merger rates & \multicolumn{2}{c}{${\rm Z}_{\rm MW}=0.0088$} & \multicolumn{2}{c}{${\rm Z}_{\rm IZw18}=0.0002$}\\
  \hline
  {\normalsize\rule{0pt}{\f@size pt}}NS-NS & $2.98^{+0.15}_{-0.24}$ & $10^{-6}$ & $2.82^{+0.16}_{-0.27}$ & $10^{-6}$\\
  {\normalsize\rule{0pt}{\f@size pt}}NS-BH & $0.00^{+0.00}_{-0.00}$ & $10^{0}$ & $1.33^{+0.13}_{-0.22}$ & $10^{-9}$\\
  {\normalsize\rule{0pt}{\f@size pt}}BH-NS & $4.05^{+0.35}_{-0.59}$ & $10^{-6}$ & $4.57^{+0.26}_{-0.37}$ & $10^{-6}$\\
  {\normalsize\rule{0pt}{\f@size pt}}BH-BH & $2.64^{+0.05}_{-0.07}$ & $10^{-7}$ & $2.96^{+0.50}_{-0.55}$ & $10^{-6}$\\
  \hline
 \end{tabular}
\end{table}

The orbit of a DCO binary shrinks over time because of GW radiation. Following \citet{pet64}, the merger time, $\tmerge$, of such a system is calculated following \EquationRef{eq:tmerge}. In \FigureRef{fig:Tmerge_LIGO_normalised_beta75}, histograms of $\tmerge$ are shown for our simulated DCO binaries. It is clearly visible that each type of system has a main peak for $\tmerge$. Multiple peaks reflect different sub-populations of a given binary type. The most important factors affecting $\tmerge$ are the semi-major axis and the eccentricity. Given that \mbox{$\tmerge\propto a^{4}$}, a factor of $10$ in semi-major axis in \mbox{\FiguresRef{fig:final_orbit_formation_MW_beta75_NSNS}--\ref{fig:final_orbit_formation_MW_beta75_BHBH}} translates into a factor of $10^{4}$ in $\tmerge$ in \FigureRef{fig:Tmerge_LIGO_normalised_beta75}. In comparison, the different masses of the DCO systems do not affect $t_{\rm merge}$ by much.

The resulting merger rates of our simulated DCO binaries are shown in \TableRef{tab:GWmergerrates_beta75}. At high metallicity (\mbox{$Z={\rm Z}_{\rm MW}=0.0088$}, \FigureRef{fig:Tmerge_LIGO_normalised_beta75} upper panels) the systems which merge within a Hubble time are dominated by double NS and BH-NS binaries. The majority of double BH systems are produced with wide orbits because at high metallicity potential progenitors of tight BH-BH binaries often coalesce during the CE~phase. The small peak in the double BH distribution with a merger time $<\unit{1}{\giga\yr}$ is caused by binaries where a massive primary BH survives the CE~phase.

The lower panels of \FigureRef{fig:Tmerge_LIGO_normalised_beta75} show the merger time of DCO merger events in a low-metallicity environment. Although there are far more double BH and mixed binaries in this case, they do not clearly dominate the total rate of systems which merge within a Hubble time (\TableRef{tab:GWmergerrates_beta75}). However, the chance of observing a merger also depends on the strain amplitude of the GW signal, making double BH observations the most likely to be observed by far because of their larger masses, cf. \TableRef{tab:detection_rates_beta75}.

The difference in the positions and the relative strength of the peaks in the GW merger time distributions between the two metallicities in \FigureRef{fig:Tmerge_LIGO_normalised_beta75} results mainly from a number of metallicity-dependent effects: the maximum stellar radius which sets how tight a system can be without mass transfer, the stellar wind mass loss and the depth of the convective envelope (determining the stability of the RLO), and finally the different mass ejections and kick magnitudes during a SN which influence the probability of disrupting the binary. In particular, the second peak (near \mbox{$\tmerge=\unit{10^{21}}{\yr}$}) in the double NS distribution at low metallicity is dominated by EC~SNe, for which we find a wider range of ZAMS progenitor masses compared to high metallicity models.

\begin{figure}
  \includegraphics[width=\columnwidth]{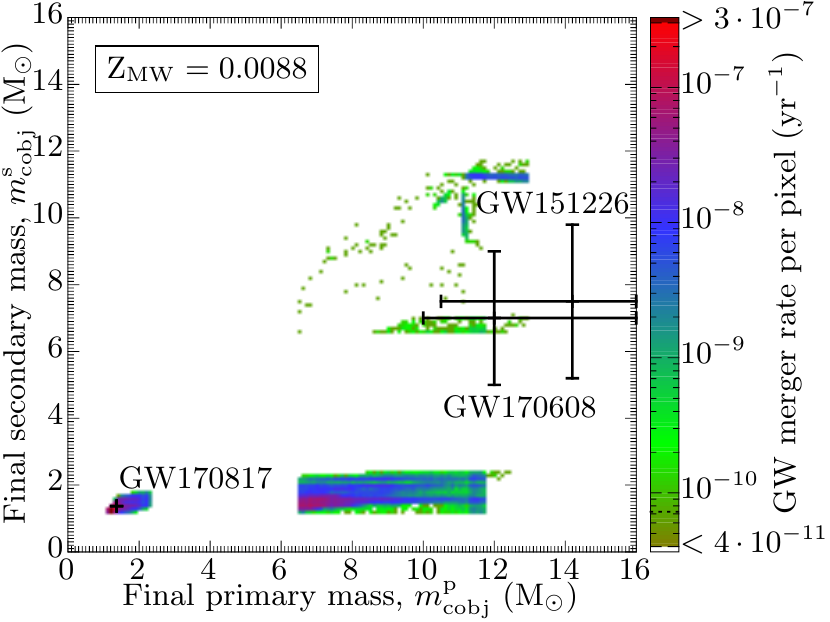}
  \includegraphics[width=\columnwidth]{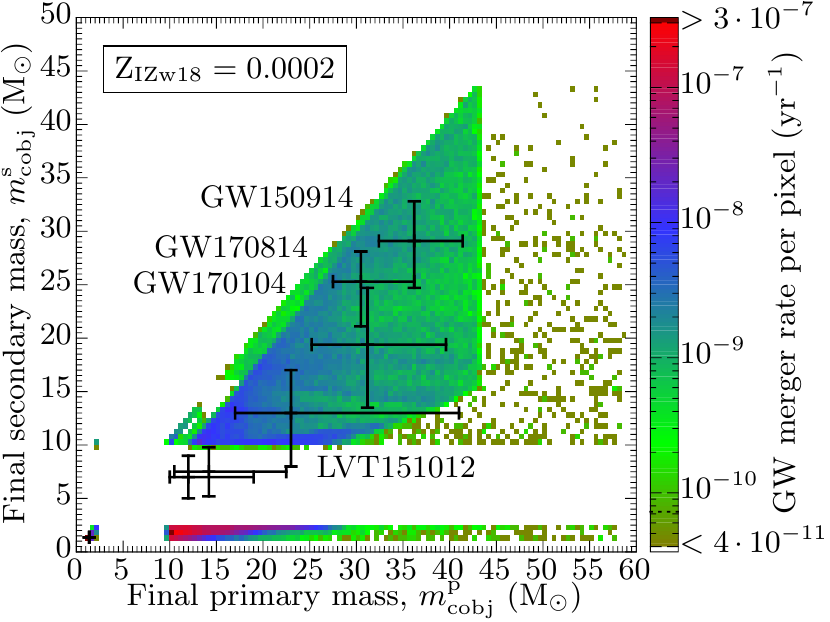}
  \caption{\label{fig:final_masses_GWmerger_beta75}Distribution of simulated DCO binaries in the mass--mass plane for \mbox{${\rm Z}_{\rm MW}=0.0088$} and \mbox{${\rm Z}_{\rm IZw18}=0.0002$} (upper and lower panel, respectively). Colour coded are their merger rates per pixel assuming that the star formation rate is the same at both metallicities. The black data points are taken from the LIGO-Virgo events reported in \citet{aaa+16c,aaa+17,aaa+17b,aaa+17c,aaa+17g}.}
\end{figure}

\begin{figure}
  \includegraphics[width=\columnwidth]{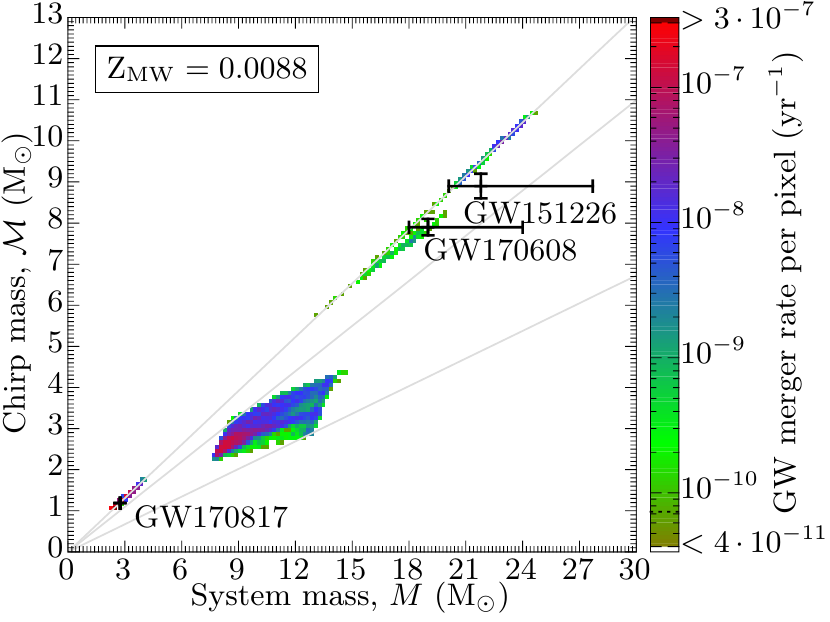}
  \includegraphics[width=\columnwidth]{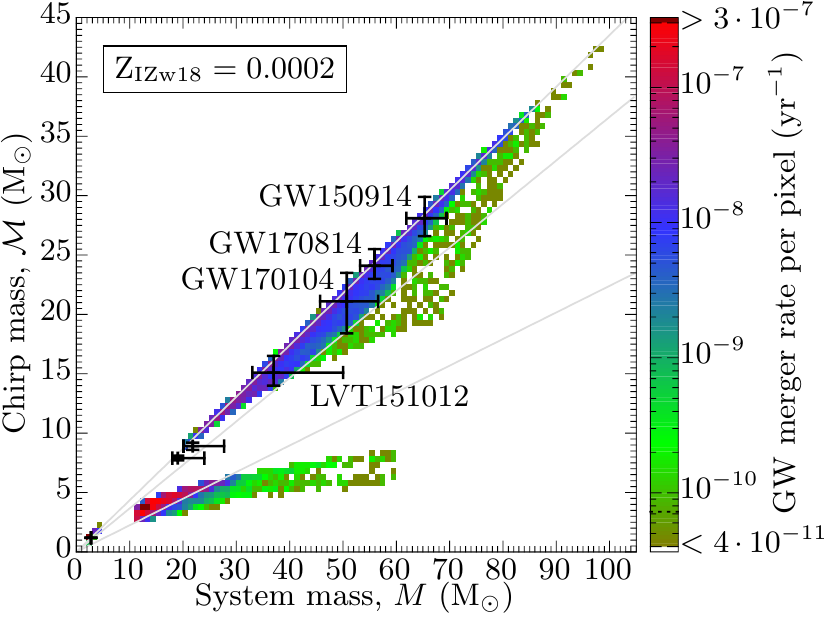}
  \caption{\label{fig:final_chirpmass_GWmerger_beta75}Distribution of simulated and observed DCO binaries in the total mass--chirp mass plane (the masses directly inferred from GW observations) for \mbox{${\rm Z}_{\rm MW}=0.0088$} and \mbox{${\rm Z}_{\rm IZw18}=0.0002$} (upper and lower panel, respectively). The three solid grey lines indicate a constant mass ratio of 1, 3 and 10 (from top to bottom). Only simulated systems which merge within a Hubble time are shown. See \FigureRef{fig:final_masses_GWmerger_beta75} for further information.}
\end{figure}

\FigureRef{fig:final_masses_GWmerger_beta75} shows DCO masses and merger rates per pixel of our simulated systems in the mass--mass plane of the two compact objects. There are, as usual, four different types of binaries located at different areas in the plot: double BHs, double NSs and the two types of mixed binaries. The six announced LIGO detections of BH-BH GWs are included in the plots. Only the events GW151226 and GW170608 could have formed in a high-metallicity environment (see upper panel). The other four LIGO events need low metallicities (see lower panel) to obtain their larger BH masses, as also concluded by many other studies \citep[e.g.][]{bhbo16,es16,rhc+16,svm+17}. Binary BHs with individual component masses above $\unit{43}{\Msun}$ enter the regime of PISNe (see \SectionRef{sec:progenitor_ZAMS_masses}), which causes the strong number density drop at this mass. As all detected LIGO events are at distances \mbox{$z=0.1\text{ to }0.2$}, they are indeed likely to originate from a slightly lower metallicity environment than that of the MW \citep[e.g.][]{plg+13}.

From GW merger observations one obtains the chirp mass and the total system mass. \FigureRef{fig:final_chirpmass_GWmerger_beta75} shows this plane for high- and low-metallicity environments (upper and lower panels, respectively). Three different areas are seen corresponding to the double NS systems, the mixed systems (NS/BH) and the double BH systems. The double NS binaries have the lowest total masses and chirp masses. The mixed systems are located in the middle with large mass ratios, while the double BH binaries occupy the upper right part of the diagram at high values.

As seen in \FigureRef{fig:final_chirpmass_GWmerger_beta75}, using \OurCode we can reproduce all the progenitor systems of the observed GW events. The extended parameter space of BH masses of our simulated systems at low metallicity covers most of the reported GW merger events, except for GW170608 where we find no solutions in a low-metallicity environment, see also \FigureRef{fig:LIGO_progenitors_beta75}). Again, we conclude that only the events GW151226 and GW170608, as well as the double NS merger GW170817, could have formed in a high-metallicity environment like that of the Milky Way. For more discussions about the LIGO events we refer to \SectionRef{sec:compare_LIGO}.

%%%%%%%%%%%%%%%%%%%%%%%%%%%%%%%%%%%%%%%%%%%%%%%%%%
\section{Discussion}\label{sec:discussion}
In the following section, we first discuss the influence of several of the input parameters applied in our simulations. We then compare our results to other binary population synthesis studies before we focus on the formation of the first seven events detected by LIGO-Virgo and give a general discussion on the inferred merger-rate densities. We also present a short discussion on the formation of recycled pulsars orbiting BHs and, finally, we briefly discuss alternative formation channels of DCO binaries.

%%%%%%%%%%%%%%%%%%%%%%%%%%%%%%%%%%%%%%%%%%%%%%%%%%
\subsection{Parameter studies}\label{sec:parameter_studies}
\begin{table*}
\caption{\label{tab:rate_variations}Variations in DCO formation and merger rates for a MW-like galaxy caused by changing the values of selected key input parameters (columns 3 to 9). The default input parameters are listed in \TableRef{tab:standard} and the resulting rates are shown in the second column. The binary types refer to the first and second compact objects formed. The pure uncertainties of Poissonian statistics are between $\unit{10^{-11}}{\yr^{-1}}$ and $\unit{10^{-8}}{\yr^{-1}}$.}
 \begin{tabular}{r<{ $(\yr^{-1})$}r@{$\times$}lr@{$\times$}lr@{$\times$}lr@{$\times$}lr@{$\times$}lr@{$\times$}lr@{$\times$}lr@{$\times$}l}
  \hline
  \multicolumn{3}{r}{{\normalsize\rule{0pt}{\f@size pt}}} & \multicolumn{2}{c}{$\alpha_{\rm CE}$} & \multicolumn{2}{c}{$\beta_{\rm min}$} & \multicolumn{2}{c}{$\alpha_{\rm RLO}$} & \multicolumn{2}{c}{$\alpha_{\rm th}$} & \multicolumn{2}{c}{$q_{\rm limit}$} & \multicolumn{2}{c}{$\alpha_{\rm IMF}$} & \multicolumn{2}{c}{$m^{\rm p}_{\rm max}=m^{\rm s}_{\rm max}$}\\
  \multicolumn{3}{r}{upper:} & \multicolumn{2}{c}{$0.80$} & \multicolumn{2}{c}{$0.79$} & \multicolumn{2}{c}{$0.24$} & \multicolumn{2}{c}{$0.70$} & \multicolumn{2}{c}{$4.0$} & \multicolumn{2}{c}{$3$} & \multicolumn{2}{c}{$\unit{150}{\Msun}$}\\
  \multicolumn{3}{r}{lower:} & \multicolumn{2}{c}{$0.20$} & \multicolumn{2}{c}{$0.50$} & \multicolumn{2}{c}{$0.15$} & \multicolumn{2}{c}{$0.30$} & \multicolumn{2}{c}{$1.5$} & \multicolumn{2}{c}{$2$} & \multicolumn{2}{c}{$\unit{\phantom{0}80}{\Msun}$}\\
  \hline
  \multicolumn{1}{r}{{\normalsize\rule{0pt}{\f@size pt}}Formation rates} & \multicolumn{2}{c}{default} & \multicolumn{2}{c}{$\alpha_{\rm CE}$} & \multicolumn{2}{c}{$\beta_{\rm min}$} & \multicolumn{2}{c}{$\alpha_{\rm RLO}$} & \multicolumn{2}{c}{$\alpha_{\rm th}$} & \multicolumn{2}{c}{$q_{\rm limit}$} & \multicolumn{2}{c}{$\alpha_{\rm IMF}$} & \multicolumn{2}{c}{$m^{\rm p}_{\rm max}=m^{\rm s}_{\rm max}$}\\
  \hline
  {\normalsize\rule{0pt}{\f@size pt}}NS-NS & $6.81$ & $10^{-6}$ & $^{+2.37}_{-1.72}$ & $10^{-6}$ & $^{-0.63}_{+3.02}$ & $10^{-6}$ & $^{-0.69}_{+1.06}$ & $10^{-6}$ & $^{+0.35}_{-1.32}$ & $10^{-7}$ & $^{+3.17}_{-1.01}$ & $10^{-6}$ & $^{-0.33}_{+2.08}$ & $10^{-5}$ & $^{-1.91}_{+1.05}$ & $10^{-8}$\\
  {\normalsize\rule{0pt}{\f@size pt}}NS-BH & $5.49$ & $10^{-9}$ & $^{+2.01}_{-0.04}$ & $10^{-8}$ & $^{+1.20}_{-0.36}$ & $10^{-8}$ & $^{+1.14}_{-0.50}$ & $10^{-8}$ & $^{-0.77}_{-1.23}$ & $10^{-9}$ & $^{+1.11}_{-0.05}$ & $10^{-7}$ & $^{-0.35}_{+3.79}$ & $10^{-8}$ & $^{-1.15}_{-1.69}$ & $10^{-9}$\\
  {\normalsize\rule{0pt}{\f@size pt}}BH-NS & $1.49$ & $10^{-5}$ & $^{+1.96}_{-3.26}$ & $10^{-6}$ & $^{+0.17}_{-1.23}$ & $10^{-5}$ & $^{+1.28}_{-2.73}$ & $10^{-6}$ & $^{+1.05}_{-0.70}$ & $10^{-6}$ & $^{+4.55}_{-1.28}$ & $10^{-5}$ & $^{-0.10}_{+1.38}$ & $10^{-4}$ & $^{+9.37}_{-9.15}$ & $10^{-7}$\\
  {\normalsize\rule{0pt}{\f@size pt}}BH-BH & $2.27$ & $10^{-6}$ & $^{+2.35}_{-0.30}$ & $10^{-6}$ & $^{+1.06}_{-0.19}$ & $10^{-6}$ & $^{+1.08}_{-0.28}$ & $10^{-6}$ & $^{+2.88}_{-1.80}$ & $10^{-7}$ & $^{+3.87}_{-0.02}$ & $10^{-5}$ & $^{-0.16}_{+2.99}$ & $10^{-5}$ & $^{+4.37}_{-1.11}$ & $10^{-6}$\\
  \hline
  \multicolumn{1}{r}{{\normalsize\rule{0pt}{\f@size pt}}GW merger rates} & \multicolumn{2}{c}{default} & \multicolumn{2}{c}{$\alpha_{\rm CE}$} & \multicolumn{2}{c}{$\beta_{\rm min}$} & \multicolumn{2}{c}{$\alpha_{\rm RLO}$} & \multicolumn{2}{c}{$\alpha_{\rm th}$} & \multicolumn{2}{c}{$q_{\rm limit}$} & \multicolumn{2}{c}{$\alpha_{\rm IMF}$} & \multicolumn{2}{c}{$m^{\rm p}_{\rm max}=m^{\rm s}_{\rm max}$}\\
  \hline
  {\normalsize\rule{0pt}{\f@size pt}}NS-NS & $2.98$ & $10^{-6}$ & $^{+7.75}_{-0.64}$ & $10^{-7}$ & $^{-0.51}_{+2.71}$ & $10^{-6}$ & $^{-5.67}_{+8.64}$ & $10^{-7}$ & $^{-2.60}_{+1.47}$ & $10^{-7}$ & $^{+0.85}_{-4.66}$ & $10^{-7}$ & $^{-1.46}_{+9.68}$ & $10^{-6}$ & $^{-3.11}_{-0.67}$ & $10^{-8}$\\
  {\normalsize\rule{0pt}{\f@size pt}}NS-BH & $0.00$ & $10^{0}$ & $^{+1.20}_{+0.01}$ & $10^{-8}$ & $^{+2.58}_{+0.00}$ & $10^{-10}$ & $^{+3.87}_{+0.00}$ & $10^{-10}$ & $^{+1.94}_{+0.00}$ & $10^{-10}$ & $^{+1.94}_{+0.00}$ & $10^{-9}$ & $^{+0.00}_{+1.34}$ & $10^{-9}$ & $^{+0.65}_{+1.93}$ & $10^{-10}$\\
  {\normalsize\rule{0pt}{\f@size pt}}BH-NS & $4.05$ & $10^{-6}$ & $^{+0.81}_{-2.09}$ & $10^{-6}$ & $^{+0.25}_{-3.56}$ & $10^{-6}$ & $^{+2.94}_{-7.65}$ & $10^{-7}$ & $^{+4.25}_{-2.49}$ & $10^{-7}$ & $^{+2.88}_{-2.73}$ & $10^{-6}$ & $^{-0.26}_{+3.56}$ & $10^{-5}$ & $^{+1.32}_{-1.61}$ & $10^{-7}$\\
  {\normalsize\rule{0pt}{\f@size pt}}BH-BH & $2.64$ & $10^{-7}$ & $^{+2.19}_{-0.25}$ & $10^{-6}$ & $^{+0.01}_{+1.91}$ & $10^{-7}$ & $^{+0.17}_{+4.45}$ & $10^{-8}$ & $^{+3.11}_{-1.41}$ & $10^{-7}$ & $^{+1.15}_{+0.10}$ & $10^{-6}$ & $^{-0.19}_{+3.84}$ & $10^{-6}$ & $^{+3.86}_{-1.96}$ & $10^{-7}$\\
  \hline
 \end{tabular}
\end{table*}

Most of our applied input parameters take values which are not known a priori and they are therefore uncertain to some degree. \TableRef{tab:rate_variations} shows how the DCO formation and merger rates are influenced by changing the different input parameters individually, compared to applying our default values given in \TableRef{tab:standard}. In the following subsections, the most important input parameters are discussed in more detail.

%%%%%%%%%%%%%%%%%%%%%%%%%%%%%%%%%%%%%%%%%%%%%%%%%%
\subsubsection{Supernova kicks}\label{sec:supernova_kicks}
\begin{table*}
 \caption{\label{tab:kicks}GW merger rates of a MW-like galaxy and their dependence on applied kicks and assumptions on EC~SNe. The binary types refer to the first and second compact objects formed.}
 \begin{tabular}{r<{ $(\yr^{-1})$}r@{$\times$}l>{\hspace{0.1cm}}r@{$\times$}l>{\hspace{0.6cm}}r@{$\times$}l>{\hspace{1.2cm}}r@{$\times$}l}
  \hline
  \multicolumn{1}{r}{{\normalsize\rule{0pt}{\f@size pt}}GW merger rates} & \multicolumn{2}{c}{default} & \multicolumn{2}{c}{small kicks$^{\ast}$} & \multicolumn{2}{c}{large EC~SN kicks$^{\ast\ast}$} & \multicolumn{2}{c}{small EC~SN mass window$^{\ast\ast\ast}$}\\
  \multicolumn{1}{r}{{\normalsize\rule{0pt}{\f@size pt}}} & \multicolumn{2}{c}{} &  \multicolumn{2}{c}{} & \multicolumn{2}{c}{$w_{\rm ECSN}=w_{\rm FeCCSN}$} & \multicolumn{2}{c}{$\unit{1.37}{\Msun}\leq m^{\rm ECSN}_{\rm CO-core}<\unit{1.38}{\Msun}$}\\
  \hline
  {\normalsize\rule{0pt}{\f@size pt}}NS-NS & $2.98$ & $10^{-6}$ & $9.34$ & $10^{-6}$ & $1.54$ & $10^{-6}$ & $2.30$ & $10^{-6}$\\
  {\normalsize\rule{0pt}{\f@size pt}}NS-BH & $0.00$ & $10^{0}$ & $1.94$ & $10^{-10}$ & $6.46$ & $10^{-11}$ & $1.29$ & $10^{-10}$\\
  {\normalsize\rule{0pt}{\f@size pt}}BH-NS & $4.05$ & $10^{-6}$ & $7.59$ & $10^{-6}$ & $4.04$ & $10^{-6}$ & $4.04$ & $10^{-6}$\\
  {\normalsize\rule{0pt}{\f@size pt}}BH-BH & $2.64$ & $10^{-7}$ & $3.05$ & $10^{-7}$ & $2.65$ & $10^{-7}$ & $2.66$ & $10^{-7}$\\
  \hline
 \end{tabular}\\
{{\normalsize\rule{0pt}{\f@size pt}}
$^{\ast}$ half of all default kick magnitudes.
$^{\ast\ast}$ similar to FeCC~SNe, see \TableRef{tab:kick_values}.
$^{\ast\ast\ast}$ the default is \mbox{$\unit{1.37}{\Msun}\leq m^{\rm ECSN}_{\rm CO-core}<\unit{1.435}{\Msun}$}.}
\end{table*}

In \SectionRef{sec:supernovae}, we outlined which kicks we apply depending on the evolutionary state of the exploding star, the type of SN and whether the compact object formed is a NS or a BH. There is some evidence to constrain these kick magnitudes, especially for double NS systems \citep{tkf+17}, however the overall kick distribution for the various SNe remains a major uncertain aspect of modelling DCO binaries \citep[see also discussions in][]{vns+18}.

To test the effects of kick magnitudes, we performed a simulation in which all kick magnitudes (all types and all remnants) were reduced by a factor of 2. The third column of \TableRef{tab:kicks} shows how, as expected, smaller kicks lead to larger merger rates compared to those simulations with standard kicks. In double NS binaries, there is even an increase in the merger rate by a factor of 3. In double BH binaries, the effect of changing the kick magnitudes is relatively small because these systems have more mass to absorb the added kick momentum.

%%%%%%%%%%%%%%%%%%%%%%%%%%%%%%%%%%%%%%%%%%%%%%%%%%
\subsubsection{Electron capture vs. iron-core collapse supernovae}\label{sec:ECvsFECCSN}
Specific to the NS-NS population, and to some extent the BH-NS and NS-BH populations, the ratio of EC~SNe to FeCC~SNe is significant in terms of the distribution of resulting NS masses. Although we can reproduce the observed distribution of NS-NS orbital separations and eccentricities (\FigureRef{fig:observed_NSNS_orbit_MW_beta75}), our distribution of NS masses, in particular the masses of the first-formed NSs, strongly peaks at $\sim\unit{1.25}{\Msun}$, caused by NSs formed via EC~SNe (\FigureRef{fig:final_masses_formation_MW_NSNS}). Using our default model, we find that for the first SNe the ratio of EC to FeCC~SNe is about $1.8$ in all the NS-NS systems formed and $1.1$ in those which merge within a Hubble time. For the second (and in close orbits ultra-stripped) SNe, we find ratios of $0.2$ and $0.07$, respectively. However, in the observed distribution of NS masses in NS-NS systems, there is no evidence for such an EC~SN peak at $\unit{1.25}{\Msun}$.

With the empirical sample of NS-NS systems in the Galactic disk with precisely measured NS masses limited to only nine systems, it is difficult to argue that {\it any} of the first-formed NSs were produced via an EC~SN. Depending on the NS equation-of-state, the gravitational mass of such a NS is expected to be about \mbox{$\unit{1.37}{\Msun}-E_{\rm bind}\product c^{-2}$}, where the gravitational binding energy is of the order \mbox{$\unit{0.10\text{ to }0.14}{\Msun}\product c^2$} \citep{ly89}. Thus EC~SNe are expected to produce NSs with a particular gravitational mass somewhere in the interval $\unit{1.23\text{ to }1.27}{\Msun}$. However, none of the known NS-NS systems have a first-formed NS with such a small mass. It is remarkable, on the other hand, that $6$ out of $9$ systems have a first-formed NS within a narrow mass range of \mbox{$\unit{1.32}{\Msun}\la m^{\rm p}_{\rm NS}\la\unit{1.34}{\Msun}$}. This implies that these NSs only originate from EC~SNe if such SNe result in more massive NSs than assumed above and/or if these NSs have accreted of the order $\unit{0.08\pm0.01}{\Msun}$ during the recycling phase. However, \citet{tkf+17} recently argued that recycled NSs in double NS systems have accreted at most $\sim\unit{0.02}{\Msun}$, thus requiring EC~SNe to produce NSs with birth masses of about $\unit{1.30\text{ to }1.32}{\Msun}$ to reconcile our simulations with observations.

To lower the ratio of NSs produced by EC to FeCC~SNe, we made two simulations: i) applying similarly large kicks to NSs produced by EC~SNe as those produced by FeCC~SNe (\mbox{$w_{\rm ECSN}=w_{\rm FeCCSN}$}, see \TableRef{tab:kick_values}); and ii) decreasing the window in CO core masses producing EC~SNe from \mbox{$\unit{1.37}{\Msun}\leq m^{\rm ECSN}_{\rm CO-core}<\unit{1.435}{\Msun}$} to \mbox{$\unit{1.37}{\Msun}\leq m^{\rm ECSN}_{\rm CO-core}<\unit{1.38}{\Msun}$}. We now discuss the outcome of these two experiments.

Applying similar kicks to NSs produced in EC~SNe as those produced in FeCC~SNe (see \TableRef{tab:kicks}) has a severe effect on the surviving population of NS-NS systems such that the ratio of EC to FeCC~SNe in the first explosion is reduced by a factor of about $20$ (e.g. the ratio for all NS-NS systems formed decreases from $1.8$ to $0.08$). The resulting simulated distribution of mass ratios in NS-NS systems is shown as a red line histogram in \FigureRef{fig:final_massratios_MW_beta75_NSNS}. Although such a small fraction of EC~SNe of about 8~per~cent is perhaps better in accordance with the distribution of NS masses observed in double NS systems (also in terms of the resulting mass ratio distribution), this ratio is possibly too small, indicating that EC~SNe might produce only slightly larger kicks than our default assumption of a flat probability distribution between $\unit{0-50}{\kilo\meter\usk\reciprocal\second}$. More measured NS masses are needed to answer this question. The reason why larger kicks lead to such a severe reduction in the ratio of EC vs. FeCC~SNe for the first-formed NSs in surviving NS-NS systems is that many of these systems will either disrupt as a consequence of the SN explosion or widen to the extent that the secondary star does not fill its Roche lobe at any time. This is a requirement for enabling CE and spiral-in evolution producing tight binaries which also survive the second SN explosion and produce NS-NS systems. In the case of the second-formed NSs, applying larger kicks in the simulations only reduces the ratio of EC vs. FeCC~SNe by a factor of 3.

Another way we can simulate a population of fewer NS-NS systems produced via EC~SNe is simply by reducing the window in core masses assumed to produce an EC~SN from \mbox{$\unit{1.37}{\Msun}\leq m^{\rm ECSN}_{\rm CO-core}<\unit{1.435}{\Msun}$} to \mbox{$\unit{1.37}{\Msun}\leq m^{\rm ECSN}_{\rm CO-core}<\unit{1.38}{\Msun}$}, see \TableRef{tab:kicks}. However, for NS-NS systems this only has the effect of decreasing the ratio of EC to FeCC~SNe by a factor of $\sim 2$ and $\sim 4$ for the first and the second SN explosion, respectively. Furthermore, \citet{pwt+17} have even argued for a wider range of progenitor star masses producing EC~SNe which would exacerbate the discrepancy between our theoretical simulations and observations.

To summarize our finding on EC vs. FeCC~SNe, we find that our default simulations produce a majority of EC~SNe for the first-formed NSs, in apparent contrast with current observations. Although the ratio of simulated EC to FeCC~SNe is strongly dependent on the kicks applied in these two types of explosions, and also on the mass window for producing EC~SNe, we still consider this an important and puzzling issue. Perhaps the answer is simply that EC~SNe produce slightly more massive NSs than usually thought. It is anticipated that the Square-Kilometre Array (SKA) will eventually increase the number of known radio pulsars by a factor of 5 to 10 \citep{kbk+15}, thus resulting in a total of 100 to 200 known NS-NS systems. A large number of these systems will have their NS masses measured accurately and it will be interesting to see if an EC~SN peak will be present in the NS mass distribution at that time.

For the second-born NSs in double NS systems, however, a couple of the observed masses are in agreement with a potential EC~SN origin according to current expectations, e.g. PSRs~J0737$-$3039B, J1756$-$2251, and J1913+1102. Our simulations (\FigureRef{fig:final_masses_formation_MW_NSNS}) show a much smaller contribution of EC~SNe to the second-born NSs as most of the systems are tightened by a CE prior to the second SN, which therefore allows for larger kick magnitudes to remain bound. An additional minor factor that increases the fraction of FeCC~SNe in the second SN is that their kick magnitudes are, in general, smaller than in the first SN, see \TableRef{tab:kick_values}.

%%%%%%%%%%%%%%%%%%%%%%%%%%%%%%%%%%%%%%%%%%%%%%%%%%
\subsubsection{Common-envelope efficiency, \texorpdfstring{$\alpha_{\rm CE}$}{alpha\_CE}}\label{sec:alpha_CE}
\begin{figure}
  \includegraphics[width=\columnwidth]{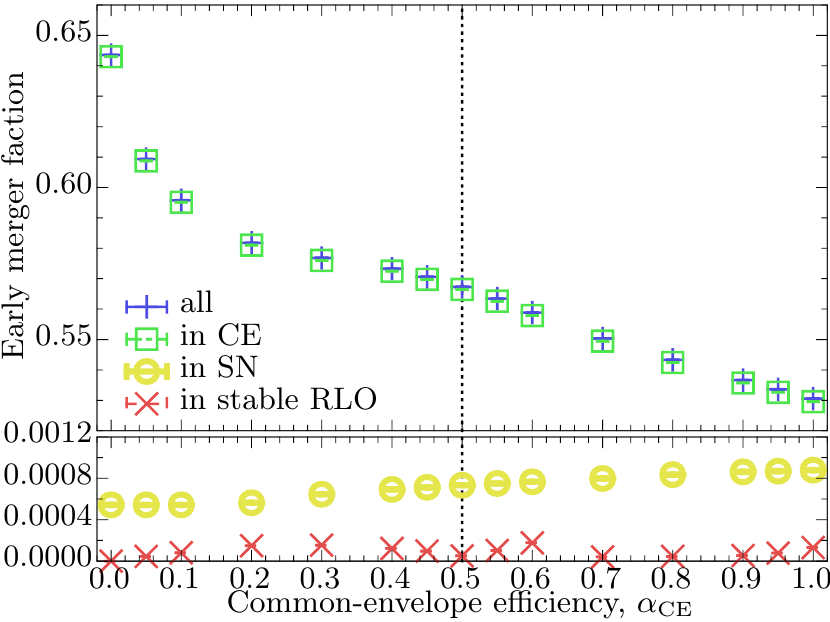}
  \caption{\label{fig:merger_alphaCE_normalized}The fraction (number of events divided by number of systems evolved) of early coalescing systems modelled with \OurCode from binaries with \mbox{$m_\mathrm{ZAMS}^\mathrm{p}\geq\unit{4}{\Msun}$} and \mbox{$m_\mathrm{ZAMS}^\mathrm{s}\geq\unit{1}{\Msun}$}, and for a MW-like metallicity (\mbox{${\rm Z}_{\rm MW}=0.0088$}), as a function of the efficiency of converting orbital energy into kinetic energy during CE evolution, $\alpha_{\rm CE}$. Also shown at the bottom is the fraction of early mergers caused by direct SN kicks and stable RLO. The vertical dotted line marks our default value of \mbox{$\alpha_{\rm CE}=0.5$}.}
\end{figure}

\FigureRef{fig:merger_alphaCE_normalized} shows the influence of the efficiency of converting released orbital energy into kinetic energy of the CE on the early binary merger fraction. Here, early binary merger refers to systems which coalesce before a DCO binary is formed. The main reasons are CE in-spiral (i.e. failed CE ejection) or direct collision of stellar components as a result of a SN changing the orbital dynamics. A very small fraction of systems undergoing RLO will also merge. The more efficient the CE energy conversion is, the more CEs are successfully ejected, and therefore the number of early mergers will decrease (as most of them happen during a CE~phase). However, for the entire DCO population as such, this effect is only modest. The early merger fraction decreases from 0.64 to 0.53 when $\alpha_{\rm CE}$ is increased from 0 to 1. These fractions are with respect to all binaries evolved with \OurCode, and thus we conclude that more than half of all massive binaries will suffer from an early merger during a CE event, cf. \citet{ipj+18} for less massive stars. For a comparison, the fraction of early mergers caused by direct SN kicks in a fine-tuned direction is only about 0.0005.

In terms of DCO merger rates, \TableRef{tab:rate_variations} shows that changing $\alpha_{\rm CE}$ from 0.5 to 0.8 (0.2) will increase (decrease) the merger rates, as expected. For double NS systems, the changes are moderate (+26 and $-2$ per~cent, respectively) as they almost always undergo a subsequent Case~BB mass-transfer phase. However, the rate of double BH mergers is very sensitive to $\alpha_{\rm CE}$ and increases by a factor 8 (decreases by a factor 19) when changing $\alpha_{\rm CE}$ from 0.5 to 0.8 (0.2), respectively.

%%%%%%%%%%%%%%%%%%%%%%%%%%%%%%%%%%%%%%%%%%%%%%%%%%
\subsubsection{Mass-transfer efficiency}\label{sec:mass-transfer_efficiency}
Among the chosen default values for the input parameters of our binary population synthesis simulations (\TableRef{tab:standard}), the minimum mass ejection fraction of the accretor during RLO is set to \mbox{$\beta _{\rm min}=0.75$}. This means that 75~per~cent of the material transferred from the donor star towards the accreting star is assumed to be re-emitted with the specific orbital angular momentum of the accretor, and an even higher fraction is ejected when the mass-transfer rate is super-Eddington. The reason we chose this low efficiency for accretion via RLO as our default value is the need to match our simulated population of double NS systems with observations -- their orbital parameters and especially their masses, cf. \FigureRef{fig:final_masses_formation_MW_NSNS}. We now discuss the mass-transfer efficiency, $\epsilon$, in light of the two input variables: $\beta_{\rm min}$ and $\alpha_{\rm RLO}$.

\subsubsection*{i) Re-emission from the accretor, $\beta_{\rm min}$}\label{sec:beta_min}
\begin{figure}
  \includegraphics[width=\columnwidth]{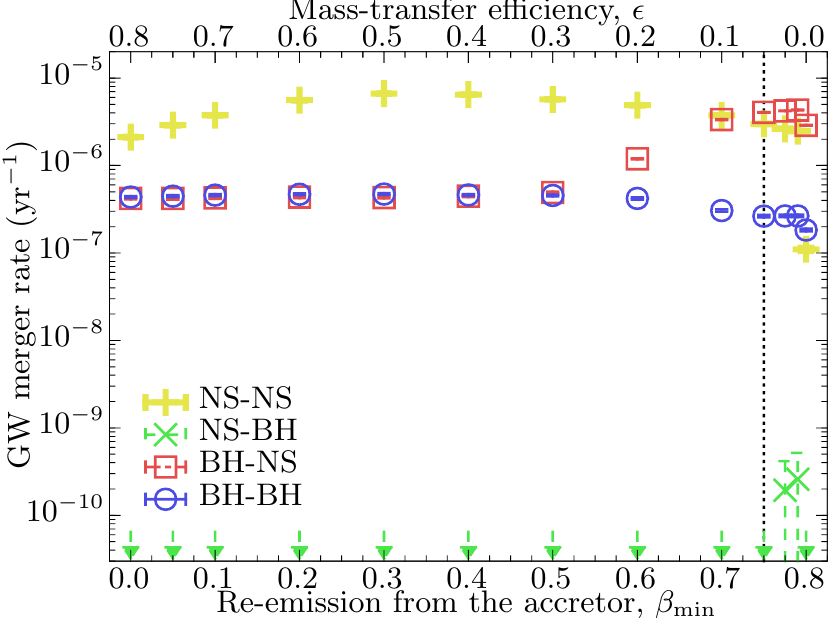}
  \caption{\label{fig:gwMergerRate_betamin}The GW merger rate in a MW-like galaxy as a function of the minimum mass ejection fraction in the vicinity of the accretor during RLO. The dotted line marks our default value of \mbox{$\beta _{\rm min}=0.75$}. The green arrows mark upper limits for simulated NS-BH systems.}
\end{figure}

The more mass a star accretes, the more massive a compact object it can produce compared to evolution in isolation as a single star. Furthermore, a more massive companion helps the binary to survive a large kick during a SN. So, naively, one would expect to produce more DCO binaries for more efficient accretion. But this is only true in wide systems. The close binaries usually evolve through a CE~phase after the first compact object is formed. In this case, a more massive companion possesses a more massive envelope which is harder to eject. Consequently, tight systems more often result in an early merger via spiral~in. This effect explains why the merger rate of our simulated BH-NS systems (\FigureRef{fig:gwMergerRate_betamin}) is smaller by about an order of magnitude for efficient accretion (\mbox{$\beta _{\rm min}=0.0$}, and \mbox{$\alpha_{\rm RLO}=0.2$} corresponding to \mbox{$\epsilon = 0.8$}) compared to inefficient accretion \mbox{($\beta _{\rm min}=0.75$)}.

In binaries producing double NS or double BH systems, however, a lower $\beta_{\rm min}$ value, i.e. more effective accretion, does not produce significantly more or fewer GW mergers -- although the merger rate for double NS systems peaks near \mbox{$\beta _{\rm min}=0.50$}.

The difference in our simulated results between using our default \mbox{$\beta_{\rm min}=0.75$} and \mbox{$\beta_{\rm min}=0.0$} is shown in \FigureRef{fig:final_masses_formation_MW_NSNS} for the final NS masses. For further comparison, all results using a high mass-transfer efficiency \mbox{($\beta _{\rm min}=0.0$)} are summarised in \AppendixRef{app:efficient_mass_transfer}.

%%%%%%%%%%%%%%%%%%%%%%%%%%%%%%%%%%%%%%%%%%%%%%%%%%
\subsubsection*{ii) Direct wind mass loss, $\alpha_{\rm RLO}$}\label{sec:alpha_RLO}
For a given $\beta_{\rm min}$, the fraction of material lost directly from the donor star in the form of an assumed fast wind, $\alpha_{\rm RLO}$, is constrained by conservation of mass. Given that \mbox{$\epsilon\geq0$} we have \mbox{$\alpha_{\rm RLO}\leq 1-\beta_{\rm min}-\delta$} (\SectionRef{sec:roche-lobe_overflow}). Hence, for \mbox{$\beta_{\rm min}=0.75$} we have \mbox{$\alpha_{\rm RLO}\in[0.0:0.25]$}. 

The dependency of the formation rate of DCO binaries on $\alpha _{\rm RLO}$ has no clear trend (lower panel of \FigureRef{fig:formationRate_alphaRLO_betamin}). Nevertheless, two main effects are at work when we increase $\alpha _{\rm RLO}$. First, stronger wind mass loss widens the orbit. Secondly, the accretor gains less mass. As a consequence, the resulting wider and lighter binary is more easily disrupted in a subsequent SN. However, if a CE follows after a SN which fails to disrupt the binary then the chance of surviving the CE~phase increases.

Formation rates calculated with $\alpha _{\rm RLO}=0.25$ are special in the case of $\beta_{\rm min}=0.75$ given that it implies $\epsilon=0.0$, i.e. no mass is accreted by the accretor and thus the initially more massive star always evolves first. This suppresses in particular the formation of NS-BH systems. 

%%%%%%%%%%%%%%%%%%%%%%%%%%%%%%%%%%%%%%%%%%%%%%%%%%
\subsubsection{Mass-ratio limit for stable mass transfer, \texorpdfstring{$q_{\rm limit}$}{q\_limit}}\label{sec:q_limit}
\begin{figure}
  \includegraphics[width=\columnwidth]{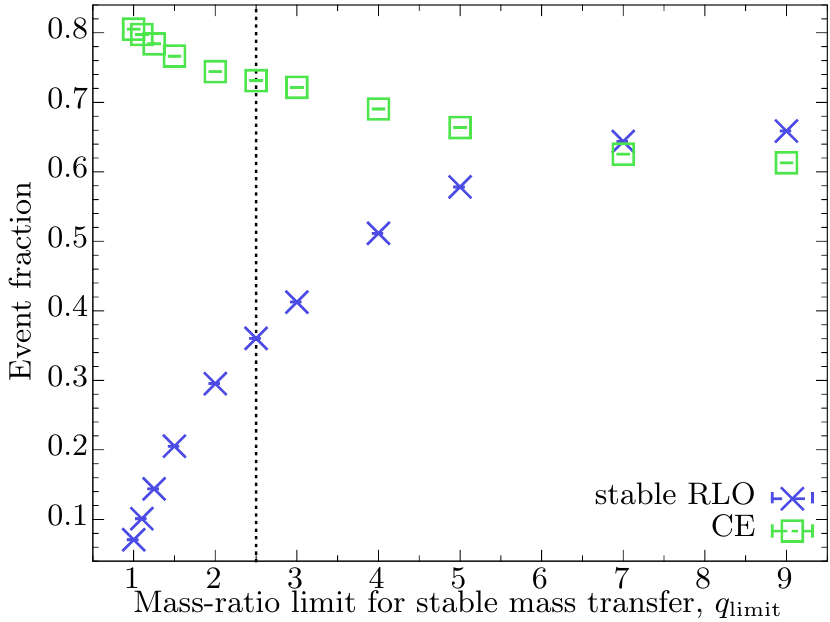}
  \caption{\label{fig:RLO-CE_qlimit_normalized}The mean count of stable RLO (blue) and CE (green) events per binary system evolved in a MW-like galaxy depending on the critical mass-ratio limit, $q_{\rm limit}$. Case~A or early Case~B RLO events with \mbox{$q<q_{\rm limit}$} are stable whereas a CE is assumed to develop for \mbox{$q\geq q_{\rm limit}$}. The dotted line marks our default \mbox{$q_{\rm limit}=2.5$}.} 
\end{figure}

One of our stability criteria of mass transfer is related to the mass ratio between the two stars at the onset of RLO. In \FigureRef{fig:RLO-CE_qlimit_normalized}, we show the number of stable RLO and CE events (unstable RLO) as a function of the chosen critical mass-ratio limit, $q_{\rm limit}$, for which \mbox{$q<q_{\rm limit}$} will lead to a stable RLO (\SectionRef{sec:roche-lobe_overflow}). The general trend of having more stable RLO events with increasing $q_{\rm limit}$ is clearly visible. Note, the added number of RLO/CE events per binary system evolved can exceed 1.0 since there are sometimes multiple stages of mass transfer between the two stars (cf. \FigureRef{fig:vdHcartoon}), and the total number of events (adding RLO and CE) increases when there are fewer episodes of unstable mass transfer. As discussed in \SectionRef{sec:alpha_CE}, early coalescence happens mainly during a CE which suppresses the possibility of subsequent mass transfer in a given system. Therefore, also the formation rates and the GW merger rates increase clearly with larger values of $q_{\rm limit}$, cf. \TableRef{tab:rate_variations}.

To further test the various stability criteria listed in \SectionRef{sec:roche-lobe_overflow}, we also performed a test run in which RLO is always assumed to be stable for giant stars with \mbox{$q<1.5$} (to avoid dynamical instabilities in systems e.g. with relatively low-mass giant stars transferring mass to a BH accretor). This results in 25~per~cent more stable RLO events compared to our default simulations, but to less than a 10~per~cent change in the formation and merger rates given in \TableRef{tab:rate_variations}.

%%%%%%%%%%%%%%%%%%%%%%%%%%%%%%%%%%%%%%%%%%%%%%%%%%
\subsubsection{Initial mass function}\label{sec:alpha_IMF}
\begin{figure}
  \includegraphics[width=\columnwidth]{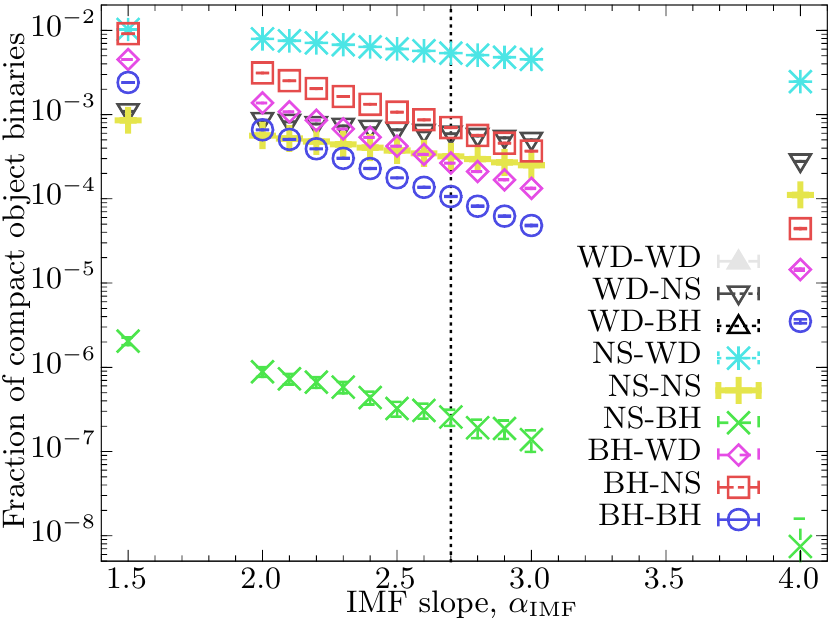}
  \caption{\label{fig:survived_formation_alphaIMF_relative}The relative fraction of DCO systems (here including WDs) formed in a MW-like galaxy depending on the slope of the IMF, $\alpha _{\rm IMF}$. There are no WD-BH binaries formed. The double WD systems are by far the main population with a relative formation fraction close to $1.0$ (outside of the plotted scale). The dotted line marks our default IMF (a Salpeter-IMF, \mbox{$\alpha_{\rm IMF}=2.7$}).}
\end{figure}

Changing the slope of the applied IMF, and therefore the relative abundance of massive stars, makes a large difference on the formation rates of DCO binaries. \FigureRef{fig:survived_formation_alphaIMF_relative} (here including WDs) shows the relative fractions of the different DCO binaries formed as a function of the slope of the IMF, \mbox{$1.5\leq\alpha_{\rm IMF}\leq4.0$}. These fractions change by an order of magnitude, or even more, when going from a steep to a more flat IMF. Independent of all the binary interactions, the change of the IMF slope has a simple monotonic effect on the formation fraction of the different types of DCO binaries: the steeper the IMF, the less binaries form with NSs or BHs.

Recently, \citet{sse+18} studied massive stars ($\unit{15\text{ to }200}{\Msun}$) in the young cluster 30~Doradus and found evidence for \mbox{$\alpha _{\rm IMF}=1.90^{+0.37}_{-0.26}$}. If such an IMF is representative for the star-formation history within the observable LIGO-Virgo volume of the local Universe, then the detection rates of DCO mergers will be substantially larger than our default results (\FigureRef{fig:survived_formation_alphaIMF_relative}).

Double WD systems are by far the most common DCOs (out of the scale shown in \FigureRef{fig:survived_formation_alphaIMF_relative}) even though our simulations do not account for the very low-mass stars below our applied mass range (\TableRef{tab:standard}). Including all low-mass stars down to e.g. $\unit{0.8}{\Msun}$ would increase the number of double WD binaries even more, but the relative ratios of DCOs without WDs would remain unaffected. The number of BH systems, however, increases by changing the upper mass limit, see \SectionRef{sec:mass_ranges}. Systems in which a WD forms before a BH are not expected to be produced in nature from isolated binaries because of the excessive mass reversal required.

%%%%%%%%%%%%%%%%%%%%%%%%%%%%%%%%%%%%%%%%%%%%%%%%%%
\subsubsection{Range of the initial primary and secondary masses}\label{sec:mass_ranges}
For the calculation of the formation and GW merger rates of DCO binaries with NSs and BHs, there are no effects when the lower ZAMS mass boundaries of the primary and the secondary stars are changed, only the number of WD progenitors changes. However, at the high-mass end of the scale, changing the maximum mass boundary influences on the rate of BH binary formation significantly more than that of NS binaries, cf. the last column of \TableRef{tab:rate_variations}.

%%%%%%%%%%%%%%%%%%%%%%%%%%%%%%%%%%%%%%%%%%%%%%%%%%
\subsubsection{Range of the initial semi-major axis}\label{sec:semi_major_axis_range}
\begin{figure}
  \includegraphics[width=\columnwidth]{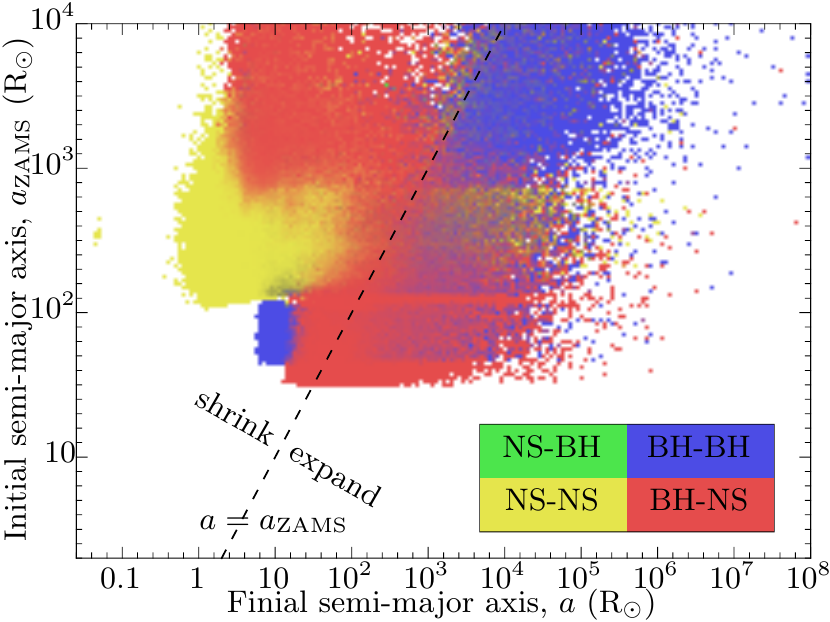}
  \caption{\label{fig:semi-major-axis_beta75}The initial ($a_{\rm ZAMS}$) and the final semi-major axis ($a$) of our simulated DCO binaries in a MW-like galaxy. Each pixel's colour shows the type of the two compact objects formed from the initial primary and secondary stars, respectively. This type is not necessarily reflecting the formation order although it is in most cases. The black dashed line indicates the binaries ending their evolution with the same semi-major axis as they started. This does not mean that they have the same orbital separation during the whole binary evolution.}
\end{figure}

The initial semi-major axis is the key parameter determining whether a system will experience binary interactions or not. \FigureRef{fig:semi-major-axis_beta75} shows how the semi-major-axis changes from the ZAMS to the time when both compact objects are formed. Because only systems which survive the binary evolution are plotted, a lack of systems in the initial separation does not mean that those systems are not simulated. For example, there are no double NS binaries from initial ZAMS separations below $\unit{100}{\Rsun}$. However, BH-NS binaries can originate from binaries with an initial orbital separation down to $\unit{30}{\Rsun}$. Most of the double NS binaries shrink their orbit during their progenitor evolution. These systems are located above the dashed line.

Binaries have a Roche-lobe filling star at birth if the initial orbital separation is too small. Thus, the minimum orbital separation needed to avoid one star filling its Roche lobe on the ZAMS is $\sim\unit{10}{\Rsun}$ and $\sim\unit{20}{\Rsun}$ for double NS and double BH progenitor binaries, respectively. Changing the maximum semi-major axis at ZAMS adds or removes systems at the top of \FigureRef{fig:semi-major-axis_beta75}. When keeping the overall number of binaries constant, adding systems simultaneously means removing some realisations from the existing distribution\footnote{Here we have ignored that changing $a_{\rm max}$ strictly speaking affects the binary fraction and that there are observational constrains on the fraction of binaries per $\log(a)$ interval, see \citet{md17}.}. Because the survival rate is lower at longer initial separations, the total formation rate of all systems decreases for a larger upper boundary of the initial orbital separation. 

Our GW merger rates are dominated by systems which have a short final separation. Therefore, the merger rate does not change in the same way as the formation rate, cf. \TableRef{tab:rate_variations2}. 

%%%%%%%%%%%%%%%%%%%%%%%%%%%%%%%%%%%%%%%%%%%%%%%%%%
\subsection{Comparison to other studies and observations}\label{sec:comparison_to_other_work}
We now compare our results to those of other recently published binary population synthesis studies on DCO binaries and to the first seven LIGO-Virgo detection events of merging double BHs and double NSs. In addition, we compare our simulated Galactic double NS merger rate with estimated constraints on core-collapse SNe (CC~SNe), short $\gamma$-ray bursts (sGRBs) and production of heavy $r$-process element events.

\subsubsection{Number of double compact object binaries present in the Milky Way}\label{sec:compare_DCO_MW}
\begin{table}
 \caption{\label{tab:MWsystems}Number of DCO systems present in the Milky Way as a function of the parameter $\beta_{\rm min}$, where the maximum RLO mass-transfer efficiency is given by \mbox{$\epsilon_{\rm max}=1-\alpha_{\rm RLO}-\beta_{\rm min}$}. The numbers are given for stellar evolution lasting $\unit{(10\pm3.81)}{\giga\yr}$.}
 \begin{tabular}{cr@{}lr@{}lr@{}l}
  \hline
  {\normalsize\rule{0pt}{\f@size pt}}Systems in MW & \multicolumn{2}{c}{$\beta_{\rm min}=0.75$} & \multicolumn{2}{c}{$\beta_{\rm min}=0.5$} & \multicolumn{2}{c}{$\beta_{\rm min}=0$}\\
  \hline
  {\normalsize\rule{0pt}{\f@size pt}}NS-NS & $38246$ & $^{+12445}_{-13100}$ & $41340$ & $^{+12624}_{-13405}$ & $1616$ & $^{+499}_{-505}$\\
  {\normalsize\rule{0pt}{\f@size pt}}NS-BH & $55$ & $^{+21}_{-21}$ & $19$ & $^{+7}_{-7}$ & $25$ & $^{+9}_{-10}$\\
  {\normalsize\rule{0pt}{\f@size pt}}BH-NS & $108845$ & $^{+36702}_{-37811}$ & $21303$ & $^{+7504}_{-7687}$ & $13603$ & $^{+4532}_{-4798}$\\
  {\normalsize\rule{0pt}{\f@size pt}}BH-BH & $20073$ & $^{+7585}_{-7602}$ & $16237$ & $^{+5934}_{-6006}$ & $21988$ & $^{+8247}_{-8267}$\\
  \hline
 \end{tabular}
\end{table}

We predict a total Galactic population (\TableRef{tab:MWsystems}) of the order $40\,000$ NS-NS binaries, $100\,000$ BH-NS binaries and $20\,000$ BH-BH binaries. Whereas the latter number is stable, the number of BH-NS and, in particular, the number of NS-NS systems present in the MW is strongly dependent on $\beta_{\rm min}$. The reason is that efficient accretion \mbox{($\beta_{\rm min}=0$)} produces relatively tighter binaries. Hence, the majority of double NS binaries formed will merge on a short timescale (thus removing them from the observable sample), while in the case of inefficient accretion \mbox{($\beta_{\rm min}=0.75$)} more than half of the double NS binaries are formed in wider systems which do not merge within a Hubble time. For BH-NS systems the decline in the Galactic population with decreasing $\beta_{\rm min}$ is caused by a combination of a significant decrease in formation rate with only a modest decrease in merger rate.

\subsubsection{Number of active Galactic radio pulsars in double neutron star and black hole companion binaries}
\label{sec:compare_pulsars_MW}
In \TableRef{tab:MWsystems}, we see that about $40\,000$ double NS systems accumulate in the MW over $\unit{10}{\giga\yr}$ (for \mbox{$\beta_{\rm min}=0.75$} or $0.50$). Assuming an active radio lifetime of $\unit{100}{\mega\yr}$ as a lower limit for the first-born (mildly recycled) pulsar \citep[i.e. slightly more than the  $\unit{10-50}{\mega\yr}$ typically expected for non-recycled radio pulsars,][]{lk04,jk17}, means that we expect at least of the order $400$ active radio pulsars in NS-NS systems in the MW. Depending on beaming effects, probably $100$ to $150$ of these double NS systems will be observable from Earth.

Radio pulsars in BH-NS systems are the second-formed compact objects and hence they are non-recycled and similar in nature to normal pulsars. Thus we expect these pulsars to have relatively slow spins and narrow beams covering only 5 to 10~per~cent of the sky \citep{lk04}. From our default simulations with \mbox{$\beta_{\rm min}=0.75$}, we find of the order $100\,000$ BH-NS binaries produced in $\unit{10}{\giga\yr}$. We then estimate about $250$ active sources based on an assumed average radio lifetime of $\unit{25}{\mega\yr}$ and a population of some 10 to 25 detectable radio pulsars (beaming in our direction) in such BH-NS binaries. If we consider our simulations with \mbox{$\beta_{\rm min}=0.50$}, we expect about $50$ active sources of which only 2 to 5 systems will be detectable. Note, current radio pulsar surveys are not able to detect the far majority of pulsars beaming in our direction. This situation will improve dramatically with the full SKA \citep{kbk+15}. For discussions on the very few NS-BH binaries (in which the NS forms first) produced in our simulations, see \SectionRef{sec:NSBH_rare}.

\subsubsection{Galactic merger rates of double compact object binaries}\label{sec:compare_merger_rates}
We first compare our new simulations to the results obtained from the old code used in \citet{vt03}. The largest difference is related to systems evolved at high metallicity. In our new study, we find that the Galactic DCO merger rate is dominated by BH-NS and double NS systems, and not by double BH binaries (\TableRef{tab:rate_variations}). Our predicted rate of Galactic double NS mergers is about $\unit{3.0}{\mega\yr^{-1}}$, using our default values for the input physics parameters, and thus within a factor of two of \citet{vt03}. Our predicted Galactic merger rate of double BHs is only $\unit{0.3}{\mega\yr^{-1}}$, which is much less than the rate of $\unit{10}{\mega\yr^{-1}}$ found in \citet{vt03}. The reason for this is a combination of our CEs being more tightly bound, thus producing fewer DCO binaries in general, and a systematic shift in our applied threshold core mass for producing BHs.

Given that our simulated NS-NS merger rate for a MW-like galaxy is about $\unit{3.0}{\mega\yr}^{-1}$, our results are among those that predict the lowest rates compared to other binary population synthesis studies \citep{aaa+10,cbkb18,vns+18}. There are many reasons for this, but one particular important issue is that many other codes model the CE~phase with a constant envelope structure parameter, $\lambda$, although it has been demonstrated that this is a poor approximation \citep{dt00,dt01,prh03} -- see also \FigureRef{fig:Lambda_MW} and \SectionRef{sec:MW_metallicity}. As an example, it was shown by \citet{vt03} that using a constant value of \mbox{$\lambda=0.5$}, instead of using realistic calculated values which depend on the evolutionary status of the star, increases the predicted merger rate by more than an order of magnitude. We confirm this by test simulations using \OurCode and also obtain similar discrepancies by combing our stellar tracks with $\lambda$-tables of a different source.

There are several binary population synthesis studies performed with the codes {\it StarTrack} \citep{bkr+08,dbf+13,dbo+15,cbkb18}, {\it binary\_c} \citep{itk+04,idk+06,igs+09,ipj+18}, {\it BPASS} \citep{es16,esx+17}, {\it COMPAS} \citep{svm+17,bmn+17,vns+18}, {\it MOBSE} \citep{mgrs17, gms17} and many more. The main differences are: the stellar evolution (i.e. applied stellar model grids and their resolution), the treatment of CE evolution, and the applied SN-kick distributions.

While \OurCode interpolates large stellar evolution grids (\SectionRef{sec:stellar_grids}) most other codes use fitting functions for the individual stars and their evolution. Such fitting functions \citep[e.g.][]{hpt00,htp02} do not recover some parameters well compared to more detailed stellar models. One example is the stellar structure parameter, $\lambda$. However, during a CE evolution such knowledge is important to determine if the system merges or survives the unstable mass transfer \citep[see e.g.][for discussions]{ktl+16}. Some other codes do use a variable stellar structure parameter, e.g. \citet{dbf+12,cpi+14,vns+18}. However, as opposed to other codes, \OurCode does this in a self-consistent way by using the same stellar evolution model for calculating $\lambda$ and e.g. the corresponding core mass, cf. \SectionRef{sec:stellar_grids}.

Furthermore, as discussed previously, the distribution of kicks received by the NSs must depend on their formation history, i.e. the remaining envelope and the core mass of the exploding star. Some observed high-velocity pulsars need large kicks while some tight and nearby circular pulsar binaries must have experienced small kicks. We account for this in our new code in a systematic way by using different kick distributions depending on the evolutionary history of the exploding star, i.e. by considering how much of the envelope is stripped prior to the SN explosion (\SectionRef{sec:supernovae}).

Finally, we note that our implementation of the more advanced numerical Case~BB RLO modelling of \citet{tlp15} increases the double NS merger rate by an order of magnitude compared to our simulations based on the results of \citet{dp03} which often lead to unstable RLO, causing a large fraction of close binaries to merge prior to the second SN.

\subsubsection{Merger-rate densities}\label{sec:compare_merger_rate_densities}
\begin{figure}
  \includegraphics[width=\columnwidth]{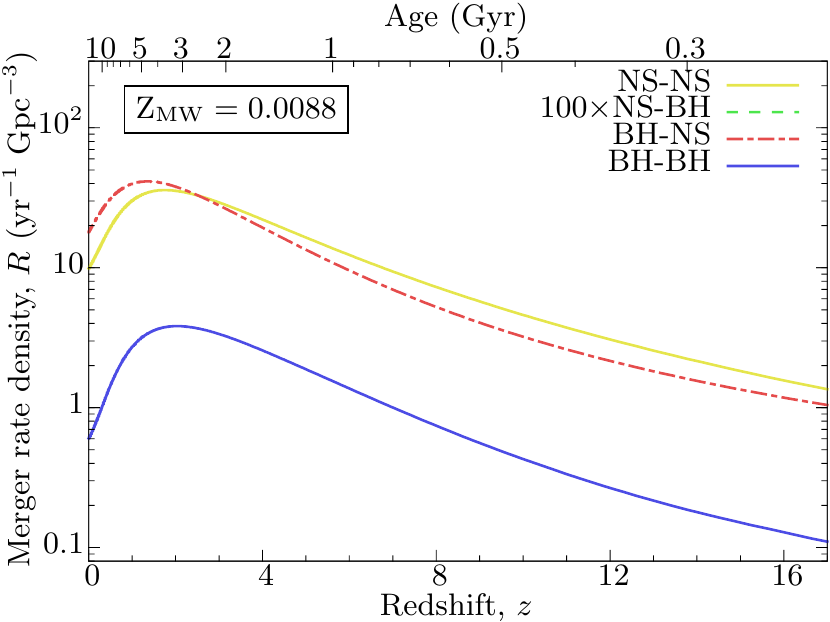}
  \includegraphics[width=\columnwidth]{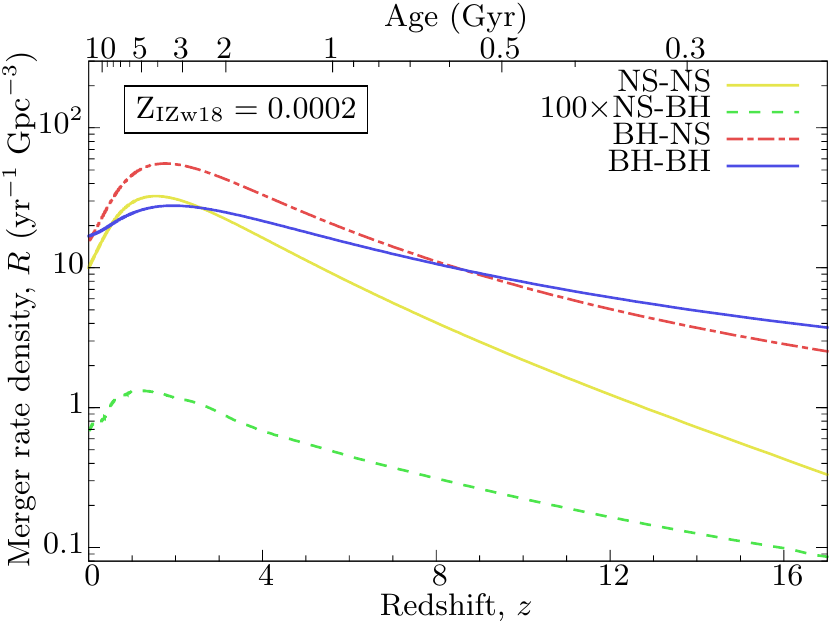}
  \caption{\label{fig:merger_rate_density_beta75} Merger-rate densities of the DCO binaries at MW-like (upper panel) and IZw18-like (lower panel) metallicity as a function of cosmological redshift and age of the Universe -- see \TableRef{tab:detection_rates_beta75} for numbers. Colour-coded is the type of the two compact objects in their formation order. The values of the NS-BH binaries are multiplied with a factor of 100 to be visible in the lower plot, while they remain absent in the upper one.}
\end{figure}

\begin{table*}
 \caption{\label{tab:detection_rates_beta75}Our simulated merger-rate densities, $R$, at redshift zero and LIGO-Virgo detection rates, $R_{\rm D}$, calculated with two different star-formation history and galaxy-density scaling methods: the central two columns are based on the procedure outlined in \citet{dbf+13} and the right two columns are based on \citet{aaa+10}. Using the unweighted average $\langle\mathcal{M}^{2.5}\rangle$, some geometrical factors and assuming a signal-to-noise threshold, \mbox{$\rho\geq8$}, we calculate the expected LIGO-Virgo detection rates $R_{\rm D}$ (fourth column) and $R_{\rm D,cSFR}$ (sixth column) following equation~(3) of \citet[][cf. their table~1]{dbo+15}. The merger-rate densities and detection rates are calculated for two different metallicity environments (\mbox{${\rm Z}_{\rm MW}=0.0088$} and \mbox{${\rm Z}_{\rm IZw18}=0.0002$}, top and central panel, respectively) applying our default input parameter settings (\TableRef{tab:standard}). The bottom panel shows our rates calculated under an ``optimistic'' setting at MW metallicity to boost the NS-NS merger rate (i.e. applying \mbox{$\alpha_{\rm IMF}=2.3$}, \mbox{$\alpha_{\rm RLO}=0.15$}, \mbox{$\beta_{\rm min}=0.5$}, \mbox{$\alpha_{\rm CE}=0.8$} and \mbox{$\alpha_{\rm TH}=0.3$}). See \SectionRef{sec:compare_LIGO} for a discussion.}
 \begin{tabular}{cr<{$\unit{}{\Msun^{2.5}}$}|r@{$\times$}l<{$\unit{}{\yr^{-1}\usk\giga{\rm pc}^{-3}}$}r<{$\unit{}{\yr^{-1}}$}|r@{$\times$}l<{$\unit{}{\yr^{-1}\usk\giga{\rm pc}^{-3}}$}r<{$\unit{}{\yr^{-1}}$}}
  \hline
  {\normalsize\rule{0pt}{\f@size pt}}${\rm Z}_{\rm MW}$ & \multicolumn{1}{c|}{$\langle\mathcal{M}^{2.5}\rangle$} & \multicolumn{2}{c}{$R_{z=0}$} & \multicolumn{1}{c|}{$R_{\rm D}$} & \multicolumn{2}{c}{$R_{\rm cSFR}$} & \multicolumn{1}{c}{$R_{\rm D,cSFR}$}\\
  \hline
  {\normalsize\rule{0pt}{\f@size pt}}NS-NS & $1.36$ & $9.85$ & $10^{0}$ & $0.28$ & $3.47$ & $10^{1}$ & $0.98$\\
  {\normalsize\rule{0pt}{\f@size pt}}NS-BH & $20.0\phantom{0}$ & $0.00$ & $10^{0}$ & $0.00$ & $0.00$ & $10^{0}$ & $0.00$\\
  {\normalsize\rule{0pt}{\f@size pt}}BH-NS & $15.7\phantom{0}$ & $1.80$ & $10^{1}$ & $5.88$ & $4.72$ & $10^{1}$ & $15.43$\\
  {\normalsize\rule{0pt}{\f@size pt}}BH-BH & $233\phantom{.00}$ & $6.01$ & $10^{-1}$ & $2.92$ & $3.08$ & $10^{0}$ & $14.95$\\
  \hline
  \hline
  {\normalsize\rule{0pt}{\f@size pt}}${\rm Z}_{\rm IZw18}$ & \multicolumn{1}{c|}{$\langle\mathcal{M}^{2.5}\rangle$} & \multicolumn{2}{c}{$R_{z=0}$} & \multicolumn{1}{c|}{$R_{\rm D}$} & \multicolumn{2}{c}{$R_{\rm cSFR}$} & \multicolumn{1}{c}{$R_{\rm D,cSFR}$}\\
  \hline
  {\normalsize\rule{0pt}{\f@size pt}}NS-NS & $1.27$ & $1.00$ & $10^{1}$ & $0.27$ & $3.28$ & $10^{1}$ & $0.87$\\
  {\normalsize\rule{0pt}{\f@size pt}}NS-BH & $32.3\phantom{0}$ & $6.61$ & $10^{-3}$ & $0.00$ & $1.55$ & $10^{-2}$ & $0.01$\\
  {\normalsize\rule{0pt}{\f@size pt}}BH-NS & $35.5\phantom{0}$ & $1.54$ & $10^{1}$ & $11.40$ & $5.32$ & $10^{1}$ & $39.34$\\
  {\normalsize\rule{0pt}{\f@size pt}}BH-BH & $1720\phantom{.00}$ & $1.68$ & $10^{1}$ & $603.02$ & $3.45$ & $10^{1}$ & $1235.27$\\
  \hline
  \hline
  {\normalsize\rule{0pt}{\f@size pt}}optimistic & \multicolumn{1}{c|}{$\langle\mathcal{M}^{2.5}\rangle$} & \multicolumn{2}{c}{$R_{z=0}$} & \multicolumn{1}{c|}{$R_{\rm D}$} & \multicolumn{2}{c}{$R_{\rm cSFR}$} & \multicolumn{1}{c}{$R_{\rm D,cSFR}$}\\
  \hline
  {\normalsize\rule{0pt}{\f@size pt}}NS-NS & $1.31$ & $7.09$ & $10^{1}$ & $1.94$ & $1.59$ & $10^{2}$ & $4.37$\\
  {\normalsize\rule{0pt}{\f@size pt}}NS-BH & $19.4\phantom{0}$ & $0.00$ & $10^{0}$ & $0.00$ & $0.00$ & $10^{0}$ & $0.00$\\
  {\normalsize\rule{0pt}{\f@size pt}}BH-NS & $21.9\phantom{0}$ & $1.34$ & $10^{1}$ & $6.11$ & $2.44$ & $10^{1}$ & $11.17$\\
  {\normalsize\rule{0pt}{\f@size pt}}BH-BH & $275\phantom{.00}$ & $4.34$ & $10^{1}$ & $248.34$ & $1.09$ & $10^{2}$ & $623.03$\\
  \hline
 \end{tabular}
\end{table*}

In \FigureRef{fig:merger_rate_density_beta75}, we plot our estimated merger-rate densities at two different metallicities. These plots are calculated from a method similar to that of figures~3 and 4 of \citet{dbf+13} and, for a better comparison, we applied the same star-formation rate function \citep{srd+04} and cosmological parameters as in \citet{dbf+13}. Furthermore, a binary fraction of $100$~per~cent is assumed. The two panels displayed are for a constant metallicity, showing the high- and low-metallicity cases (${\rm Z}_{\rm MW}=0.0088$ and ${\rm Z}_{\rm IZw18}=0.0002$) used in our work. At high metallicity, our simulations yield merger-rate densities at redshift zero of at least $\unit{10}{\yr^{-1}\usk\giga{\rm pc}^{-3}}$ and $\unit{0.6}{\yr^{-1}\usk\giga{\rm pc}^{-3}}$, for double NS and double BH mergers respectively, depending on the applied galaxy-density scaling. See \TableRef{tab:detection_rates_beta75} for detailed numbers.

When considering mergers in low-metallicity environments and at low redshift \mbox{($z<1$)}, we find that the three dominant binary types (BH-BH, BH-NS and NS-NS) are more or less equally frequent (i.e. \mbox{$R_{z=0}\simeq\unit{10-17}{\yr^{-1}\usk\giga{\rm pc}^{-3}}$}). However, for NS-BH binaries we obtain \mbox{$R_{z=0}<\unit{0.01}{\yr^{-1}\usk\giga{\rm pc}^{-3}}$} (see \SectionRef{sec:NSBH_rare} for a discussion on these systems). Applying efficient RLO makes the NS binaries less dominant, see \AppendixRef{app:merger_rate_density}. As our applied inefficiency of RLO is motivated by measurements of NS masses in double NS binaries in the Milky Way (\SectionRef{sec:compact_object_masses_double_neutron_star_binaries}), it can not be excluded that the mass-transfer efficiency is larger at lower metallicities.

In terms of anticipated LIGO-Virgo detection rates, it is evident from \TableRef{tab:detection_rates_beta75} that LIGO-Virgo should mainly detect GW mergers of binary BHs originating from low-metallicity environments. At a much smaller rate, LIGO-Virgo will detect mixed NS/BH and double NS mergers originating from both high- and low-metallicity environments. It should be noted that we do not expect the GW merger data analysis to be able to distinguish between NS-BH and BH-NS systems. It will not be possible to determine the formation order of the compact objects, despite potential differences in the NS spin rates depending on whether the NSs are (mildly) recycled. While the merger-rate densities of the BH-NS and double BH binaries are similar at lower metallicity, their detection rate is different by more than an order of magnitude. This difference originates from their different chirp masses which are typically \mbox{$\mathcal{M}_{\rm BH/NS}\approx\unit{4}{\Msun}$} and \mbox{$\mathcal{M}_{\rm BH-BH}\approx\unit{20}{\Msun}$}, respectively (\FigureRef{fig:final_chirpmass_GWmerger_beta75}, lower panel). That of the double NS systems is only about $\unit{1.1}{\Msun}$. 

\subsubsection{Comparison to reported LIGO-Virgo detections}\label{sec:compare_LIGO}
\begin{figure*}
  \includegraphics[width=\textwidth]{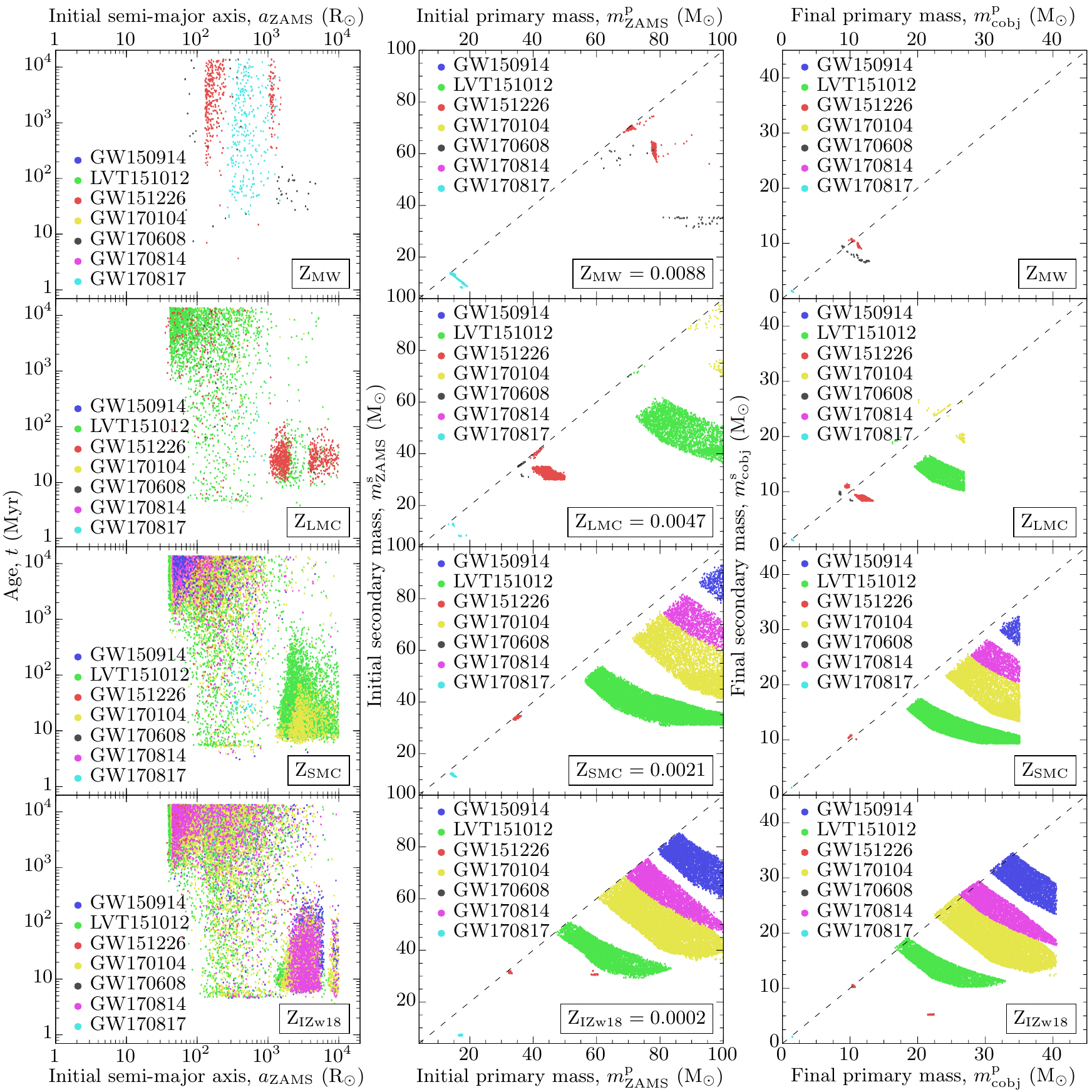}
  \caption{\label{fig:LIGO_progenitors_beta75}Simulated binary progenitor properties of the first seven LIGO-Virgo detections \citep[GW150914 blue, LVT151012 green, GW151226 red, GW170104 yellow, GW170608 grey, GW170814 purple and GW170817 teal,][]{aaa+16c,aaa+17,aaa+17b,aaa+17c}. The four rows (top to bottom) correspond to metallicities of the Milky Way, the LMC, the SMC and IZw18, respectively. The dashed diagonal lines indicate similar masses.}
\end{figure*}

In \FigureRef{fig:LIGO_progenitors_beta75}, we plot the properties of the progenitor binaries of the first seven LIGO-Virgo events according to our simulations. Shown here are the distributions of the ZAMS masses and of the semi-major axis, as well as that of the final BH masses and ages of the systems at the time of the merger events. The systems selected, plotted in seven different colours, are those which will merge within a Hubble time and match the observations in both chirp mass and system mass within the uncertainties given in \citet{aaa+16c,aaa+17,aaa+17b,aaa+17c}. At low metallicity, possible progenitors are found in our simulations for six of the reported LIGO-Virgo events -- the exception is GW170608. The more massive the merging BHs are, the more likely the event happened in a low-metallicity environment. Although GW151226 could have formed at all metallicities we investigated, it is more likely that it originates from LMC-like or higher metallicity, as it is the case for GW170608. The double NS merger GW170817 could also have formed at all the metallicities we investigated.

\citet{svm+17} investigated the progenitors of the first three LIGO detections, and the second and third columns in \FigureRef{fig:LIGO_progenitors_beta75} are similar to their figure~1. The results of our work and that of \citet{svm+17} are similar. However, our lowest metallicity case (${\rm Z}_{\rm IZw18}=0.0002$) is lower than theirs at \mbox{$Z=0.001$}. Our SMC metallicity simulation is similar to their \mbox{$Z=0.002$} model. The first column of \FigureRef{fig:LIGO_progenitors_beta75} shows the initial semi-major axis and the age of the binaries when they merge via GW radiation. At low metallicity, there is a much larger spread in initial separations and ages of the potential progenitor systems compared to high metallicity.

The empirical merger-rate density of BH-BH systems as determined by LIGO-Virgo detections is currently reported to be \mbox{$R\simeq\unit{12-213}{\yr^{-1}\usk\giga{\rm pc}^{-3}}$} \citep{aaa+17}. This empirical range is consistent with our default simulations at low-metallicity environments \mbox{(${\rm Z}_{\rm IZw18}=0.0002$)} where we obtain \mbox{$R_{z=0}=\unit{16.8}{\yr^{-1}\usk\giga{\rm pc}^{-3}}$}, but higher than our simulations of high-metallicity environment \mbox{(${\rm Z}_{\rm MW}=0.0088$)} where we only obtain \mbox{$R_{z=0}=\unit{0.6}{\yr^{-1}\usk\giga{\rm pc}^{-3}}$}. However, as discussed in \SectionRef{sec:parameter_studies}, variations in the assumed values of key input physics parameters can increase or decrease the merger rate significantly. As an example, the effect of solely increasing the CE ejection efficiency parameter from \mbox{$\alpha_{\rm CE}=0.5$} (our default value) to \mbox{$\alpha_{\rm CE}=0.8$} is to increase our estimated BH-BH merger rate by an order of magnitude (\TableRef{tab:rate_variations}). Our simulated merger-rate density also increases by a factor of typically 2 to 4 when considering redshifts near \mbox{$z\simeq 2$} in comparison to \mbox{$z\simeq 0$} (\FigureRef{fig:merger_rate_density_beta75}). Finally, we note that using the galaxy-density scaling method of \citet{aaa+10} and a constant star-formation rate yields a default simulated local merger-rate density of BH-BH systems between \mbox{$R_{\rm cSFR}=\unit{3-35}{\yr^{-1}\usk\giga{\rm pc}^{-3}}$}, depending on metallicity (see the fifth column in \TableRef{tab:detection_rates_beta75}).

An astonishing empirical NS-NS merger-rate density of \mbox{$R=\unit{1540^{+3200}_{-1220}}{\yr^{-1}\usk\giga{\rm pc}^{-3}}$} was recently reported by LIGO-Virgo based on the first GW detection of such a system \mbox{\citep[GW170817,][]{aaa+17c}}. This NS-NS merger event, the loudest GW signal ever recorded with a S/N ratio of 32.4, was also detected as a sGRB \citep{fermi+17,aaa+17d} and a multiwavelength kilonova \citep[e.g.][]{sha+17,cfk+17,aaa+17e,scj+17,dps+17}. The combined signal seen in all three LIGO-Virgo detectors enabled an electromagnetic follow-up campaign which identified a counterpart near the galaxy NGC~4993, consistent with the localization and distance ($\sim\unit{40}{\mega{\rm pc}}$) inferred from GWs.

From our default simulations, we find a Galactic NS-NS merger rate of about $\unit{3.0}{\mega\yr^{-1}}$, translating into a merger-rate density of \mbox{$R_{z=0}\simeq\unit{10}{\yr^{-1}\usk\giga{\rm pc}^{-3}}$}, or a LIGO-Virgo detection rate of only \mbox{$R_{\rm D}\simeq\unit{0.3}{\yr^{-1}}$} at full design sensitivity. Hence, to better match our simulated rates with that inferred from observations, we calculate an ``optimistic'' simulation for MW metallicity. This optimizes our rates by changing the values of selected input physics parameters to increase our predicted NS-NS merger rate (bottom part of \TableRef{tab:detection_rates_beta75}). To further increase the merger-rate density, we apply a local galaxy-density scaling of $0.01$~MW-equivalent galaxies per $\mega{\rm pc}^{3}$ \citep[][and references therein]{aaa+10}. Thus we are able to obtain an ``optimistic'' merger-rate density of about \mbox{$R_{\rm cSFR}\simeq \unit{159}{\yr^{-1}\usk\giga{\rm pc}^{-3}}$}, corresponding to a LIGO-Virgo detection rate of double NS systems of about $\unit{4}{\yr^{-1}}$ at design sensitivity. By applying smaller kicks (\TableRef{tab:kicks}), we increase our predicted merger-rate density of NS-NS binaries, and thus their detection rates, by an additional factor of a few. We confirmed this by additional simulations combining the ``optimistic'' setting with applying small kicks. 

To conclude, we find that only under rather optimistic circumstances are we able to produce NS-NS merger-rate densities of up to \mbox{$R_{\rm max}\simeq \unit{400}{\yr^{-1}\usk\giga{\rm pc}^{-3}}$} in the local Universe. Although this value is within the error bar of the rate reported by \citet{aaa+17c}, we emphasize that the empirical rate of such double NS mergers is so far only based on one GW detection. From our simulations we predict that near-future GW detections, or rather non-detections, of double NS mergers from LIGO-Virgo runs O3 and O4 will decrease the empirical merger-rate density of double NSs to \mbox{$R\simeq \unit{10-400}{\yr^{-1}\usk\giga{\rm pc}^{-3}}$}.

It is also evident from \TableRef{tab:detection_rates_beta75} that assuming $\beta_{\rm min}=0.75$ we expect significantly more detections of BH-NS systems compared to double NS systems, which again illustrates GW170817 as being somewhat unexpected. The detection ratio between these two populations of DCO mergers depends strongly on $\beta_{\rm min}$ (\FigureRef{fig:gwMergerRate_betamin}) and thus statistics from future LIGO-Virgo detections may, in principle, help to constrain $\beta_{\rm min}$ outside the MW. For example, we expect the detection rate of BH-NS and NS-NS systems to be similar within a factor of a few if \mbox{$\beta_{\rm min}=0.50$}. In general, more studies also on Galactic binaries at various evolutionary stages are needed to better constrain the mass transfer efficiency and thus $\beta_{\rm min}$.

We note that also other recent binary population synthesis studies \citep{baa+17,cbkb18} predict NS-NS merger rate densities much smaller than the empirical rate of \mbox{$R=\unit{1540^{+3200}_{-1220}}{\yr^{-1}\usk\giga{\rm pc}^{-3}}$}. However, we stress again that modelling the merger-rate density alone is far from the only success criterion of binary population synthesis. A model should also be able to explain other observable systems during the binary evolution. It is particularly important to be able to reproduce the characteristics of the Galactic population of double NS systems (cf. \SectionRef{sec:orbital_parameters} and \FigureRef{fig:observed_NSNS_orbit_MW_beta75}), since these are {\it clean} systems with very well-determined parameters.

For double BH and BH-NS systems, we note that under the optimistic settings discussed above (and which optimise the merger rate of double NS systems at MW metallicity), we obtain local merger-rate densities of \mbox{$\unit{109}{\yr^{-1}\usk\giga{\rm pc}^{-3}}$} and $\unit{24}{\yr^{-1}\usk\giga{\rm pc}^{-3}}$, respectively. These theoretical values can only be increased by a factor of 2 by applying small kicks (\TableRef{tab:kicks}). Our optimistic merger-rate density for double BH binaries is therefore in good agreement with the current empirical upper limit of \mbox{$R\simeq \unit{213}{\yr^{-1}\usk\giga{\rm pc}^{-3}}$} reported in \citep{aaa+17}. This upper limit might still be biased somewhat by GW150914 and may decrease in LIGO-Virgo observation runs O3 and O4.

%%%%%%%%%%%%%%%%%%%%%%%%%%%%%%%%%%%%%%%%%%%%%%%%%%
\subsubsection{The progenitor system of GW170817}\label{sec:GW170817}
\begin{figure}
  \includegraphics[width=\columnwidth]{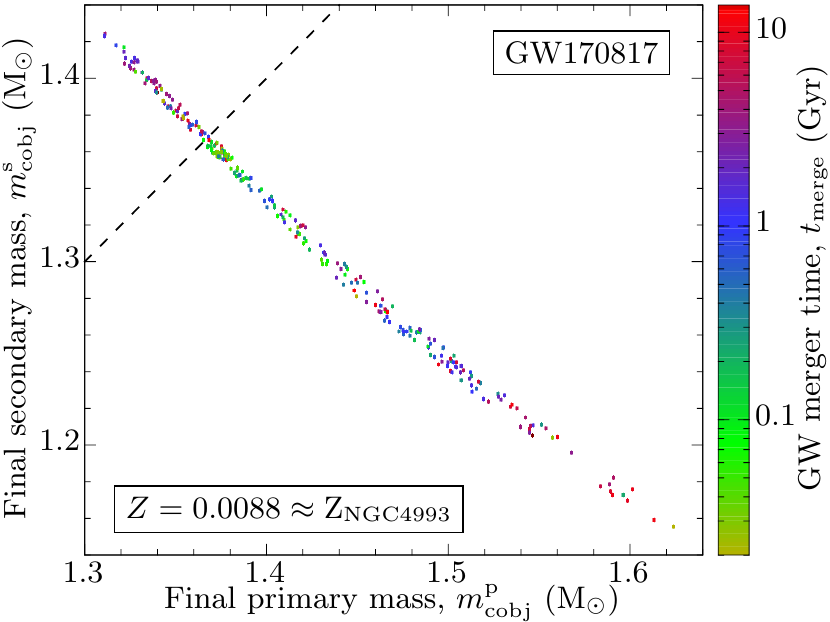}
  \caption{\label{fig:GW170817_progenitors_beta75}NS masses of simulated binary progenitors of GW170817 \citep[][]{aaa+17c}. The merger time (delay time) of the binary mergers is colour coded. The dashed diagonal line indicates equal masses.}
\end{figure}

\begin{figure}
  \includegraphics[width=\columnwidth]{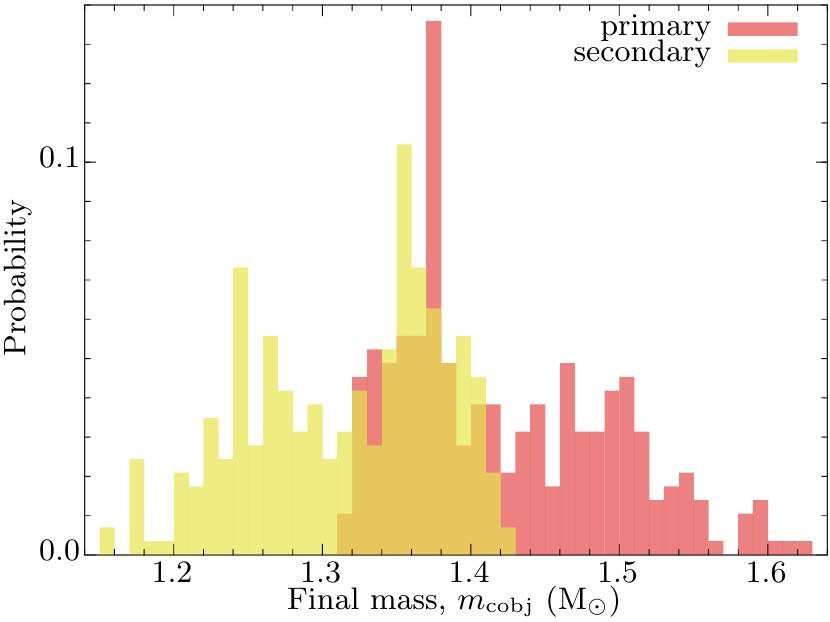}
  \caption{\label{fig:GW170817_masses_beta75}Histograms of our simulated primary (red) and secondary (yellow) NS masses which are solutions to the progenitor binary of GW170817 \citep[][]{aaa+17c}.}
\end{figure}

\begin{figure}
  \includegraphics[width=\columnwidth]{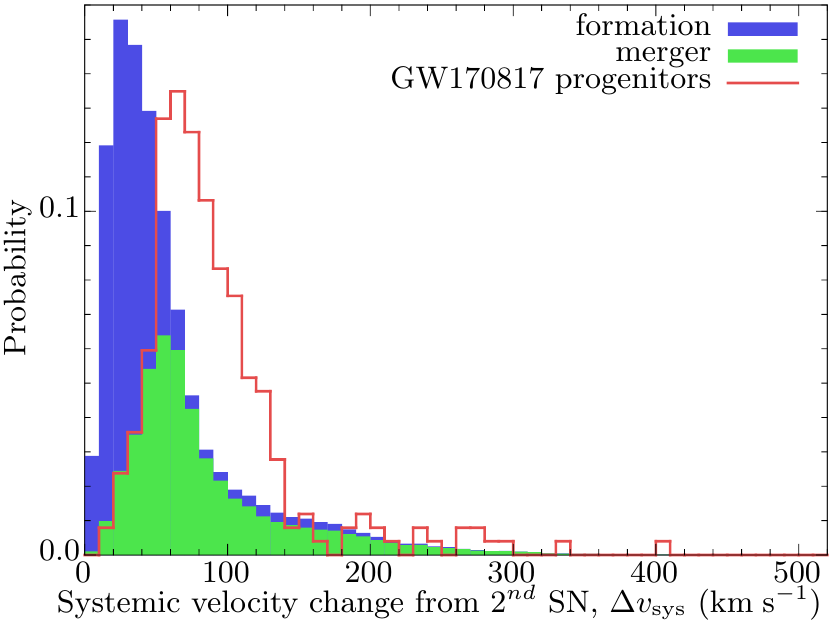}
  \caption{\label{fig:GW170817_kick_beta75}Histograms of our simulated changes of the systemic velocities in double NS systems caused by the kick of the second SN. The blue distribution is for all double NS systems formed, whereas the green distribution is only for those systems which merge within a Hubble time. The red line shows the histogram of possible progenitor binaries of GW170817 \citep[][]{aaa+17c}. The median velocities are about $\unit{44}{\kilo\meter\usk\reciprocal\second}$ (blue), $\unit{67}{\kilo\meter\usk\reciprocal\second}$ (green) and $\unit{79}{\kilo\meter\usk\reciprocal\second}$ (red).}
\end{figure}

The relatively massive S0~galaxy NGC~4993 is identified as the host galaxy of GW170817 \citep[e.g.][]{sha+17,cfk+17}. It is located about $\unit{40}{\mega\rm pc}$ away and has a metallicity between $0.2$ and $\unit{1.0}{{\rm Z}_{\sun}}$, i.e. similar to that of the MW \citep[e.g.][]{iyl+17}. NGC~4993 is less massive than the MW (\mbox{$M\approx\unit{10^{10.5}}{\Msun}$} and \mbox{$M\approx\unit{10^{11.8}}{\Msun}$}, respectively) and shows negligible recent star formation \citep[e.g.][]{pks+17}. Therefore, GW170817 is expected to have an old progenitor system and a delay time of at least a few $\giga\yr$.

In \FigureRef{fig:GW170817_progenitors_beta75}, we show our simulations of double NS mergers in an environment with a MW metallicity of \mbox{$Z=0.0088$}, similar to that of NGC~4993, and which have NS masses identical to those inferred for GW170817. We conclude that we can easily reproduce the progenitor system of GW170817 which had a chirp mass of $\unit{1.188^{+0.004}_{-0.002}}{\Msun}$ and a total mass of $\unit{2.74^{+0.04}_{-0.01}}{\Msun}$ \citep[][]{aaa+17c}. In \FigureRef{fig:GW170817_masses_beta75}, we plot a histogram of the NS component masses of our solutions to the progenitor system of GW170817. These masses are in good agreement with the typical masses of known Galactic double NS systems \citep{tkf+17}.

We find solutions to the double NS progenitor of GW170817 which have ages from less than $\unit{100}{\mega\yr}$ to more than $\unit{10}{\giga\yr}$. This large spread in ages implies that we cannot constrain the age of the progenitor binary of GW170817 and also demonstrates that the delay times of sGRBs span a rather large range. Thus one cannot rule out sGRBs in regions with active star formation less than $\unit{100}{\mega\yr}$ ago. In extreme cases, a merger (and thus a sGRB) may occur less than $\unit{1}{\mega\yr}$ after the formation of a double NS system, see figure~40 in \citet{tkf+17}.

The sGRB associated with GW170817 is located within the effective radius of its host galaxy \citep[e.g.][]{bbf+17}. This is not surprising given that in a massive galaxy like the MW or NGC~4993 the systemic velocities of double NS systems (resulting from NS kicks at SN birth) mainly spread out their distribution rather than ejecting them from their host galaxy. \citet{fb13} found a median sGRB projected physical offset of $\unit{4.5}{\kilo\pc}$. For NGC~4393, the escape velocity at the location of GW170817 is about $\unit{350}{\kilo\meter\usk\reciprocal\second}$ \citep{pks+17}, much larger than the typical systemic velocities we obtain in our simulations.

In our simulated Galactic double NS systems which merge within a Hubble time, we find a spread in typical systemic velocities from a few $\kilo\meter\usk\reciprocal\second$ up to about $\unit{300}{\kilo\meter\usk\reciprocal\second}$ (with a very few systems reaching almost $\unit{500}{\kilo\meter\usk\reciprocal\second}$) with respect to the centre-of-mass reference frame of the binary system prior to the second SN (\FigureRef{fig:GW170817_kick_beta75}). The systemic velocities from the first SN are, in the vast majority, less than $\unit{50}{\kilo\meter\usk\reciprocal\second}$ and the final systemic velocities are therefore strongly dominated by the kinematic effect of the second SN.

\subsubsection{Comparison to the rates of core-collapse supernovae, short \texorpdfstring{$\gamma$}{gamma}--ray bursts and \texorpdfstring{$r$}{r}-process element events}\label{sec:compare_SN-sGRB_rates}
In the following, we compare our simulated Galactic double NS merger rates with estimated constraints on CC~SNe, sGRBs and heavy $r$-process element events.

Our default and ``optimistic'' estimates of a double NS merger rate of $\unit{3.0}{\mega\yr^{-1}}$ and $\unit{14}{\mega\yr^{-1}}$ for a MW-like galaxy translate into relative merger rates of about $3.0\times 10^{-4}$ and $1.4\times 10^{-3}$ per CC~SN, assuming a Galactic CC~SN rate of about $\unit{0.01}{\yr^{-1}}$. These rates can be compared to that of \mbox{$5.0\times 10^{-4}$} to \mbox{$2.0\times 10^{-3}$}, obtained by \citet{bhp16b} based on an analysis of production of heavy $r$-process elements. Hence, we conclude that our simulated double NS merger rates agree with the results of \citet{bhp16b} based on a completely different method.

The local sGRB rate density is estimated to be $4.1^{+2.3}_{-1.9}\product f_{\rm b}^{-1}\unit{}{\yr^{-1}\usk\giga{\rm pc}^{-3}}$ \citep{wp15}, where $f_{\rm b}^{-1}$ is a beaming factor in the range \mbox{$1<f_{\rm b}^{-1}<100$}. Besides double NS mergers, it is expected that also mixed NS/BH mergers produce sGRBs. Adding our simulated mixed NS/BH merger-rate densities to our double NS merger-rate density, we estimate a total of \mbox{$R_{\rm sGRB}\simeq\unit{25\text{ to }86}{\yr^{-1}\usk\giga{\rm pc}^{-3}}$} (at \mbox{$z\simeq 0$}, i.e. using $R_{z=0}$ or $R_{\rm cSFR}$ in \TableRef{tab:detection_rates_beta75}), based on our default values and almost independent on metallicity. This number agrees with that of \citet{wp15} for any beaming factor of \mbox{$f_{\rm b}^{-1}\simeq 4\text{ to }40$}, consistent with observations \citep{mb12,fbmz15}. It should be noted that the luminosity function of sGRBs is fairly unknown and that the sGRB opening angles may well be smaller than estimated in \citet{fbmz15}. These effects would increase the beaming factor.

%%%%%%%%%%%%%%%%%%%%%%%%%%%%%%%%%%%%%%%%%%%%%%%%%%
\subsection{Recycled pulsars orbiting black holes}\label{sec:NSBH_rare}
The detection of a recycled radio pulsar orbiting a BH would be of great interest, for example, to test theories of gravity \citep{wk99,lewk14,ssa+15,ys16}. Unfortunately, the rarest DCO systems produced in our simulations are exactly those mixed binaries where the NS forms before the BH companion. As shown in \TableRef{tab:rate_variations}, however, changing some of the input parameters can boost the number of these otherwise rare systems. The parameters related to the onset and outcome of CE evolution have the biggest influence on the formation rate of such NS-BH systems. The $q_{\rm limit}$ differentiates between stable and unstable mass transfer. The larger this limit, the more progenitor systems transfer mass in a stable manner. This also includes an increase in stable RLO from a naked He-star to its companion. As a result, the number of NS-BH binaries increases by avoiding a second CE~phase after Case~BB RLO which often leads to an early coalescence prior to the formation of a DCO binary. Similarly, an increase of the CE efficiency, $\alpha_{\rm CE}$, and/or the internal energy contribution, $\alpha_{\rm th}$, also helps NSs to successfully eject the envelope of the BH progenitor, thus producing more NS-BH systems.

The number of produced NS-BH systems depends on the accretion efficiency (and thus $\beta_{\rm min}$) in a complex manner (\FigureRef{fig:formationRate_alphaRLO_betamin}). Applying a larger mass-transfer accretion efficiency compared to our default simulation, the number of NS-BH systems reaches a minimum before increasing again. In the most efficient case, more NS-BH systems are formed which would otherwise have become double NS systems, given the enhanced amount of material accreted by the secondary star. Reducing the mass-transfer accretion efficiency creates more NS-BH binaries as well. Because the progenitor of the BH, the initially less massive of the two ZAMS stars, accretes under such inefficient conditions, it can only be slightly less massive than the progenitor of the NS. The reason is that it must accumulate sufficient material to pass the mass threshold for producing a BH. At the same time, the NS is probably relatively massive because its progenitor star is also close to, but below, the BH formation threshold. Both of these mass conditions help the binary survive and eject its CE \citep{ktl+16}.

Smaller SN kicks reduce the number of disrupted binaries and thus increase the formation rate of DCO binaries in general, including NS-BH binaries. Finally, a flatter, or top heavy, IMF increases the number of binaries produced which contain BHs and therefore also the number NS-BH binaries.

Combining our input physics parameters to optimize the formation of NS-BH binaries, we find that the relative number of tight NS-BH binaries (with a mildly recycled pulsar) can reach $\sim 0.01$ times the number of tight double NS systems. Given that the total radio pulsar population of known Galactic double NS systems is anticipated to reach a number between 100 and 200 with the completion of (the full) SKA \citep{kbk+15}, we therefore only expect detection of about one such NS-BH system.

In the above discussion, care must be taken to distinguish between NS-BH systems depending on whether or not the NS is recycled after its formation. Recycling requires accretion and thus only changes in input parameters which promote the survival of the CE evolution, leading to tight binaries and subsequent Case~BB RLO, boost the population of recycled pulsars with BH companions. However, most of the above-mentioned modifications only increase the total number of NS-BH binaries and often only add wide-orbit systems to the sample which will end up containing a non-recycled pulsar anyway (cf. first panels of \FiguresRef{fig:final_orbit_formation_MW_beta75_mixed} and \ref{fig:final_orbit_formation_MW}).

Another issue to be investigated further is that NS-BH progenitor binaries often avoid Case~BB RLO from the massive helium star progenitor of the BH \citep{tlp15}. Hence, the NS only accretes inefficiently from wind accretion during the previous HMXB stage and thus the pulsar will hardly be recycled at all. This should be kept in mind when looking at the number of NS-BH binaries (and double NS systems) expected from our simulations compared to the observed populations, i.e. many of the NS-BH and double NS systems may either be in very wide-orbit systems which never merge in a Hubble time and/or only contain non-recycled radio pulsars. Such non-recycled pulsars are short-lived and their radio emission fades away after typically $\unit{10-50}{\mega\yr}$ \citep{lk04,jk17}, which limits the chances of discovering such a system as a radio pulsar binary. 

%%%%%%%%%%%%%%%%%%%%%%%%%%%%%%%%%%%%%%%%%%%%%%%%%%
\subsection{Alternative formation channels}\label{sec:alternative_scenario}
The first LIGO event GW150914 has been suggested to form following the standard formation scenario by CE evolution \citep{bhbo16}. However, for massive stellar-mass BHs there are other formations channels in which a progenitor binary may evolve to become a tight pair of BHs. The three main formation channels to produce such a BH-BH pair are:
\begin{itemize}
 \item[i)] the CE formation channel (i.e. standard channel) 
 \item[ii)] the chemically homogeneous evolution (CHE) channel with or without a massive overcontact binary (MOB) 
 \item[iii)] the dynamical channel in dense stellar environments 
\end{itemize}
In the following we discuss the latter two alternative formation channels not included in this work.

The dynamical formation channel \citep[e.g.][]{pm00,bbk10,rcr16,crkr17,ban17} produces DCO mergers via three-body and binary-binary encounter interactions in dense stellar clusters and thereby circumvents the need for mass transfer and CE evolution. In analogy to the other production channels mentioned above, the rate of DCO mergers from the dynamical formation channel is also difficult to constrain. Some recent studies predict that this channel might account for the order of only a few percent, possibly less, of all DCO mergers \citep[e.g.][]{bkl14,rcr16,baa+17}.

It has been suggested \citep[e.g.][]{fsm+17,sbm17} that the small effective inspiral spin parameters, $\chi_{\rm eff}$, inferred for the first four LIGO events \citep{aaa+17} could be evidence for isotropic misalignment angles and thus a dynamical formation origin. This reasoning is based on the hypothesis that the standard formation channel, due to mass transfer, leads to aligned spins of the BHs and the orbital angular momentum vector. However, as argued by \citet{tkf+17}, it cannot be ruled out that the spin axis of the collapsing star is tossed in a new, possibly random, direction as a result of the SN, similar to what has been suggested for NSs \citep{sp98,fklk11}. If this is the case, then all past memory from mass transfer is lost. For NSs, there is evidence for such spin axis tossing in both of the two known young pulsars found in double NS systems: PSR~J0737$-$3039 \citep{bkk+08} and J1906+0746 \citep[][in prep.]{des+17}.

An alternative and novel formation channel of relative massive BH-BH binaries, which avoids the CE~phase altogether, is the the CHE channel \citep[e.g.][]{dcl+09,mlp+16,md16b}. In this scenario, the stars avoid the usual strong post-main sequence expansion as a result of effective mixing enforced through the rapidly rotating stars via tidal interactions. Therefore, the CHE scenario works only for massive stars at low metallicity in which strong angular-momentum loss due to stellar winds can be avoided. \citet{mlp+16} presented the first detailed CHE models leading to the formation of BH-BH systems and demonstrated that massive-overcontact binaries are particularly suited for this channel. Very massive stellar-mass BH-BH mergers can form, in agreement with the detections of GW150914, GW170104 and GW170814. Double NS mergers (GW170817) and lower-mass BH-BH mergers like GW151226 ($\unit{14+8}{\Msun}$) and GW170608 ($\unit{12+7}{\Msun}$), however, cannot form in this scenario because the chemical mixing is insufficient in less massive progenitor stars.

Finally, it should be mentioned that in addition a ``double core scenario'' \citep{bro95,bk01,dps06} has been proposed in which CE evolution with a NS is also avoided. In this scenario, two stars with an initial mass ratio close to unity evolve in parallel and reach the giant stage roughly at the same time. Therefore, when the CE forms it will embed both stars in their giant stages, or as a giant star and a helium core, thereby avoiding the formation of a CE with a NS. The double core scenario was originally proposed \citep{che93} at a time when it was thought that a NS in a CE might suffer from hypercritical accretion leading to its collapse into a BH. Thus to explain the observed double NS systems, this alternative scenario without CE evolution was invented. This formation channel, however, only works for the evolution of two stars with a mass ratio close to $1$ and is thus not suited to explain the observations of tight binaries with, for example, a NS orbited by a WD companion \citep{tlk12}.

%%%%%%%%%%%%%%%%%%%%%%%%%%%%%%%%%%%%%%%%%%%%%%%%%%
\section{Summary and conclusions}\label{sec:conclusions}
We have developed a new grid-based binary stellar populations synthesis code, \OurCode. The main motivation is to better understand the formation process of binaries with compact stars and the empirical merger rates reported by LIGO-Virgo based on recent GW detections of colliding double BH and double NS systems. Because the dynamical formation channel is anticipated to produce a minority of the detected DCO mergers, and the CHE channel cannot produce double NSs nor low-mass double BH systems like GW151226 and GW170608, in this work we have investigated the standard (or CE) formation channel (\FigureRef{fig:vdHcartoon}) for the different types of DCO binaries and succeeded in reproducing all GW merger events detected so far.

Our code is based on the earlier work by \citet{vt03} and simulates the evolution of typically one billion binary stars from the ZAMS until two compact objects form. In each system, the two stars are carefully followed in terms of their stellar evolution and mutual interactions via the so-called standard scenario. These interactions include stellar wind mass loss, RLO mass transfer and accretion, CEs and SNe. We apply self-consistent analyses of the binding energies of CEs and implement the results of recent numerical modelling of the subsequent Case~BB RLO with a compact object accretor. In addition, we scale SN kicks according to the stripping of the exploding stars and we show that the simulated merger rates are particularly dependent on the treatment of these three interaction phases.

We demonstrate that all currently detected double BH mergers: GW150914, LVT151012, GW151226, GW170104, GW170608 and GW170814, as well as the recently reported double NS merger GW170817, can be accounted for in our models, depending on the metallicity of the progenitor stars (\FiguresRef{fig:final_chirpmass_GWmerger_beta75} and \ref{fig:LIGO_progenitors_beta75}). For MW-equivalent galaxies \mbox(${\rm Z}_{\rm MW}=0.0088$), and applying default values of our input physics parameters (\TableRef{tab:rate_variations}), we find a double NS merger rate of about $\unit{3.0}{\mega\yr^{-1}}$. A similar merger rate is found for mixed BH/NS binaries, among which we predict very few, systems with a recycled pulsar orbiting a BH. The relative merger rate of double BH systems is lower by an order of magnitude at high metallicity. At low metallicity \mbox(${\rm Z}_{\rm IZw18}=0.0002$), however, we predict the formation of double BH systems with total masses up to $\sim \unit{100}{\Msun}$ and their merger rate is similar to the merger rates of double NSs and mixed BH/NS systems (the latter two of which remain close to the values obtained for a MW-like metallicity).

The corresponding merger-rate densities in the local Universe (\mbox{$z=0$}) for all types of systems combined is about \mbox{$R\simeq\unit{30-120}{\yr^{-1}\usk\giga{\rm pc}^{-3}}$}, depending on the galaxy-density scaling (\TableRef{tab:detection_rates_beta75} and \FigureRef{fig:merger_rate_density_beta75}). We caution that all above-quoted rates are easily changed by more than an order of magnitude when adopting other values for some of the input physics parameters. More specifically, we find an ``optimistic'' merger-rate density for double NS systems of up to \mbox{$R_{\rm max}\simeq\unit{400}{\yr^{-1}\usk\giga{\rm pc}^{-3}}$} when optimizing our input physics parameters within reasonable limits, including the use of relatively small kicks. Our upper limit is still on the lower side compared to the recent empirical double NS merger-rate density of \mbox{$R=\unit{1540^{+3200}_{-1220}}{\yr^{-1}\usk\giga{\rm pc}^{-3}}$} which was recently reported based on the first GW detection of such a system \citep[GW170817,][]{aaa+17c}. Based on our simulations we predict that near-future GW detections (or non-detections) of double NS mergers from LIGO-Virgo runs O3 and O4 will decrease the derived empirical merger-rate density of double NSs to a level of the order \mbox{$R\simeq \unit{10-400}{\yr^{-1}\usk\giga{\rm pc}^{-3}}$}. Such a range also seems in good agreement with comparison to the rates estimated from CC~SNe, sGRBs and the production of heavy $r$-process elements (\SectionRef{sec:comparison_to_other_work}). We predict a NS-NS detection rate of at most 1 to 4 events per year at LIGO-Virgo design sensitivity (\TableRef{tab:detection_rates_beta75}).

Our double BH merger simulations yield local (\mbox{$z=0$}) merger-rate densities spanning the entire interval of \mbox{$R\simeq \unit{0.6-109}{\yr^{-1}\usk\giga{\rm pc}^{-3}}$}, depending on the input physics parameters, the metallicity distribution among the sources and the applied galaxy-density scaling (\TablesRef{tab:rate_variations}, \ref{tab:kicks} and \ref{tab:detection_rates_beta75}). This range is in agreement with the current empirical LIGO-Virgo rate of \mbox{$R=\unit{12-213}{\yr^{-1}\usk\giga{\rm pc}^{-3}}$} \citep{aaa+17}. Finally, for mixed BH/NS binaries we predict a local (\mbox{$z=0$}) merger-rate density \mbox{$R\simeq \unit{13-53}{\yr^{-1}\usk\giga{\rm pc}^{-3}}$}. We cannot lower the large uncertainty intervals in our predicted values due to uncertain input physics. Once the future LIGO-Virgo empirical merger rates converge, we can use these to constrain and gain new insight to the physics of binary evolution \citep[e.g.][]{duvs17,bgn+17}. In addition, the possibility to measure NS spins from the GW signals of double NS mergers \citep{zto+17} will enable us to compare with current models for NS spin and B-field evolution.

Our binary interaction parameters are calibrated to match the observed properties of Galactic double NS systems. Any binary population synthesis on DCO binaries must be able to reproduce the masses and orbital characteristics of binary pulsars (\FigureRef{fig:final_masses_formation_MW_NSNS} top panel and \FigureRef{fig:final_orbit_formation_MW_beta75_NSNS}). To match observational data with our simulations, we generally must invoke a low accretion efficiency during RLO. Only with this assumption are we able to match the distribution of NS masses and orbital parameters simultaneously.

Finally, we find a discrepancy between our simulated distribution of NS masses and those inferred from observations of binary radio pulsars in double NS systems, unless the formation channel of EC~SNe is somehow significantly suppressed compared to that of low-mass FeCC~SNe or EC~SNe produce more massive NSs (about \mbox{$\unit{1.30\text{ to }1.32}{\Msun}$}) than usually thought (\SectionRef{sec:ECvsFECCSN}).

We conclude that the \OurCode binary population synthesis code is working well based on its initial application to GW sources and binary pulsars. The grid interpolation allows a fast and consistent use of data from detailed and up-to-date stellar evolution models. Other applications with this code are planned, see also the recent publication on the formation of WD-NS binaries in which the WD forms first \citep{nkt+18}.

%%%%%%%%%%%%%%%%%%%%%%%%%%%%%%%%%%%%%%%%%%%%%%%%%%
\section*{Acknowledgements}
We are indebted to the referee for the many sound and critical comments which significantly improved the reading of this manuscript and stimulated thoughts on future improvements on the code. We thank Philipp Podsiadlowski, Pablo Marchant, Ilya Mandel, Norbert Wex, Sambaran Bannerjee for discussions and, not least, Chris Belczynski for many useful and detailed comments. MUK acknowledges financial support by the DFG Grant: TA 964/1-1 awarded to TMT. MK acknowledges financial support by the European Research Council for the ERC Synergy Grant BlackHoleCam under contract no.~610058. RGI thanks the STFC for funding his Rutherford fellowship under grant ST/L003910/1.

\bibliographystyle{mnras}
\bibliography{kruckow_refs.bib}

\clearpage
\appendix

%%%%%%%%%%%%%%%%%%%%%%%%%%%%%%%%%%%%%%%%%%%%%%%%%%
\section{Interpolation of stellar grids}\label{app:stellar_grid}
The stellar grids described in \SectionRef{sec:stellar_grids} are based on evolutionary tracks of stars with a given initial mass and provide full evolutionary data. Each of these tracks have a number of supporting points (grid points), and the two dimensions of each grid are the initial mass and the age of the star evolved.

%%%%%%%%%%%%%%%%%%%%%%%%%%%%%%%%%%%%%%%%%%%%%%%%%%
\subsection{Grid structure and interpolation}\label{app:interpolation}
\begin{table}
 \caption{\label{tab:index}Short notations in \EquationsRef{eq:uppermass} to \eqref{eq:F}.}
 \begin{tabular}{ccc}
  \hline
  index & $i$ & $j$\\
  \hline
  \rule{0pt}{12pt}$_{11}$ & $\up{i}$  & $\up{j}-1$  \\
  \rule{0pt}{12pt}$_{12}$ & $\up{i}$  & $\up{j}$    \\
  \rule{0pt}{12pt}$_{21}$ & $\low{i}$ & $\low{j}-1$ \\
  \rule{0pt}{12pt}$_{22}$ & $\low{i}$ & $\low{j}$   \\
  \hline
  \rule{0pt}{12pt}$\up{~}_{,\rm max}$  & $\up{i}$  & $\maximum{j}(\up{i})$  \\
  \rule{0pt}{12pt}$\low{~}_{,\rm max}$ & $\low{i}$ & $\maximum{j}(\low{i})$ \\
  \hline
 \end{tabular}
\end{table}

The grid points are two-dimensional matrices in each stellar variable: age $t_{i,j}$, mass $m_{i,j}$, core mass $c_{i,j}$, radius $R_{i,j}$, luminosity $L_{i,j}$, effective temperature $T_{{\rm eff,}\,i,j}$ and the structure parameter of the envelope $\lambda _{i,j}$. The index $i$ indicates the initial (ZAMS) mass and the index $j$ expresses the relative time evolved. \OurCode allows for individual sets in \mbox{$j\in\left\lbrace0,1,...,\maximum{j}(i)\right\rbrace$} at a given $i$. Therefore, any track at a given $i$ is scaled by the fraction of total time of the track to interpolate between two neighbouring tracks indicated by $\up{i}$ and \mbox{$\low{i}=\up{i}-1$}. For a given initial mass, $\mini$, there is an upper track and a lower track such that \mbox{$\up{m}_{,0}\geq\mini\geq\low{m}_{,0}$}\footnote{Any variable $x_{i,j}$ at \mbox{$i=\up{i}$} or \mbox{$i=\low{i}$} is shortened to $\up{x}_{,j}$ or $\low{x}_{,j}$, respectively.}. In this way, the neighbouring mass indices $\up{i}$ and $\low{i}$ are determined for a given $\mini$. To interpolate on the grid, a ratio in mass is defined by,
\begin{equation}
  \ratio{m}=\frac{\up{m}_{,0}-\mini}{\up{m}_{,0}-\low{m}_{,0}},
  \label{eq:mratio}
\end{equation}
such that $0\le \ratio{m} \le 1$. As an example, the mass\footnote{In the same way, the other variables (age: $t$; radius: $R$; core mass, $c$; luminosity, $L$; effective temperature, $T_{\rm eff}$; and the structure parameter of the envelope, $\lambda$) are interpolated using the same ratios $\ratio{m}$, $\up{r}$ and $\low{r}$.} at a given time is found by,
\begin{equation}
  m=\up{m}-\ratio{m}\product\left(\up{m}-\low{m}\right),
  \label{eq:currentmass}
\end{equation}
where $\up{m}$ and $\low{m}$ are the masses at the same relative time at the upper and lower track, respectively. The definition of the short notations in the following equations are found in \TableRef{tab:index}. The current masses on the neighbouring tracks are interpolated as,
\begin{equation}
  \up{m}=m_{12}-\up{r}\product\left(m_{12}-m_{11}\right),
  \label{eq:uppermass}
\end{equation}
and
\begin{equation}
  \low{m}=m_{22}-\low{r}\product\left(m_{22}-m_{21}\right),
  \label{eq:lowermass}
\end{equation}
with
\begin{equation}
  \up{r}=\frac{t_{12}-\up{t}}{t_{12}-t_{11}}\qquad\text{and}\qquad\low{r}=\frac{t_{22}-\low{t}}{t_{22}-t_{21}},
  \label{eq:tratio}
\end{equation}
where $t_{12}\geq\up{t}\geq t_{11}$ and $t_{22}\geq\low{t}\geq t_{21}$ holds and determines $\up{j}$ and $\low{j}$. Effectively there is only one time ratio because $\up{r}$ and $\low{r}$ depend on each other via:
\begin{equation}
  \trel=\frac{\up{t}}{\up{t}_{,\rm max}}=\frac{\low{t}}{\low{t}_{,\rm max}}=\frac{t}{\maximum{t}}.
  \label{eq:trel}
\end{equation}
$\maximum{t}$ is the lifetime of the star (from the ZAMS until a compact object is formed) and $\trel$ is the relative age. The two variables $\mini$ and $\trel$ determine a position in the grid and the current values of this position can be calculated with \EquationRef{eq:currentmass} and using \EquationsRef{eq:mratio} to \eqref{eq:trel}.

%%%%%%%%%%%%%%%%%%%%%%%%%%%%%%%%%%%%%%%%%%%%%%%%%%
\subsection{Placing a star on the stellar grid}\label{app:place_into_grid}
In principle, one can replace the two determining dimensions $\mini$ and $\trel$ by other quantities. For example, after mass transfer the two stars evolve differently compared to the evolutionary tracks they followed before the interaction. Therefore, one needs two independent quantities to place them on the new tracks interpolated from the stellar grids after the interaction. \OurCode uses the current stellar mass and the current core mass (\SectionRef{sec:core-envelope_boundary}) of the star as these two quantities. The combination of the two masses gives the amount of already burned material in the core and available fuel for future burning in the envelope. Furthermore, the stellar mass determines how massive a single star progenitor with such a core mass would have been on the ZAMS.

To get the new track interpolated from the stellar grids and the current position in it, a mapping from the current stellar mass, $m$, and the core mass, $c$, to the initial mass, $\mini(m,c)$ and the evolved time, $\trel(m,c)$, is needed. This mapping is not necessarily unique. Therefore, the mapping implemented in \OurCode uses the solution with the shortest age of the star as the accretor is usually a less evolved star. The two dimensions are fixed when $\up{i}$, $\up{j}$, $\low{j}$, $\ratio{m}$, $\up{r}$ and $\low{r}$ are known. The interpolated ratios $\ratio{m}$, $\up{r}$ and $\low{r}$ are limited to be within $\left[0:1\right]$ and this determines the indices $\up{i}$, $\up{j}$ and $\low{j}$.

Computationally, the indices $\up{i}$, $\up{j}$ and $\low{j}$ are fixed first. Solutions are only possible if \mbox{$\max\left\lbrace m_{11},m_{12},m_{21},m_{22}\right\rbrace\geq m\geq\min\left\lbrace m_{11},m_{12},m_{21},m_{22}\right\rbrace$} and \mbox{$\max\left\lbrace c_{11},c_{12},c_{21},c_{22}\right\rbrace\geq c\geq\min\left\lbrace c_{11},c_{12},c_{21},c_{22}\right\rbrace$} are simultaneously fulfilled. The definition of the short notation is given in \TableRef{tab:index}. One can then solve \EquationsRef{eq:mratio} to \eqref{eq:trel} for the stellar mass, $m$, and core mass, $c$, to get the ratios $\ratio{m}$, $\up{r}$ and $\low{r}$. The two solutions are,
\begin{equation}
  \ratio{m}^{\pm} = \frac{-0.5\product (C-B-A)\pm \sqrt{0.25\product (C-B-A)^2-B\product A}}{A},
  \label{eq:r}
\end{equation}
where,
\begin{equation}
 \begin{split}
  A& = (l_{22}-l_{21})\product (c_{11}\product m_{12}-c_{12}\product m_{11})\\
   & + (l_{12}-l_{22})\product (c_{11}\product m_{21}-c_{21}\product m_{11})\\
   & + (l_{22}-l_{11})\product (c_{12}\product m_{21}-c_{21}\product m_{12})\\
   & + (l_{12}-l_{21})\product (c_{22}\product m_{11}-c_{11}\product m_{22})\\
   & + (l_{21}-l_{11})\product (c_{22}\product m_{12}-c_{12}\product m_{22})\\
   & + (l_{12}-l_{11})\product (c_{21}\product m_{22}-c_{22}\product m_{21}),
 \end{split}
  \label{eq:A}
\end{equation}
\begin{equation}
 \begin{split}
  B = (l_{22}-l_{21})\product [c\product (m_{11}-m_{12})+&c_{11}\product (m_{12}-m)\\
                                                        +&c_{12}\product (m-m_{11})],
 \end{split}
  \label{eq:B}
\end{equation}
and
\begin{equation}
 \begin{split}
  C = (l_{12}-l_{11})\product [c\product (m_{21}-m_{22})+&c_{21}\product (m_{22}-m)\\
                                                        +&c_{22}\product (m-m_{21})]
 \end{split}
  \label{eq:C}
\end{equation}
with
\begin{equation}
  l_{i,j}\equiv\frac{t_{i,j}}{t_{i,{\rm max}}}.
  \label{eq:l}
\end{equation}
For any ratio in mass, $r$, one then finds the ratios in time,
\begin{equation}
  \up{r} = \frac{D}{F}
  \label{eq:rup}
\end{equation}
and
\begin{equation}
  \low{r} = \frac{E}{F},
  \label{eq:rlow}
\end{equation}
where,
\begin{equation}
 \begin{split}
  D& = [m-m_{11}\product (1-r)]\product (l_{22}-l_{21})\\
   & + r\product [m_{22}\product (l_{21}-l_{11})-m_{21}\product (l_{22}-l_{11})],
 \end{split}
  \label{eq:D}
\end{equation}
\begin{equation}
 \begin{split}
  E& = (m-m_{21}\product r)\product (l_{12}-l_{11})\\
   & + (1-r)\product [m_{12}\product (l_{11}-l_{21})-m_{11}\product (l_{12}-l_{21})]
 \end{split}
  \label{eq:E}
\end{equation}
and
\begin{equation}
 \begin{split}
  F& = (m_{22}-m_{21})\product r\product (l_{12}-l_{11})\\
   & + (m_{12}-m_{11})\product (1-r)\product (l_{22}-l_{21}).
 \end{split}
  \label{eq:F}
\end{equation}
The solutions for $\ratio{m}$, $\up{r}$ and $\low{r}$ are accepted if all values are in their domain $\left[0:1\right]$. Otherwise the next possible group of $\up{i}$, $\up{j}$, $\low{j}$ is checked.

%%%%%%%%%%%%%%%%%%%%%%%%%%%%%%%%%%%%%%%%%%%%%%%%%%
\section{Efficient mass transfer}\label{app:efficient_mass_transfer}
This section gives additional details of our simulations using efficient mass transfer (i.e. little re-emission of the transferred material reaching the accretor, \mbox{$\beta_{\rm min}=0$}, and thus a high accretion efficiency, \mbox{$\epsilon=0.8$}). All the other input parameters have the same values as in \TableRef{tab:standard} in \SectionRef{sec:results}.

%%%%%%%%%%%%%%%%%%%%%%%%%%%%%%%%%%%%%%%%%%%%%%%%%%
\subsection{Progenitor zero-age main-sequence masses}\label{app:progenitor_ZAMS_masses}
\begin{figure}
  \includegraphics[width=\columnwidth]{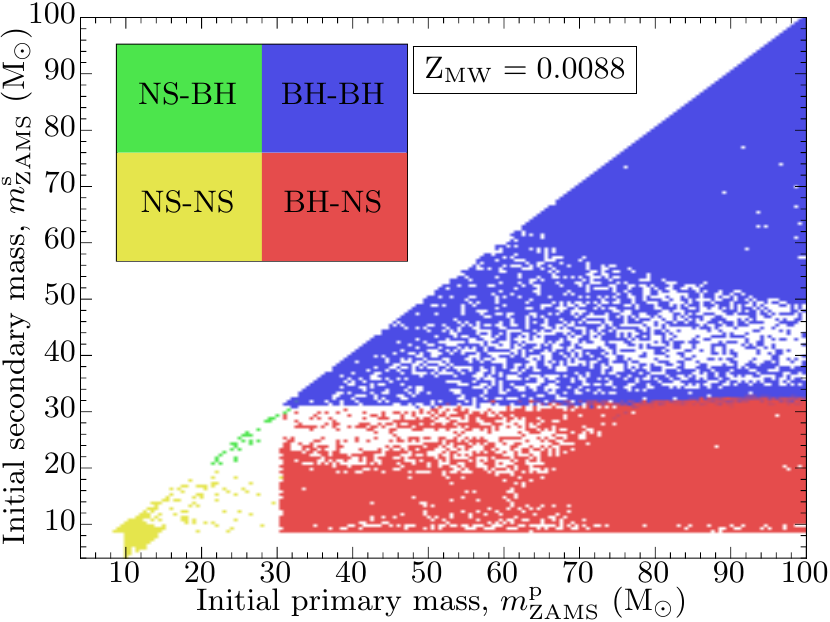}
  \includegraphics[width=\columnwidth]{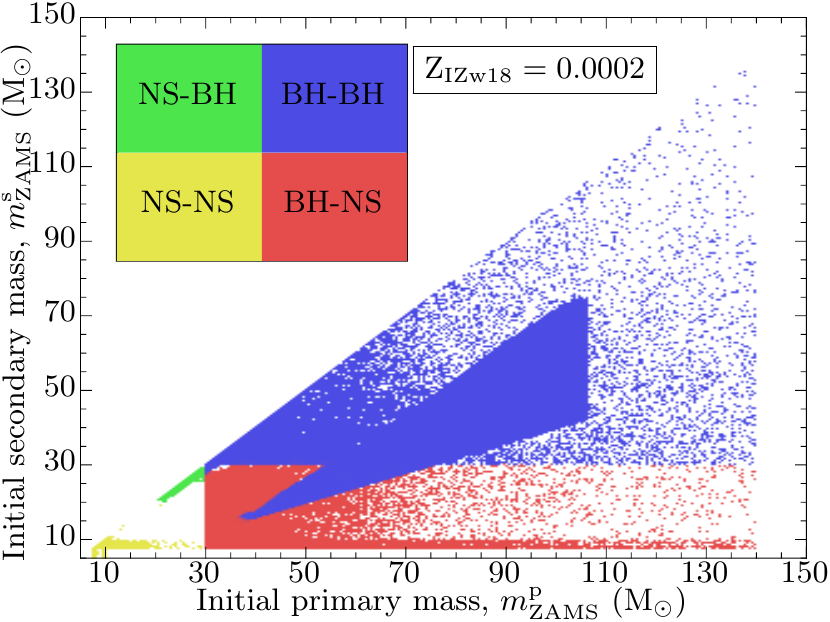}
  \caption{\label{fig:initial_masses}The upper and lower panels are our high- and low-metallicity cases, as in \FigureRef{fig:initial_masses_beta75}, but with efficient mass transfer \mbox{($\beta_{\rm min}=0$)}.}
\end{figure}

\FigureRef{fig:initial_masses} shows the initial (ZAMS) progenitor star masses of the different types of binaries formed in our simulation with efficient mass transfer \mbox{($\beta_{\rm min}=0$)}. The plot looks similar to our simulations with low mass-transfer efficiency although double NS systems are more suppressed at both metallicities, cf. \FigureRef{fig:initial_masses_beta75}. Furthermore, a triangular region of double BH systems appears to penetrate into the BH-NS binaries. This region is shifted in secondary mass compared to \FigureRef{fig:initial_masses_beta75}.

%%%%%%%%%%%%%%%%%%%%%%%%%%%%%%%%%%%%%%%%%%%%%%%%%%
\subsection{Compact object masses}\label{app:compact_object_masses}
\begin{figure}
  \includegraphics[width=\columnwidth]{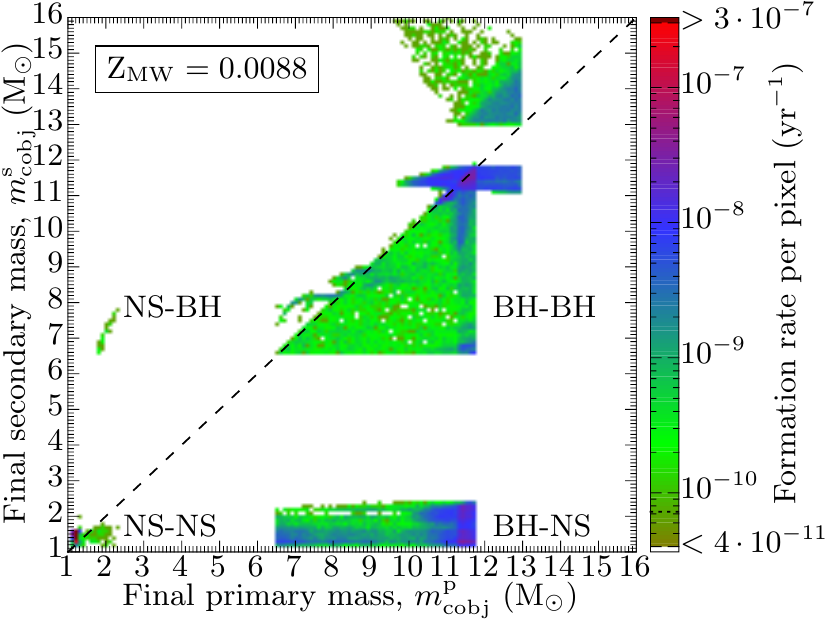}
  \includegraphics[width=\columnwidth]{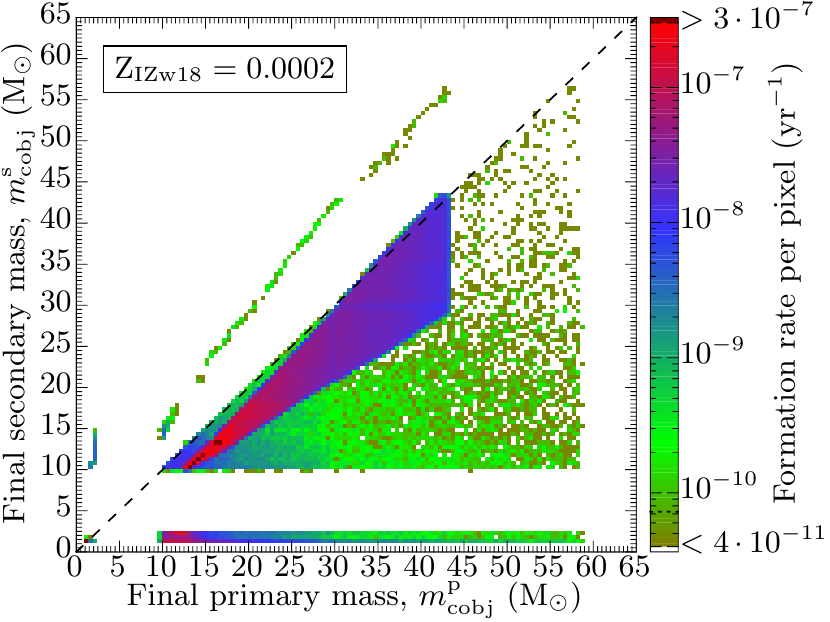}
  \caption{\label{fig:final_masses_formation}The upper and lower panels are our high- and low-metallicity cases, as in \FigureRef{fig:final_masses_formation_beta75}, but with efficient mass transfer \mbox{($\beta_{\rm min}=0$)}.}
\end{figure}

\begin{figure}
  \includegraphics[width=\columnwidth]{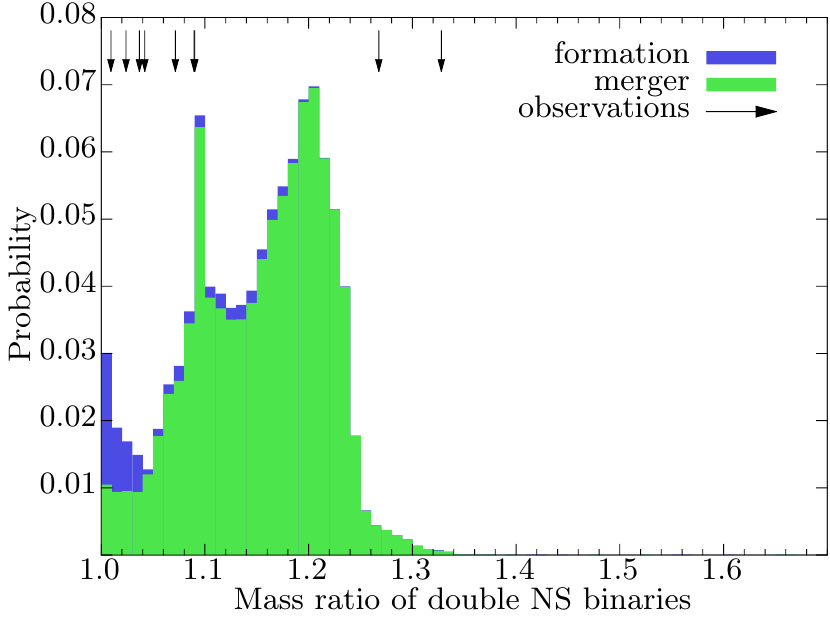}
  \caption{\label{fig:final_massratios_MW_NSNS}As \FigureRef{fig:final_massratios_MW_beta75_NSNS} but with efficient mass transfer \mbox{($\beta_{\rm min}=0$)}.}
\end{figure}

\begin{table}
 \caption{\label{tab:formationrates}Formation rates of DCO binaries in a MW-like galaxy with efficient mass transfer (\mbox{$\beta_{\rm min}=0$}), cf. \TableRef{tab:formationrates_beta75}. The binary types refer to the first and second formed compact object.}
 \begin{tabular}{cr@{$\times$}l<{$\unit{}{\yr^{-1}}$}r@{$\times$}l<{$\unit{}{\yr^{-1}}$}}
  \hline
   Formation rates & \multicolumn{2}{c}{${\rm Z}_{\rm MW}=0.0088$} & \multicolumn{2}{c}{${\rm Z}_{\rm IZw18}=0.0002$}\\
  \hline
  {\normalsize\rule{0pt}{\f@size pt}}NS-NS & $2.27$ & $10^{-6}$ & $7.64$ & $10^{-6}$\\
  {\normalsize\rule{0pt}{\f@size pt}}NS-BH & $2.45$ & $10^{-9}$ & $2.58$ & $10^{-8}$\\
  {\normalsize\rule{0pt}{\f@size pt}}BH-NS & $1.78$ & $10^{-6}$ & $2.51$ & $10^{-6}$\\
  {\normalsize\rule{0pt}{\f@size pt}}BH-BH & $2.64$ & $10^{-6}$ & $3.93$ & $10^{-5}$\\
  \hline
 \end{tabular}
\end{table}

From \FigureRef{fig:final_masses_formation} we see that compared to our standard case (\SectionRef{sec:compact_object_masses}), several things change in the distribution of final compact object masses when applying efficient mass transfer. Mainly the formation rates are smaller (compare \TableRef{tab:formationrates} with \TableRef{tab:formationrates_beta75}).

In the following, we first consider the results of our MW metallicity study. Here, the population of double BH systems with large secondary BH masses is frequent and it is further extended in mass. Therefore, fewer double BH binaries are produced with inverted masses and a mass ratio close to $1$. The region with most massive primary BHs and the least massive secondary BHs disappears. While the NS-BH binaries look similar, the dominant diagonal region of BH-NS systems disappears. The double NS binaries have less massive primary NSs, as shown in \FigureRef{fig:final_masses_formation_MW_NSNS}. This results in a distribution of mass ratios of double NS binaries which disagrees with observations, see \FigureRef{fig:final_massratios_MW_NSNS}.

We now consider our low-metallicity simulations. Secondary BHs become more massive when their progenitors accrete more mass, cf. \FigureRef{fig:final_masses_formation_beta75}. Therefore, a new line of double BH binaries appears where the secondary is more massive than the primary. The DCO systems in which the secondary becomes a NS are less common compared to our default simulation.

%%%%%%%%%%%%%%%%%%%%%%%%%%%%%%%%%%%%%%%%%%%%%%%%%%
\subsection{Orbital parameters}\label{app:orbital_parameters}
\begin{figure}
  \includegraphics[width=\columnwidth]{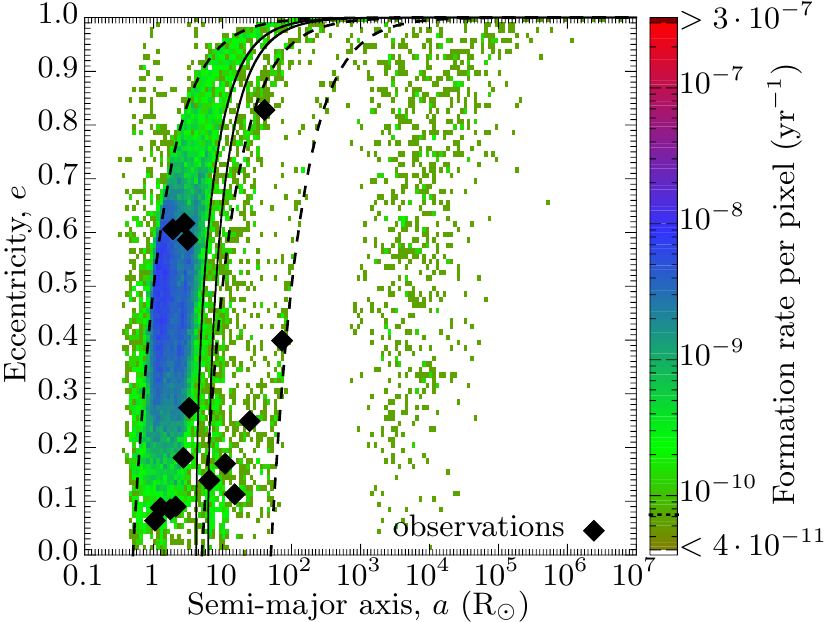}
  \includegraphics[width=\columnwidth]{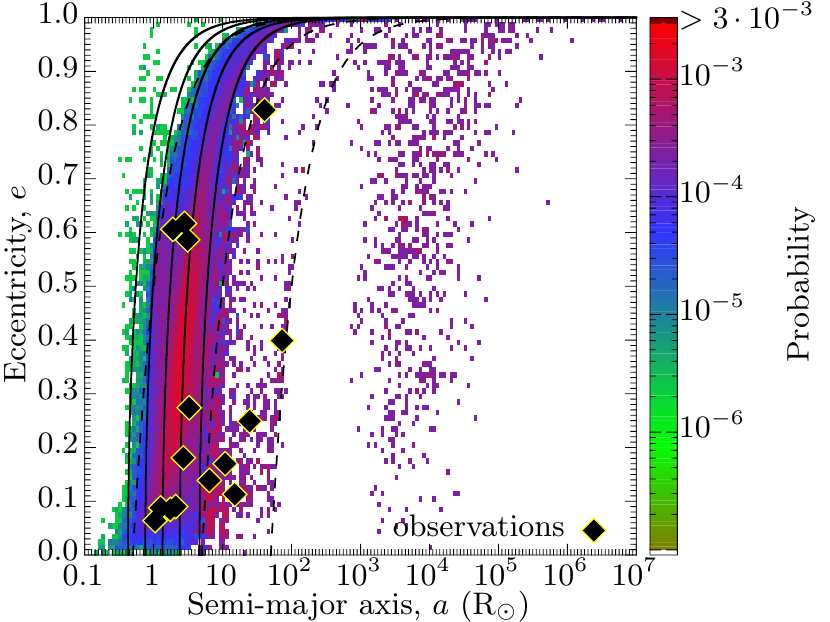}
  \caption{\label{fig:final_orbit_MW_NSNS}The upper panel shows, like \FigureRef{fig:final_orbit_formation_MW_beta75_NSNS}, our double NS binaries at birth with efficient mass transfer \mbox{($\beta_{\rm min}=0$)}. In the lower panel, GW evolution is considered, which changes the observable parameters compared to their values at formation, cf. \FigureRef{fig:observed_NSNS_orbit_MW_beta75}.}
\end{figure}

\begin{figure}
  \includegraphics[width=\columnwidth]{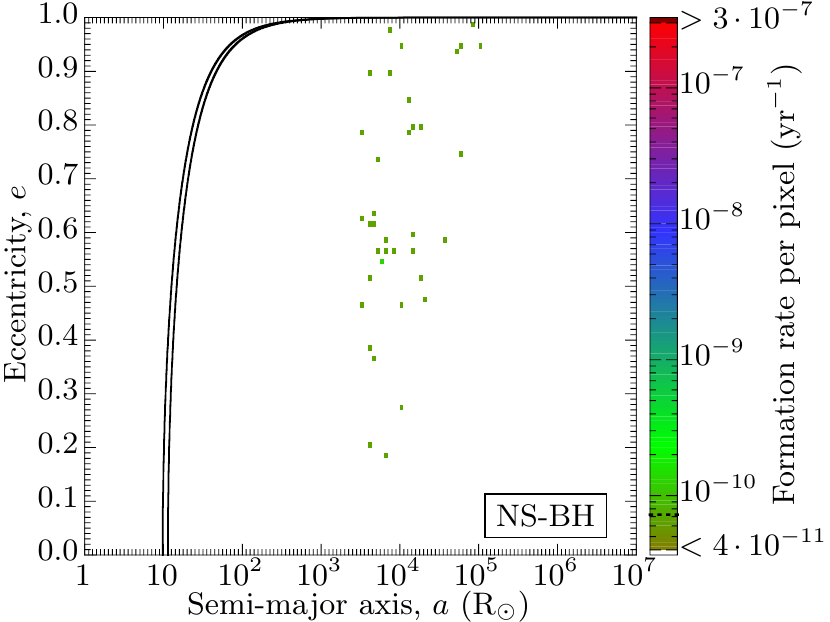}
  \includegraphics[width=\columnwidth]{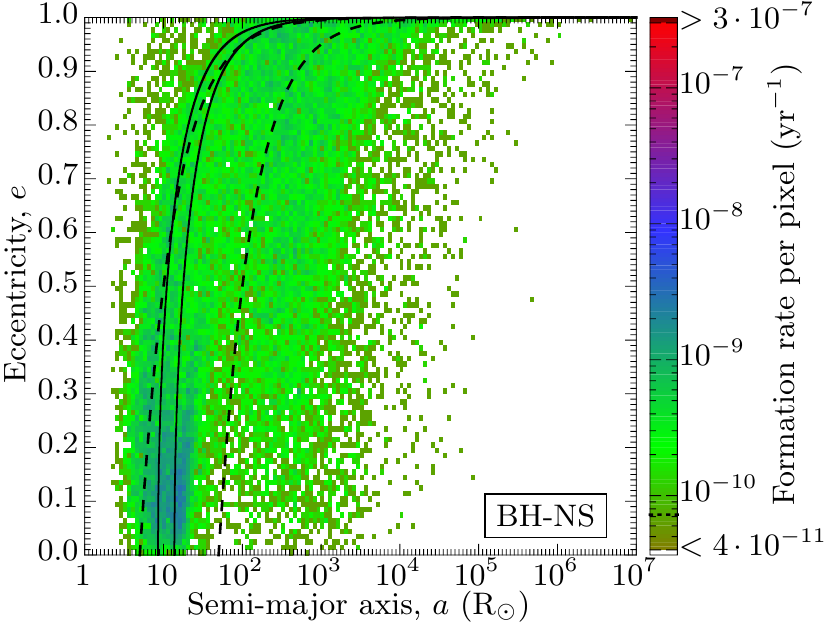}
  \includegraphics[width=\columnwidth]{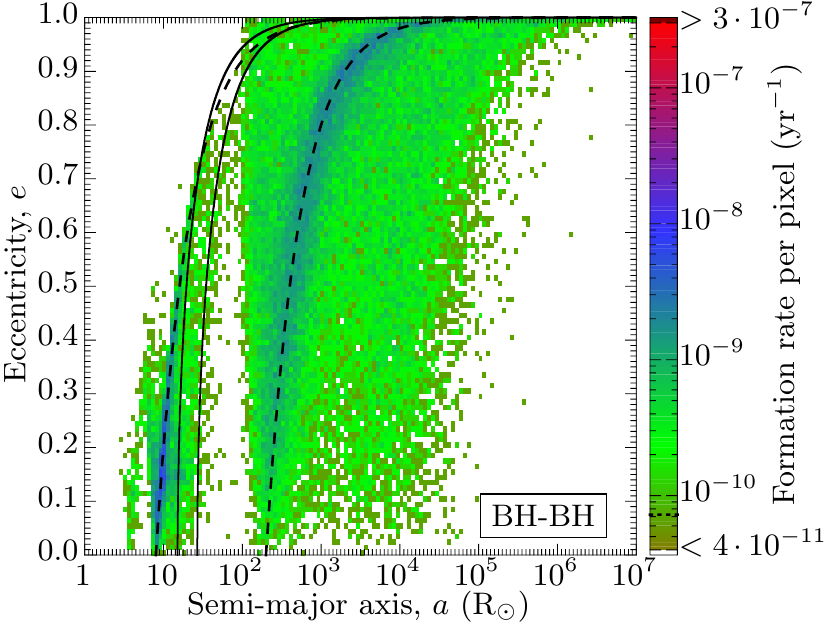}
  \caption{\label{fig:final_orbit_formation_MW}Similar to the upper panel of \FigureRef{fig:final_orbit_formation_MW_beta75_mixed} (NS-BH) and the lower panels of \FigureRef{fig:final_orbit_formation_MW_beta75_mixed} (BH-NS) and \FigureRef{fig:final_orbit_formation_MW_beta75_BHBH} (BH-BH), but here simulated with efficient mass transfer \mbox{($\beta_{\rm min}=0$)}.}
\end{figure}

Simulating the formation of double NS systems at MW metallicity assuming efficient mass transfer (\FigureRef{fig:final_orbit_MW_NSNS} -- instead of an ineffective mass transfer as in \FiguresRef{fig:final_orbit_formation_MW_beta75_NSNS} and \ref{fig:observed_NSNS_orbit_MW_beta75}) leads to a discrepancy compared to observations. Furthermore, in this case fewer double NS systems are produced. By assuming efficient RLO, all systems are a bit wider on average. This makes it more likely to disrupt a given binary by a SN kick. The orbital parameters of the mixed binary types (BH-NS and NS-BH in \FigureRef{fig:final_orbit_formation_MW}) look very similar to the standard case with less efficient RLO. However, because the double BH binaries (\FigureRef{fig:final_orbit_formation_MW}) become more massive in average for efficient RLO, more such systems stay bound after a SN.

%%%%%%%%%%%%%%%%%%%%%%%%%%%%%%%%%%%%%%%%%%%%%%%%%%
\subsection{Gravitational wave-driven merger rates}\label{app:gravitational_wave_radiation_and_merger}
\begin{figure*}
  \framebox{$\mathrm{Z}_\mathrm{MW}=0.0088$}\\
  \includegraphics[width=\columnwidth]{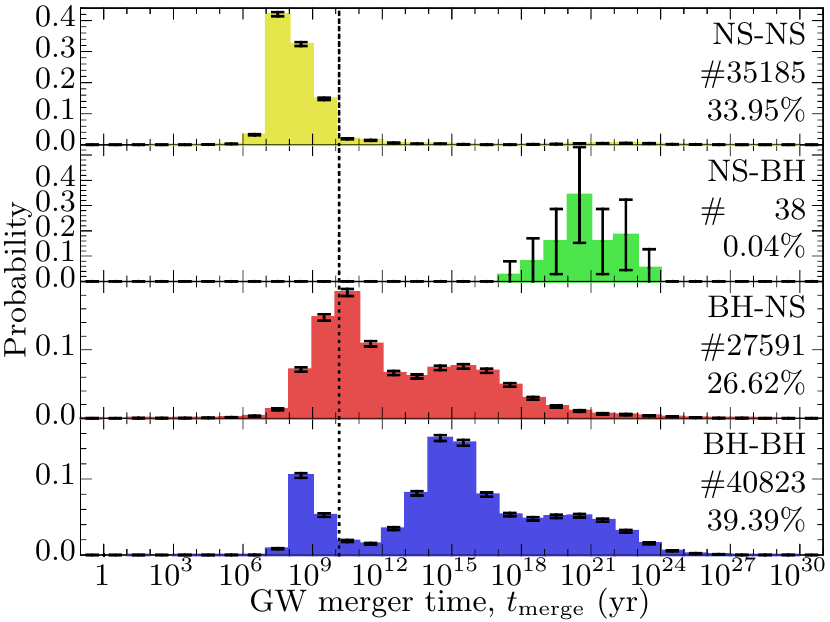}
  \hfill
  \includegraphics[width=\columnwidth]{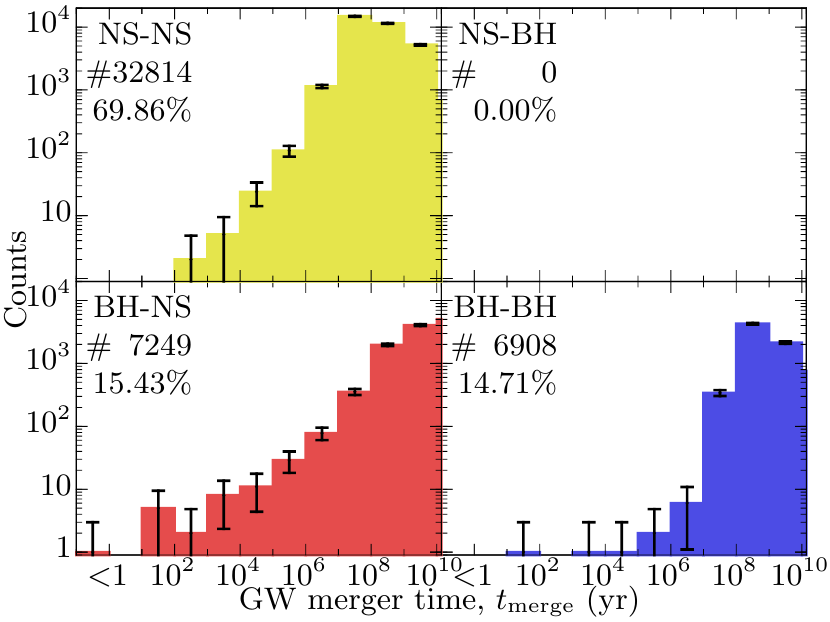}\\
  \framebox{$\mathrm{Z}_\mathrm{IZw18}=0.0002$}\\
  \includegraphics[width=\columnwidth]{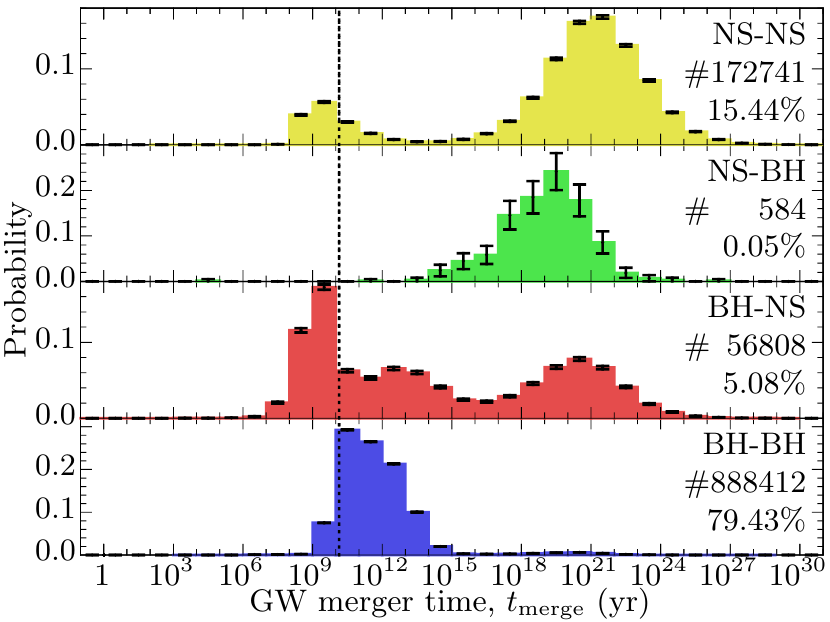}
  \hfill
  \includegraphics[width=\columnwidth]{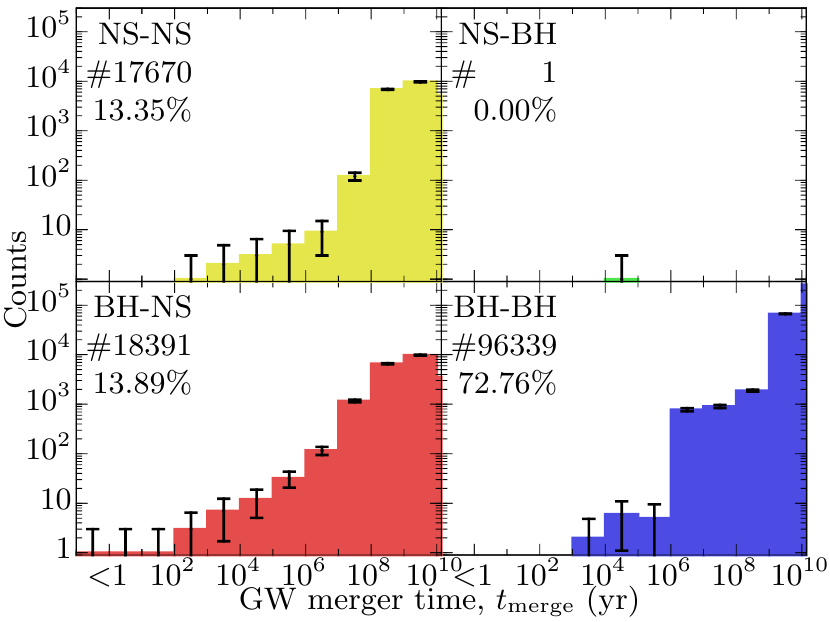}\\
  \caption{\label{fig:Tmerge_LIGO_normalised}As \FigureRef{fig:Tmerge_LIGO_normalised_beta75}, but here with efficient mass transfer \mbox{($\beta_{\rm min}=0$)}.}
\end{figure*}

\begin{table}
 \caption{\label{tab:GWmergerrates}GW merger rates of a MW-like galaxy with efficient mass transfer \mbox{($\beta_{\rm min}=0$)}, cf. \TableRef{tab:GWmergerrates_beta75}. The binary types refer to the first and second formed compact object.}
 \begin{tabular}{cr@{$\times$}l<{$\unit{}{\yr^{-1}}$}r@{$\times$}l<{$\unit{}{\yr^{-1}}$}}
  \hline
  {\normalsize\rule{0pt}{\f@size pt}}GW-merger rates & \multicolumn{2}{c}{${\rm Z}_{\rm MW}=0.0088$} & \multicolumn{2}{c}{${\rm Z}_{\rm IZw18}=0.0002$}\\
  \hline
  {\normalsize\rule{0pt}{\f@size pt}}NS-NS & $2.11^{+0.01}_{-0.02}$ & $10^{-6}$ & $7.37^{+0.44}_{-0.74}$ & $10^{-7}$\\
  {\normalsize\rule{0pt}{\f@size pt}}NS-BH & $0.00^{+0.00}_{-0.00}$ & $10^{0}$ & $4.42^{+0.00}_{-0.00}$ & $10^{-11}$\\
  {\normalsize\rule{0pt}{\f@size pt}}BH-NS & $4.21^{+0.47}_{-0.62}$ & $10^{-7}$ & $7.84^{+0.29}_{-0.58}$ & $10^{-7}$\\
  {\normalsize\rule{0pt}{\f@size pt}}BH-BH & $4.36^{+0.09}_{-0.18}$ & $10^{-7}$ & $3.12^{+1.14}_{-1.27}$ & $10^{-6}$\\
  \hline
 \end{tabular}
\end{table}

Because binaries containing at least one NS are partly suppressed when mass transfer is more efficient, double BH binaries are more frequent in the overall DCO population (\FigureRef{fig:Tmerge_LIGO_normalised}). While double NS binaries still dominate the merger rate in the high-metallicity regime, double BH systems dominate binaries which merge within a Hubble time at low metallicity, cf. \TableRef{tab:GWmergerrates}.

\begin{figure}
  \includegraphics[width=\columnwidth]{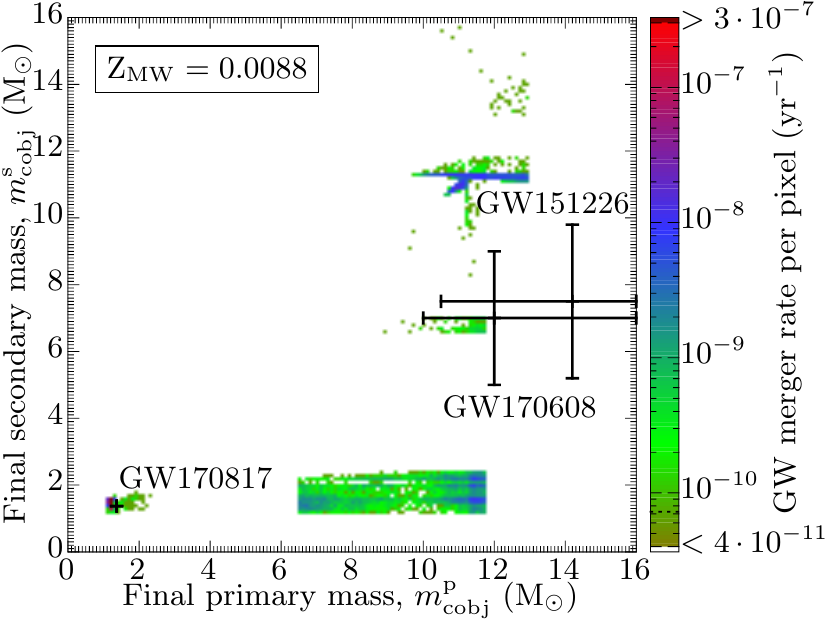}
  \vspace{-\baselineskip}
  \caption{\label{fig:final_masses_GWmerger}Our high-metallicity case as in \FigureRef{fig:final_masses_GWmerger_beta75} but with efficient mass transfer \mbox{($\beta_{\rm min}=0$)}.}
\end{figure}

\begin{figure}
  \includegraphics[width=\columnwidth]{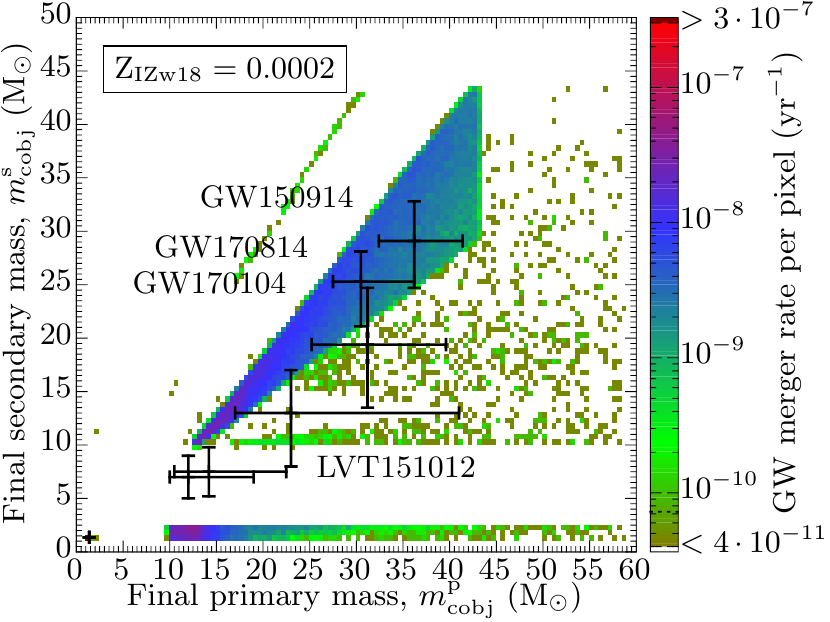}
  \contcaption{\label{fig:final_masses_GWmerger2}. Our low-metallicity case as in \FigureRef{fig:final_masses_GWmerger_beta75} but with efficient mass transfer \mbox{($\beta_{\rm min}=0$)}.}
\end{figure}

\begin{figure}
  \includegraphics[width=\columnwidth]{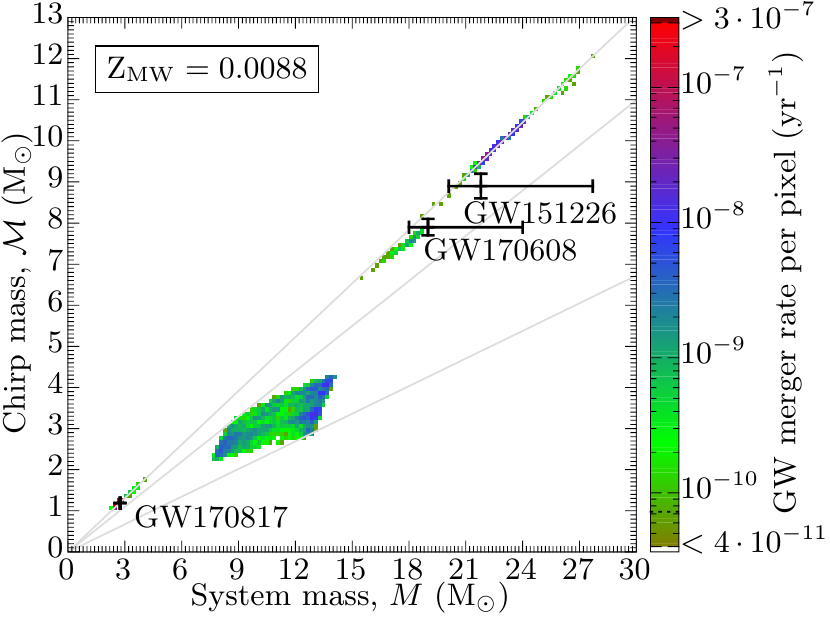}
  \includegraphics[width=\columnwidth]{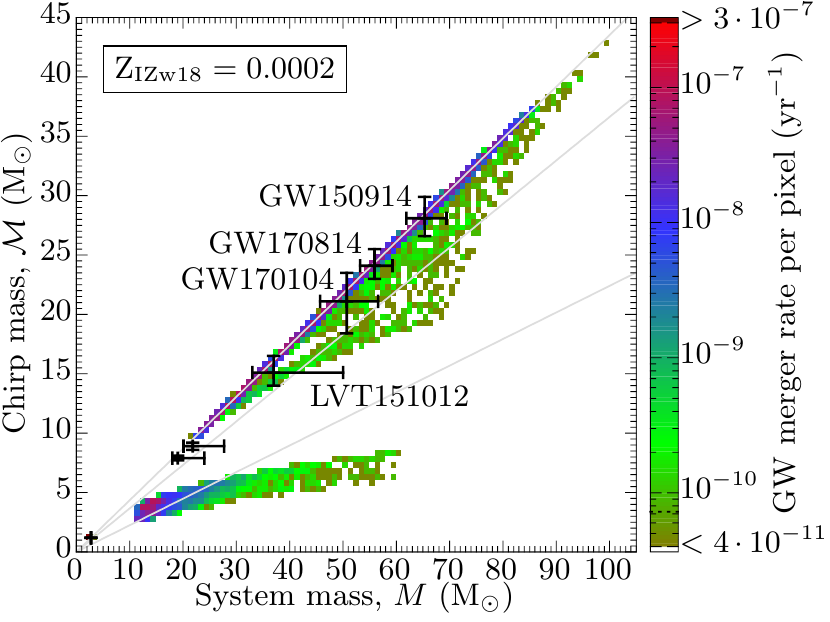}
  \caption{\label{fig:final_chirpmass_GWmerger}The upper and lower panels are our high- and low-metallicity cases, as in \FigureRef{fig:final_chirpmass_GWmerger_beta75}, but with efficient mass transfer \mbox{($\beta_{\rm min}=0$)}.}
\end{figure}

The component masses, total masses and chirp masses of DCO binaries simulated with efficient mass transfer are shown in \FiguresRef{fig:final_masses_GWmerger} and \ref{fig:final_chirpmass_GWmerger}. The distributions look similar to our default case with inefficient mass transfer as discussed in \SectionRef{sec:gravitational_wave_radiation_and_merger}.

%%%%%%%%%%%%%%%%%%%%%%%%%%%%%%%%%%%%%%%%%%%%%%%%%%
\subsection{Merger-rate density}\label{app:merger_rate_density}
\begin{figure}
  \includegraphics[width=\columnwidth]{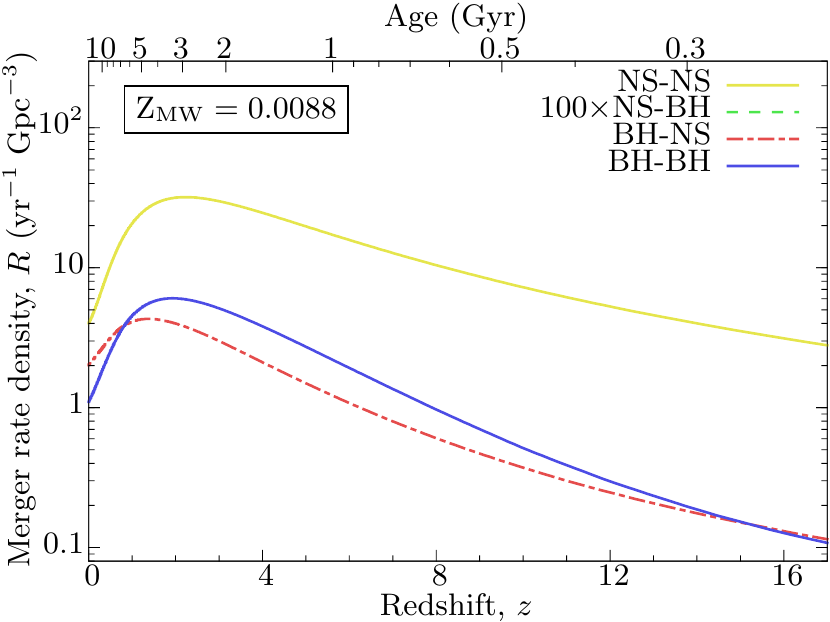}
  \includegraphics[width=\columnwidth]{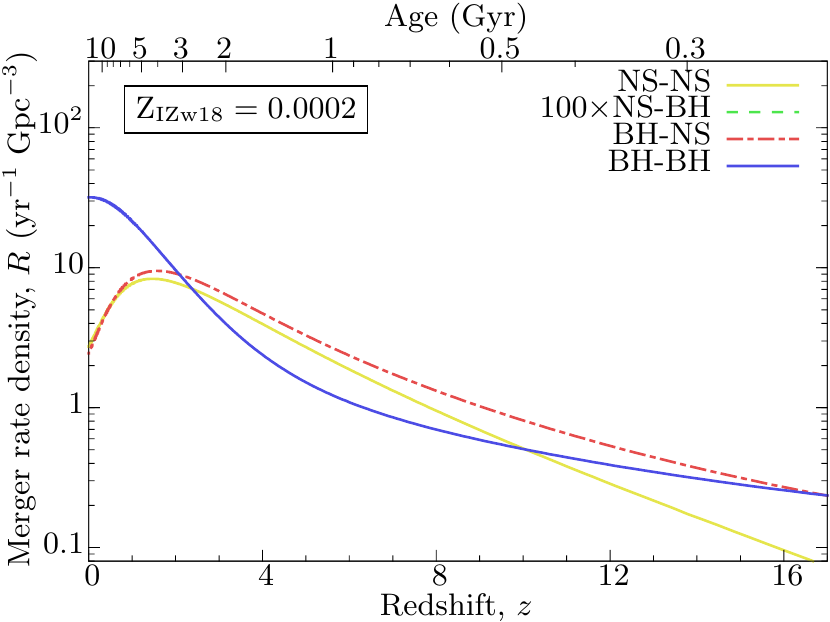}
  \caption{\label{fig:merger_rate_density}The upper and lower panels are our high- and low-metallicity cases, as in \FigureRef{fig:merger_rate_density_beta75}, but with efficient mass transfer \mbox{($\beta_{\rm min}=0$)}.}
\end{figure}

The merger-rate density changes clearly when using a more efficient mass transfer. Systems containing NSs are less common while double BH binaries are more frequent. \FigureRef{fig:merger_rate_density} shows that the influence at low metallicity is even stronger than at MW metallicity. Therefore, further observations of GW mergers in low-metallicity environments could constrain the mass-transfer efficiency there.

\newpage
\begin{table*}
 \caption{\label{tab:detection_rates}Merger-rate densities and detection rates, as in \TableRef{tab:detection_rates_beta75}, but here with efficient mass transfer \mbox{($\beta_{\rm min}=0$)}.}
 \begin{tabular}{cr<{$\unit{}{\Msun^{2.5}}$}|r@{$\times$}l<{$\unit{}{\yr^{-1}\usk\giga{\rm pc}^{-3}}$}r<{$\unit{}{\yr^{-1}}$}|r@{$\times$}l<{$\unit{}{\yr^{-1}\usk\giga{\rm pc}^{-3}}$}r<{$\unit{}{\yr^{-1}}$}}
  \hline
  {\normalsize\rule{0pt}{\f@size pt}}${\rm Z}_{\rm MW}$ & \multicolumn{1}{c|}{$\langle\mathcal{M}^{2.5}\rangle$} & \multicolumn{2}{c}{$R_{z=0}$} & \multicolumn{1}{c|}{$R_{\rm D}$} & \multicolumn{2}{c}{$R_{\rm cSFR}$} & \multicolumn{1}{c}{$R_{\rm D,cSFR}$}\\
  \hline
  {\normalsize\rule{0pt}{\f@size pt}}NS-NS & $1.47$ & $4.03$ & $10^{0}$ & $0.12$ & $2.46$ & $10^{1}$ & $0.75$\\
  {\normalsize\rule{0pt}{\f@size pt}}NS-BH & $17.9\phantom{0}$ & $0.00$ & $10^{0}$ & $0.00$ & $0.00$ & $10^{0}$ & $0.00$\\
  {\normalsize\rule{0pt}{\f@size pt}}BH-NS & $21.3\phantom{0}$ & $2.01$ & $10^{0}$ & $0.89$ & $4.90$ & $10^{0}$ & $2.18$\\
  {\normalsize\rule{0pt}{\f@size pt}}BH-BH & $295\phantom{.00}$ & $1.10$ & $10^{0}$ & $6.77$ & $5.08$ & $10^{0}$ & $31.31$\\
  \hline
  \hline
  {\normalsize\rule{0pt}{\f@size pt}}${\rm Z}_{\rm IZw18}$ & \multicolumn{1}{c|}{$\langle\mathcal{M}^{2.5}\rangle$} & \multicolumn{2}{c}{$R_{z=0}$} & \multicolumn{1}{c|}{$R_{\rm D}$} & \multicolumn{2}{c}{$R_{\rm cSFR}$} & \multicolumn{1}{c}{$R_{\rm D,cSFR}$}\\
  \hline
  {\normalsize\rule{0pt}{\f@size pt}}NS-NS & $1.23$ & $2.73$ & $10^{0}$ & $0.07$ & $8.59$ & $10^{0}$ & $0.22$\\
  {\normalsize\rule{0pt}{\f@size pt}}NS-BH & $34.8\phantom{0}$ & $7.42$ & $10^{-5}$ & $0.00$ & $5.15$ & $10^{-4}$ & $0.00$\\
  {\normalsize\rule{0pt}{\f@size pt}}BH-NS & $29.8\phantom{0}$ & $2.45$ & $10^{0}$ & $1.52$ & $9.13$ & $10^{0}$ & $5.67$\\
  {\normalsize\rule{0pt}{\f@size pt}}BH-BH & $2370\phantom{.00}$ & $3.19$ & $10^{1}$ & $1577.37$ & $3.63$ & $10^{1}$ & $1796.42$\\
  \hline
 \end{tabular}
\end{table*}

~
\newpage
~
\newpage
~
%%%%%%%%%%%%%%%%%%%%%%%%%%%%%%%%%%%%%%%%%%%%%%%%%%
\section{Formation channels}\label{app:formation_channels}
\begin{table*}
 \caption{\label{tab:formationchannel_MW_beta75}Formation channels of DCO binaries (NSs or BHs) in our high-metallicity case \mbox{(${\rm Z}_{\rm MW}=0.0088$)}. The first formed compact object is quoted first, even if its progenitor was the secondary star on the ZAMS. For each kind of binary the relative channel fraction is given and its main channel is marked.}
 \begin{tabular}{R{1.3cm}R{1.75cm}>{\normalsize\rule{0pt}{\f@size pt}}c@{: }L{6.6cm}R{1.15cm}R{1.15cm}R{1.15cm}R{1.15cm}}
  \hline
  relative & GW merger & \multicolumn{2}{l}{channel index:} & \multicolumn{4}{c}{channel fraction of}\\
  frequency & rates ($\mega\yr^{-1}$) & \multicolumn{2}{c}{formation channel} & BH-BH & BH-NS & NS-BH & NS-NS\\
  \hline
  $36.35\%$ & $0.422$ & A & RLO from primary to secondary, primary~SN, RLO from secondary to primary, secondary~SN & $0.53\%$ & $99.47\%$ main ch.&  & \\
  $36.24\%$ & $4.918$ & B & RLO from primary to secondary, primary~SN, CE from secondary, He-RLO from secondary, ultra-stripped secondary~SN &  & $30.27\%$ &  & $69.73\%$ main ch.\\
  \hline
  $ 9.56\%$ & $1.273$ & C & RLO from primary to secondary, primary~SN, CE from secondary, secondary~SN & $7.27\%$ & $86.44\%$ &  & $6.29\%$\\
  $ 3.98\%$ & $0.002$ & D & primary~SN, secondary~SN & $89.37\%$ main ch. & $5.27\%$ &  & $5.35\%$\\
  $ 2.50\%$ & $0.013$ & E & primary~SN, RLO from secondary to primary, secondary~SN & $8.70\%$ & $91.30\%$ &  & \\
  $ 2.49\%$ & $0.327$ & F & primary~SN, CE from secondary, secondary~SN & $7.99\%$ & $91.88\%$ &  & $0.13\%$\\
  $ 2.37\%$ & $0.002$ & G & RLO from primary to secondary, primary~SN, secondary~SN & $70.77\%$ & $0.35\%$ &  & $28.88\%$\\
  $ 2.14\%$ & $0.146$ & H & primary~SN, CE from secondary, He-RLO from secondary, ultra-stripped secondary~SN &  & $90.26\%$ &  & $9.74\%$\\
  $ 1.58\%$ & $0.003$ & I & RLO from primary to secondary, RLO from secondary back to primary, primary~SN, secondary~SN & $98.52\%$ &  & $1.43\%$ main ch. & $0.05\%$\\
  $ 1.23\%$ & $0.143$ & J & RLO from primary to secondary, He-RLO from primary, ultra-stripped primary~SN, CE from secondary, He-RLO from secondary, ultra-stripped secondary~SN &  &  &  & $100.00\%$\\
  \hline
  $ 0.49\%$ & $<0.001$ & K & RLO from secondary to primary, RLO from primary back to secondary, secondary~SN, primary~SN & $99.95\%$ &  & $0.05\%$ & \\
  $ 0.41\%$ & $0.001$ & L & RLO from secondary to primary, primary~SN, secondary~SN & $100.00\%$ &  &  & \\
  $ 0.23\%$ & $<0.001$ & M & RLO from primary to secondary, secondary~SN, primary~SN & $99.65\%$ & $0.35\%$ &  & \\
  $ 0.18\%$ & $0.032$ & N & RLO from secondary to primary, CE from primary, secondary~SN, primary~SN & $100.00\%$ &  &  & \\
  \hline
  $ 0.10\%$ & $0.004$ & O & RLO from primary to secondary, CE from secondary, He-RLO from primary, ultra-stripped primary~SN, He-RLO from secondary, ultra-stripped secondary~SN &  &  &  & $100.00\%$\\
  $ 0.08\%$ & $0.000$ & P & primary~SN, RLO from secondary to primary, He-RLO from secondary, ultra-stripped secondary~SN &  & $100.00\%$ &  & \\
  $ 0.05\%$ & $0.006$ & Q & RLO from primary to secondary, CE from secondary, primary~SN, secondary~SN & $94.41\%$ &  &  & $5.59\%$\\
  $ 0.02\%$ & $<0.001$ & R & RLO from primary to secondary, CE from secondary, He-RLO from primary, ultra-stripped primary~SN, secondary~SN &  &  &  & $100.00\%$\\
  \hline
  $<0.019\%$ & $0.001$ & S & 6 other channels & $>0\%$ & $>0\%$ &  & $>0\%$\\
  \multicolumn{8}{L{17.0cm}}{\begin{itemize}\item RLO from secondary to primary, secondary~SN, primary~SN \item RLO from secondary to primary, CE from primary, primary~SN, secondary~SN \item RLO from primary to secondary, primary~SN, RLO from secondary to primary, He-RLO from secondary, ultra-stripped secondary~SN \item RLO from primary to secondary, He-RLO from primary, primary~SN, secondary~SN \item RLO from primary to secondary, primary~SN, CE from secondary, He-CE from secondary, ultra-stripped secondary~SN \item RLO from secondary to primary, secondary~SN, CE from primary, primary~SN
\end{itemize}}\\
  \hline
 \end{tabular}
\end{table*}

All types of DCO binaries follow different formation channels. The most common ones which produce NSs or BHs at MW metallicity are listed in \TableRef{tab:formationchannel_MW_beta75}.

\newpage
~
\newpage
~
%%%%%%%%%%%%%%%%%%%%%%%%%%%%%%%%%%%%%%%%%%%%%%%%%%
\section{Further parameter variations}\label{app:further_parameter_variations}
\begin{table*}
 \caption{\label{tab:rate_variations2}Formation and merger rates of DCO binaries and the effect of changing the initial ranges of stellar mass or semi-major axis, see also \TableRef{tab:rate_variations}.} 
 \begin{tabular}{r@{ $(\yr^{-1})$ }r@{$\times$}lr@{$\times$}lr@{$\times$}lr@{$\times$}lr@{$\times$}lr@{$\times$}lr@{$\times$}l}
  \hline
  \multicolumn{3}{r}{{\normalsize\rule{0pt}{\f@size pt}}} & \multicolumn{2}{c}{$m^{\rm p}_{\rm min}$} & \multicolumn{2}{c}{$m^{\rm p}_{\rm max}$} & \multicolumn{2}{c}{$m^{\rm s}_{\rm min}$} & \multicolumn{2}{c}{$m^{\rm s}_{\rm max}$} & \multicolumn{2}{c}{$a_{\rm min}$} & \multicolumn{2}{c}{$a_{\rm max}$}\\
  \multicolumn{3}{r}{upper:} & \multicolumn{2}{c}{$\unit{10}{\Msun}$} & \multicolumn{2}{c}{$\unit{150}{\Msun}$} & \multicolumn{2}{c}{$\unit{4.0}{\Msun}$} & \multicolumn{2}{c}{$\unit{150}{\Msun}$} & \multicolumn{2}{c}{$\unit{20.0}{\Rsun}$} & \multicolumn{2}{c}{$\unit{10^5}{\Rsun}$}\\
  \multicolumn{3}{r}{lower:} & \multicolumn{2}{c}{$\unit{\phantom{0}1}{\Msun}$} & \multicolumn{2}{c}{$\unit{\phantom{0}80}{\Msun}$} & \multicolumn{2}{c}{$\unit{0.5}{\Msun}$} & \multicolumn{2}{c}{$\unit{\phantom{0}80}{\Msun}$} & \multicolumn{2}{c}{$\unit{\phantom{0}0.2}{\Rsun}$} & \multicolumn{2}{c}{$\unit{10^3}{\Rsun}$}\\
  \hline
  \multicolumn{1}{r}{{\normalsize\rule{0pt}{\f@size pt}}Formation rates} & \multicolumn{2}{c}{default} & \multicolumn{2}{c}{$m^{\rm p}_{\rm min}$} & \multicolumn{2}{c}{$m^{\rm p}_{\rm max}$} & \multicolumn{2}{c}{$m^{\rm s}_{\rm min}$} & \multicolumn{2}{c}{$m^{\rm s}_{\rm max}$} & \multicolumn{2}{c}{$a_{\rm min}$} & \multicolumn{2}{c}{$a_{\rm max}$}\\
  \hline
  {\normalsize\rule{0pt}{\f@size pt}}NS-NS & $6.81$ & $10^{-6}$ & $^{-9.01}_{+4.42}$ & $10^{-8}$ & $^{+0.06}_{+1.05}$ & $10^{-8}$ & $^{+4.90}_{-0.55}$ & $10^{-6}$ & $^{-1.91}_{-1.71}$ & $10^{-8}$ & $^{+5.00}_{+0.00}$ & $10^{-7}$ & $^{-1.72}_{+2.89}$ & $10^{-6}$\\
  {\normalsize\rule{0pt}{\f@size pt}}NS-BH & $5.49$ & $10^{-9}$ & $^{-2.82}_{+6.69}$ & $10^{-10}$ & $^{+0.21}_{-1.69}$ & $10^{-9}$ & $^{+8.39}_{-5.16}$ & $10^{-10}$ & $^{-1.15}_{-0.45}$ & $10^{-9}$ & $^{-1.36}_{+0.00}$ & $10^{-9}$ & $^{-1.81}_{+4.00}$ & $10^{-9}$\\
  {\normalsize\rule{0pt}{\f@size pt}}BH-NS & $1.49$ & $10^{-5}$ & $^{-0.08}_{-1.02}$ & $10^{-7}$ & $^{+1.07}_{-0.91}$ & $10^{-6}$ & $^{+2.06}_{-0.36}$ & $10^{-6}$ & $^{+9.37}_{+0.30}$ & $10^{-7}$ & $^{-4.62}_{+0.00}$ & $10^{-8}$ & $^{-3.98}_{+1.76}$ & $10^{-6}$\\
  {\normalsize\rule{0pt}{\f@size pt}}BH-BH & $2.27$ & $10^{-6}$ & $^{-1.02}_{-8.86}$ & $10^{-8}$ & $^{+3.75}_{-1.11}$ & $10^{-6}$ & $^{+1.90}_{-0.40}$ & $10^{-7}$ & $^{+4.37}_{-0.40}$ & $10^{-6}$ & $^{-1.47}_{+0.00}$ & $10^{-8}$ & $^{-5.03}_{+1.61}$ & $10^{-7}$\\
  \hline
  \multicolumn{1}{r}{{\normalsize\rule{0pt}{\f@size pt}}GW-merger rates} & \multicolumn{2}{c}{default} & \multicolumn{2}{c}{$m^{\rm p}_{\rm min}$} & \multicolumn{2}{c}{$m^{\rm p}_{\rm max}$} & \multicolumn{2}{c}{$m^{\rm s}_{\rm min}$} & \multicolumn{2}{c}{$m^{\rm s}_{\rm max}$} & \multicolumn{2}{c}{$a_{\rm min}$} & \multicolumn{2}{c}{$a_{\rm max}$}\\
  \hline
  {\normalsize\rule{0pt}{\f@size pt}}NS-NS & $2.98$ & $10^{-6}$ & $^{-6.50}_{+5.28}$ & $10^{-9}$ & $^{+1.70}_{-0.67}$ & $10^{-8}$ & $^{+1.96}_{-0.24}$ & $10^{-6}$ & $^{-3.11}_{-2.08}$ & $10^{-8}$ & $^{+1.91}_{+0.00}$ & $10^{-7}$ & $^{-0.76}_{+1.28}$ & $10^{-6}$\\
  {\normalsize\rule{0pt}{\f@size pt}}NS-BH & $0.00$ & $10^{0}$ & $^{+9.37}_{+0.00}$ & $10^{-11}$ & $^{+0.00}_{+1.93}$ & $10^{-10}$ & $^{+1.29}_{+1.29}$ & $10^{-10}$ & $^{+0.65}_{+1.29}$ & $10^{-10}$ & $^{+6.46}_{+0.00}$ & $10^{-11}$ & $^{+0.65}_{+1.94}$ & $10^{-10}$\\
  {\normalsize\rule{0pt}{\f@size pt}}BH-NS & $4.05$ & $10^{-6}$ & $^{+0.52}_{-6.13}$ & $10^{-8}$ & $^{+1.36}_{-1.61}$ & $10^{-7}$ & $^{+5.85}_{-0.94}$ & $10^{-7}$ & $^{+1.32}_{+0.01}$ & $10^{-7}$ & $^{-1.78}_{+0.00}$ & $10^{-8}$ & $^{-1.05}_{-1.79}$ & $10^{-6}$\\
  {\normalsize\rule{0pt}{\f@size pt}}BH-BH & $2.64$ & $10^{-7}$ & $^{+1.56}_{+3.37}$ & $10^{-9}$ & $^{+3.93}_{-1.96}$ & $10^{-7}$ & $^{+2.25}_{-0.16}$ & $10^{-8}$ & $^{+3.86}_{+0.12}$ & $10^{-7}$ & $^{-1.29}_{+0.00}$ & $10^{-9}$ & $^{-0.77}_{+1.52}$ & $10^{-7}$\\
  \hline
 \end{tabular}
\end{table*}

\TableRef{tab:rate_variations2} shows the changes in formation and merger rates when varying the ranges of the initial stellar masses and the semi-major axis.

%%%%%%%%%%%%%%%%%%%%%%%%%%%%%%%%%%%%%%%%%%%%%%%%%%
\subsection{Iron-core collapse supernovae}\label{app:FeCCSN}
\begin{figure}
  \includegraphics[width=\columnwidth]{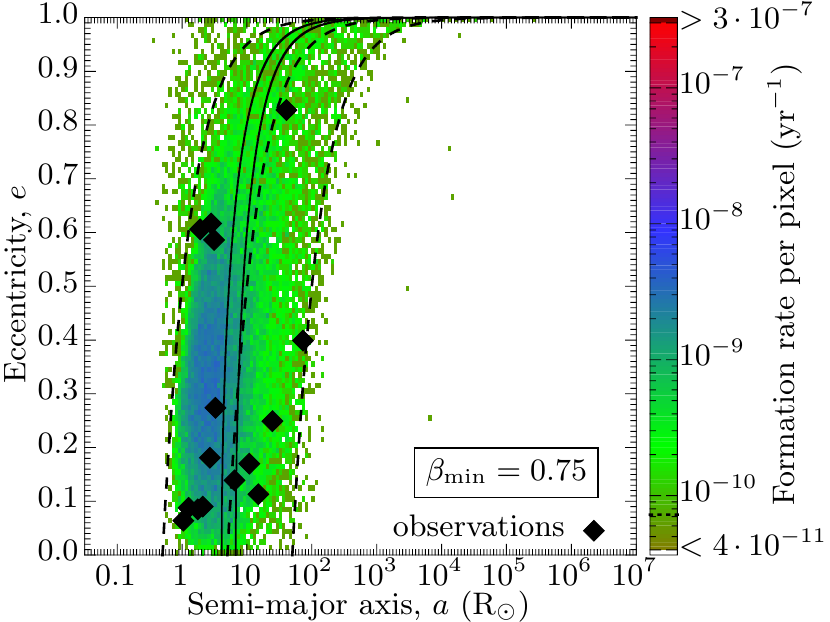}
  \includegraphics[width=\columnwidth]{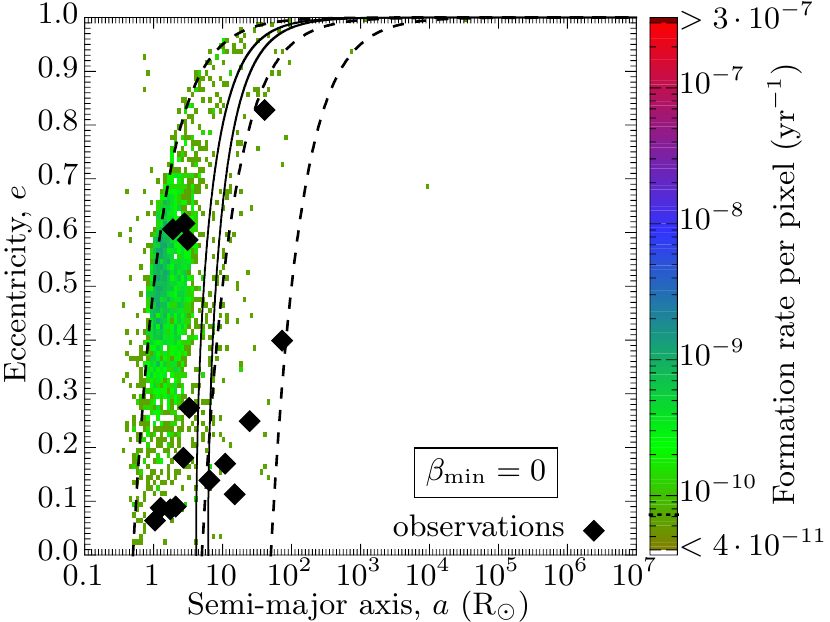}
  \caption{\label{fig:final_orbit_formation_MW_NSNS_FeCC}Our simulated orbital parameters when both NSs form by FeCC~SNe. The upper and lower plots show inefficient and efficient mass transfer, respectively (cf. \FigureRef{fig:final_orbit_formation_MW_beta75_NSNS} and the upper panel of \FigureRef{fig:final_orbit_MW_NSNS}).}
\end{figure}

There is no evidence that any observed double NS binaries contain NSs which are produced by EC~SNe. \FigureRef{fig:final_orbit_formation_MW_NSNS_FeCC} shows our simulated orbital parameters of systems in which both stars undergo FeCC~SN. It is even clear that it is necessary to assume inefficient mass transfer to explain the observational data, cf. \FigureRef{fig:final_orbit_formation_MW_beta75_NSNS} and the upper panel of \FigureRef{fig:final_orbit_MW_NSNS}.

%%%%%%%%%%%%%%%%%%%%%%%%%%%%%%%%%%%%%%%%%%%%%%%%%%
\subsection{Mass-transfer efficiency}\label{app:mass-transfer_efficiency}
\begin{figure}
  \includegraphics[width=\columnwidth]{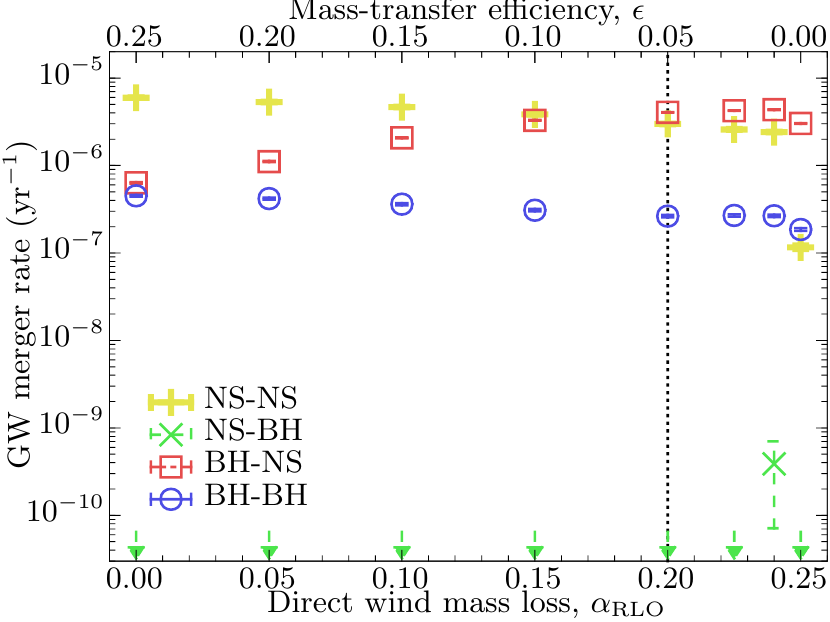}
  \caption{\label{fig:gwMergerRate_alphaRLO}As \FigureRef{fig:gwMergerRate_betamin}, but instead showing the dependence on the direct wind mass loss parameter during RLO.}
\end{figure}

\begin{figure}
  \includegraphics[width=\columnwidth]{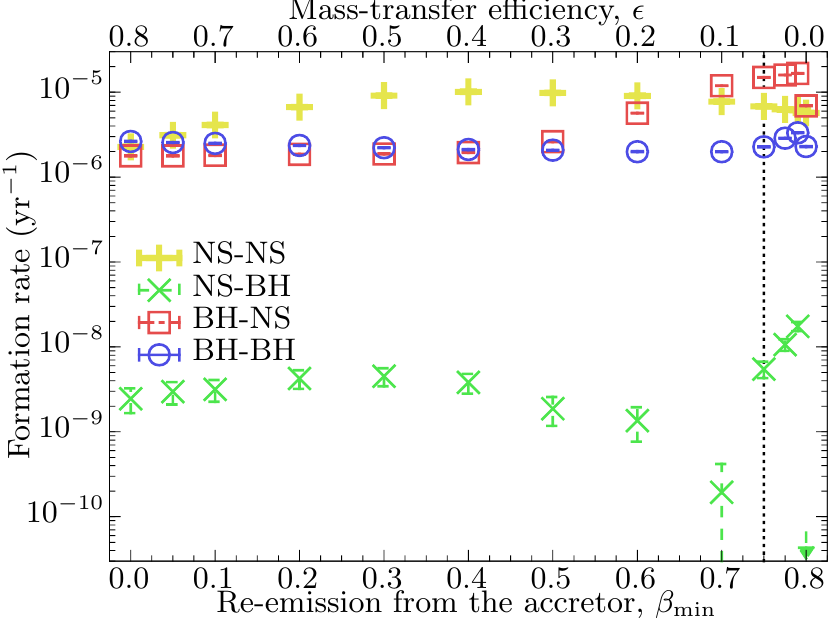}
  \includegraphics[width=\columnwidth]{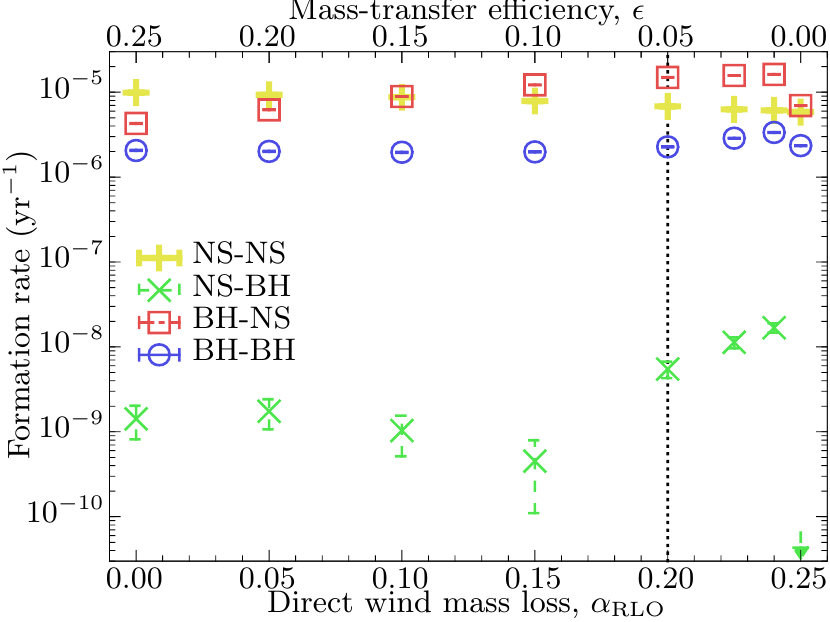}
  \caption{\label{fig:formationRate_alphaRLO_betamin}Formation rate of DCO binaries in a MW-like galaxy as a function of the re-emission fraction in the vicinity of the accretor, $\beta_{\rm min}$, (upper panel) and the direct wind mass loss parameter, $\alpha_{\rm RLO}$, (lower panel) during RLO. The dotted lines mark our default values of $\beta_{\rm min}$ and $\alpha _{\rm RLO}$, respectively.}
\end{figure}

The influences on the GW merger rate of the parameters $\beta_{\rm min}$ and $\alpha _{\rm RLO}$ are similar (cf. \FiguresRef{fig:gwMergerRate_alphaRLO} and \ref{fig:gwMergerRate_betamin}). The same holds true when comparing the formation rates in the two panels of \FigureRef{fig:formationRate_alphaRLO_betamin}. Both parameters change the mass-transfer efficiency, $\epsilon$, by ejecting more or less material during stable RLO. The only difference is the specific angular momentum which is carried away by the ejecta.

%%%%%%%%%%%%%%%%%%%%%%%%%%%%%%%%%%%%%%%%%%%%%%%%%%
\subsection{Internal energy parameter, \texorpdfstring{$\alpha_{\rm th}$}{alpha\_th}}\label{app:alpha_th}
\begin{figure}
  \includegraphics[width=\columnwidth]{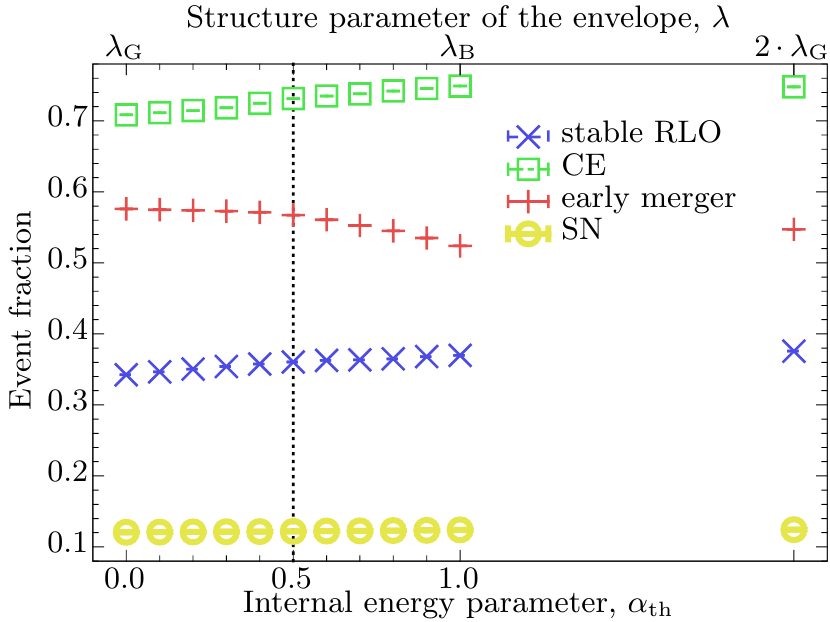}
  \caption{\label{fig:events_alphaTH_normalized}Mean counts of evolutionary events from relatively massive binaries modelled with \OurCode at MW metallicity as a function of the fraction of the internal envelope energy, $\alpha_{\rm th}$, included in the total binding energy of the envelope. The vertical dotted line marks our default value. The symbols to the very right \mbox{($\lambda=2\product\lambda_{\rm G}$)} are from a simulation in which it is assumed that the stellar envelopes are in virial equilibrium.}
\end{figure}

When more internal energy is included in determining the total binding energy of the envelope (i.e. larger value of $\alpha_{\rm th}$, cf. \EquationRef{eq:Ebind}) more systems survive the CE~phase. This naturally increases the number of subsequent evolutionary phases (\FigureRef{fig:events_alphaTH_normalized}) and thus the number of final mergers. Assuming the envelope is in virial equilibrium, i.e. the total binding energy is half the gravitational binding energy \mbox{($\lambda=2\product\lambda_{\rm G}$)}, we find similar results as with $\alpha_{\rm th}$ between $0.7$ and $0.8$.

\label{lastpage}
\end{document}